\documentclass[%
 sor,
 amsmath,amssymb,
 preprint,
 onecolumn,
]{revtex4-2}

\usepackage{graphicx,color}% Include figure files
\usepackage{dcolumn}% Align table columns on decimal point
\usepackage{bm}% bold math
\begin{document}

\title{The distorting lens of human mobility data
}

\author{Riccardo Gallotti}
 \email{rgallotti@fbk.eu}
 \affiliation{CoMuNe Lab, Fondazione Bruno Kessler, Via Sommarive 18 38123 Povo (TN), Italy}

\author{Davide Maniscalco}
\affiliation{CoMuNe Lab, Fondazione Bruno Kessler, Via Sommarive 18 38123 Povo (TN), Italy}

\author{Marc Barthelemy}
\affiliation{Université Paris-Saclay, CNRS, CEA, Institut de Physique Théorique, 91191, Gif-sur-Yvette, France}
\affiliation{CAMS (CNRS/EHESS) 190-198, avenue de France, 75244 Paris Cedex 13, France}

\author{Manlio De Domenico}
\affiliation{CoMuNe Lab, Fondazione Bruno Kessler, Via Sommarive 18 38123 Povo (TN), Italy}
\affiliation{Department of Physics and Astronomy ``Galileo Galilei'', University of Padua, Italy}

%%%%%%%%%%%%%%%%%%%%%%%%%%%%%%%%%%%%%%%%%%%%%%%%%%%%%
\begin{abstract}
The description of complex human mobility patterns is at the core of many important applications ranging from urbanism and transportation to epidemics containment. Data about collective human movements, once scarce, has become widely available thanks to new sources such as Phone CDR, GPS devices, or Smartphone apps. Nevertheless, it is still common to rely on a single dataset by implicitly assuming that it is a valid instance of universal dynamics, regardless of factors such as data gathering and processing techniques.
Here, we test such an overarching assumption on an unprecedented scale by comparing human mobility datasets obtained from $7$ different data-sources, tracing over $500$ millions individuals in $145$ countries. We report wide quantifiable differences in the resulting mobility networks and, in particular, in the displacement distribution previously thought to be universal. These variations -- that do not necessarily imply that the human mobility is not universal -- also impact processes taking place on these networks, as we show for the specific case of epidemic spreading. Our results point to the crucial need for disclosing the data processing and, overall, to follow good practices to ensure the robustness and the reproducibility of the results.
\end{abstract}

\maketitle

%%%%%%%%%%%%%%%%%%%%%%%%%%%%%%%%%%%%%%%%%%%%%%%%%%%%%
\section*{Introduction}
%%%%%%%%%%%%%%%%%%%%%%%%%%%%%%%%%%%%%%%%%%%%%%%%%%%%%

In the past two decades, access to the rich high-resolution datasets was exclusive to a rather limited research community, which nevertheless produced a series of important results about universal features of mobility behavior~\cite{gonzalez2008understanding, song2010modelling, pappalardo2015returners, gallotti2016stochastic, alessandretti2020scales, moro2021mobility, schlapfer2021universal}. The potential impact of human mobility studies is enormous, as exemplified by most COVID-19 studies which rely heavily on mobility data~\cite{buckee2020aggregated,nanni2021give}. The recent need for tackling promptly the problems associated with this pandemics both from the epidemiological and socio-economical perspective has been answered by many IT companies, which activated a number of `Data For Good' programs and collaborations. This unprecedented access to mobility data by a large number of research groups led to the publication of a series of important results about human behavior and, more recently, dynamics of interests such as the spreading of the COVID-19~\cite{bonaccorsi2020economic,rader2020crowding,kissler2020reductions,Hazarie:2021jc,Lemey:2021ch,chang2021mobility,mena2021socioeconomic,mazzoli2021interplay}. However, even if this sudden abundance of dataset provided scientists with many research opportunities, it also comes with new challenges. Without any doubt, an easier access to data favored the possibility of performing innovative data-informed analysis, in some cases across multiple countries and with a high temporal resolution. These recent studies also allowed in many cases their full reproducibility by providing their codes and datasets. However, for companies to be able to provide large amounts of data, in almost real time, they had to overcome multiple limitations, and since the methods used are proprietary, the data pipeline cannot be openly released. As a consequence, important details might not be disclosed, often leaving to the final user the task to use data under uncertain, or even unknown, gathering methodology. Consequently, the lack of full knowledge about the measurement and processing details might prevent scientists from correctly interpreting the outcome of their analysis. Moreover, not being able to be in control of the data elaboration from the raw to their final, analyzable, form, limits the ability of researchers to tailor the data to the research question they have in mind. Even more importantly, each data provider extracts information using a different gathering and processing methodology, raising the question to which extent different data sets are in agreement with each other when used to model human mobility and derived dynamical processes, such as traffic or epidemics. The implicit assumption made is that the characteristics of the shared data is marginally affected by the details of the underlying methodology. Such an assumption is often explicitly acknowledged as a study limitation~\cite{ruktanonchai2020assessing,Kraemer:2020hy, rader2020crowding}, but it has been never systematically tested or verified, to date, and consequently, the extent of these limitations underlying our understanding of human mobility and their impact on modeling dynamics with high societal impact are essentially unknown.

Here, we illustrate both the opportunities provided and the limitations faced while studying mobility data coming from seven different providers (including both openly accessible and closed data): Google, Facebook, GPS based applications (Cuebiq, Safegraph), Mobile phones, vehicle GPS blackboxes and public census. We first provide a short description of these datasets and more details and discussions  -- in particular their limitations -- can be found in the Methods section.

%%%%%%%%%%%%%%%%%%%%%%%%%%%%%%%%%%%%%%%%%%%%%%%%%%%%%
\section*{Results}
%%%%%%%%%%%%%%%%%%%%%%%%%%%%%%%%%%%%%%%%%%%%%%%%%%%%%

%%%%%%%%%%%%%%%%%%%%%%%%%%%%%%%%%%%%%%%%%%%%%%%%%%%%%
\subsection*{Overview of the data}

We considered a diverse set of mobility datasets, accounting for the movements of over 500 millions individual in 145 countries. Specifically, we used census data which provide statistical information about the urban and inter-urban commuters' flows. We collected and analyzed official census data for the US at the county level~\cite{uscensus} and for Italy aggregated at the province level~\cite{itcensus}. We also used data coming from mobile phone data call detail records (CDR), used to reconstruct the displacement distribution function for the US on the basis of published fitting parameters~\cite{gonzalez2008understanding}, and those for Portugal, Spain and France, from the data published for reproducibility purposes~\cite{Tizzoni:2014cl}. Another source of mobility data comes from small blackboxes equipped with GPS trackers and accelerometers that are installed in private vehicles for insurance reasons and record trajectory data, and here we analyzed the urban and inter-urban displacement statistics of Italian drivers as in~\cite{gallotti2016stochastic}. We studied the Google datasets that are elaborated from the opt-in service `Google location history', where the trajectory data captures the movement of Android smartphone users between pairs of stop locations. We used the displacement statistics released by the authors of~\cite{Kraemer:2020hy} and which covers almost all countries. Another important source of data is provided by Facebook and obtained by tracking the movements of mobile phone users that opted-in to the Location History and Background Location collection services~\cite{iyer2020large}. Here, we use the mobility network where the nodes represent administrative areas and the edges correspond to the long-range mobility flows between these areas. The flows are aggregated every 8 hours and we further aggregated them on a weekly basis. If a flow is below a certain threshold (that we estimate being of order 10 users), it is not reported in the data. For epidemic simulations, we also used, as a proxy for population, the baseline values describing the number of Facebook active users in a given node, similarly averaged over a whole week. The fifth dataset studied here comes from Cuebiq Inc., which collects mobility data of anonymized mobile app users who opted-in to a large number of different location based services in different countries including the US, the UK, Italy, Spain, France and Germany. Here, we use Cuebiq-HDR mobility data for Italy elaborated by Pepe and collaborators from the device-level data, and aggregated weekly at the province scale~\cite{pepe2020covid}. We also use flows describing the movements across the US, computed directly by Cuebiq and provided within the framework of the Data4Good program. The flows are here also aggregated weekly and at the county scale. We computed the baseline flows by aggregating them over the weeks preceding the beginning of lockdowns in Italy and the US, rescaled the flows by the country total population and used the associated census population for epidemic modelling. Finally, the last dataset used in our study is provided by Safegraph which collects statistics about visiting patterns of different points of interests (PoI) by aggregating anonymized location data from mobile applications. The data covers the US only, where the home location is identified for each user at the level of census block group. Here, we will use the flows aggregated weekly and at the county level~\cite{kang2020multiscale}, and a two month period before the US lockdown to define the baseline flows.x

%%%%%%%%%%%%%%%%%%%%%%%%%%%%%%
\begin{figure}[ht!]
\begin{center}
\includegraphics[angle=0,width=0.9\textwidth]{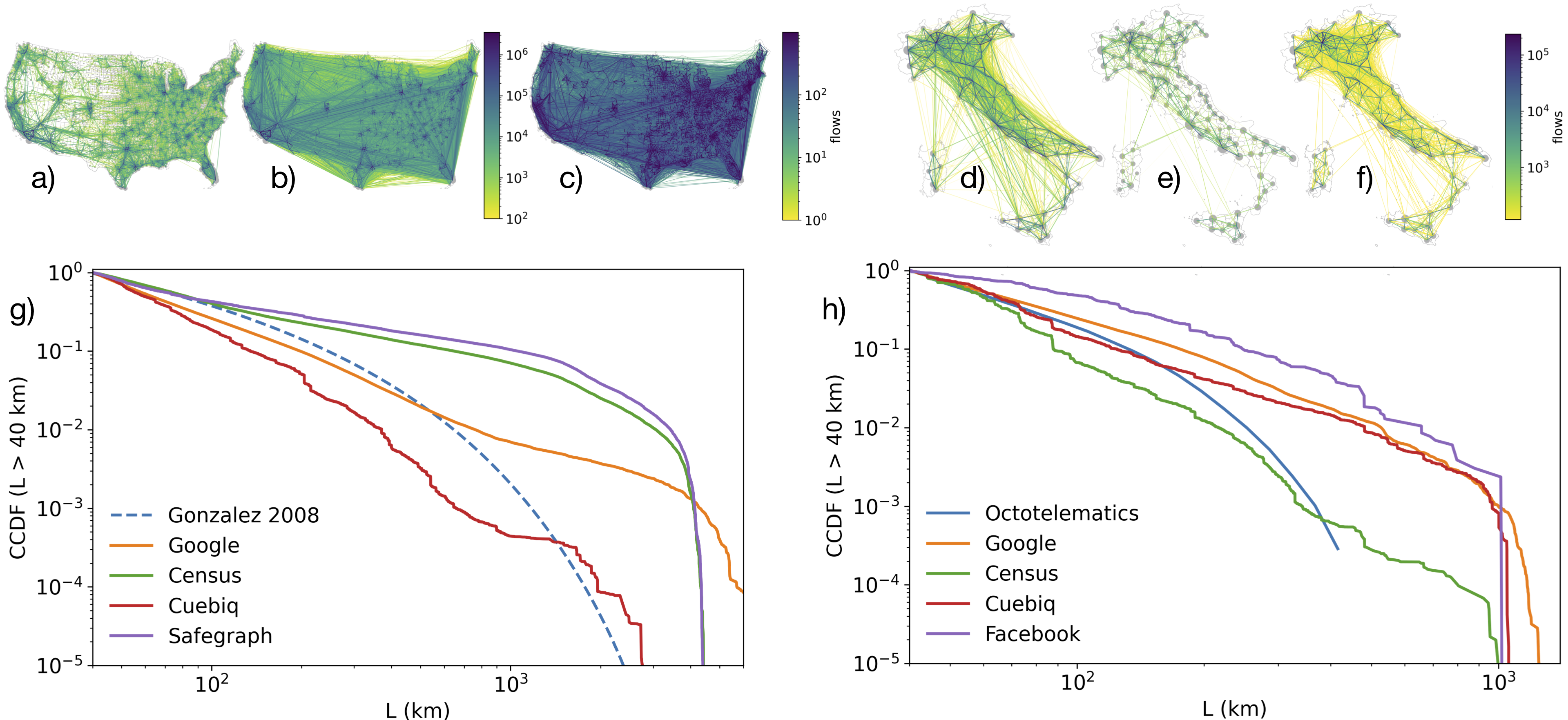}
\end{center}
\caption{
{\bf The baseline mobility networks and displacement distributions for the US and Italy.} The baseline mobility networks captured for the US using (a) Cuebiq mobility insight data, (b) Safegraph, or (c) official census commuting flows display radically different structures. Whereas in the Cuebiq data the data mostly comprises short-mid distance displacements characteristics of daily mobility, both Safegraph and the Census data include a large fraction of long trips that can be associated with the movements between different regions. The baseline mobility network for the Italy using Cuebiq HDR data (d), Facebook data (e) or official census commuting flows (f) also displays remarkable differences: the Census data includes a larger fraction of short trips while the network described by Facebook has a significantly smaller number of connections.
In the bottom panels we illustrate the Complementary Cumulative Density Function (CCDF) of the displacement distribution $P(L)$ for the US (g) and Italy (h), computed for the datasets discussed above, and compare them with public flows derived from Google Location History data~\cite{Kraemer:2020hy} and to what is observed for the US using mobile phone CDR~\cite{gonzalez2008understanding}. In Italy we use measures provided by Octotelematics' private vehicle's GPS blackboxes~\cite{gallotti2016stochastic}. Since these datasets have different granularity, all CCDF have been computed only for $L > 40$ km. The resulting distributions display a huge variability between datasets, with Cuebiq and Mobile data characterised by a shorter tail in the US data, while Google data displays a curve suggesting the presence of two separate scales. In addition, Safegraph and the US census clearly display a totally different mobility scale. In Italy, the shorter tail is instead associated to Census and Octotelematics data, and Google, Cuebiq and Facebook are displaying a similar long-tail trend. }
\label{fig1}
\end{figure}
%%%%%%%%%%%%%%%%%%%%%%%%%%%%%%

%%%%%%%%%%%%%%%%%%%%%%%%%%%%%%%%%%%%%%%%%%%%%%%%%%%%%
\subsection*{The weak universality of the displacement distribution}

We first consider the distribution $P(L)$ of displacement distances $L$ computed with various datasets. This distribution has been intensively studied during the last 15 years~\cite{gonzalez2008understanding,song2010modelling,gallotti2016stochastic,Kraemer:2020hy}, and represents one of the facets of human behaviour which have been many times referred to as universal according to available empirical evidence~\cite{gonzalez2008understanding,song2010modelling,bazzani2010statistical}. Moreover, studying $P(L)$ allows for comparing our results with published data where the analysis has been done on the raw data.~\cite{lenormand2014cross,gallotti2016stochastic,alessandretti2017multi}. The tail behaviour of $P(L)$ varies across countries~\cite{Kraemer:2020hy}, as it is product of the combination of mobility acting at multiple scales~\cite{gallotti2016stochastic,alessandretti2020scales} while also combining individual contributions that differ by both by gender \cite{alessandretti2020scales} and socio-economic group \cite{moro2021mobility,chang2021mobility}. These differences reflected in the specific parameters of the functional forms taken by these distributions, that are  typically analysed in aggregated form (see \cite{gallotti2016stochastic,alessandretti2017multi,Kraemer:2020hy} and references therein). 

In Fig.\ref{fig1}, we show the mobility networks and displacement distributions for the US and Italy computed using the different datasets described above. In panels {\bf a-f} we display the networks obtained for different datasets describing the baseline mobility across US and Italy. Due to the processes used for creating these datasets, the mobility networks are characterized by very different edge densities (see Table~\ref{table_networks} in Methods) which is visually manifest here (additional measures and details can be found in the Supplementary Fig.~\ref{figSI_maps_same_density}).
We then compute the complementary cumulative (CCDF) of the displacement distribution $P(L)$ (see Fig~\ref{fig1} {\bf g,h}). This distribution has the advantage of capturing simultaneously information about the distances covered and the associated flows. 

The CCDF of $P(L)$ is evaluated by progressively summing up the fraction of flows up to a distance $\leq L$. Depending on the type of datasets, these summations are made either by counting the number of individual displacements shorter than $L$ or by summing up flows passing through edges connecting distances shorter than $L$. The latter is the case of all the data analyzed here, with the exception of the curves computed in other papers directly from mobile phone CDR~\cite{gonzalez2008understanding} or GPS blackboxes~\cite{gallotti2016stochastic}.

%%%%%%%%%%%%%%%%%%%%%%%%%%%%%%
\begin{figure}[ht!]
\begin{center}
  \includegraphics[angle=0,width=0.8\textwidth]{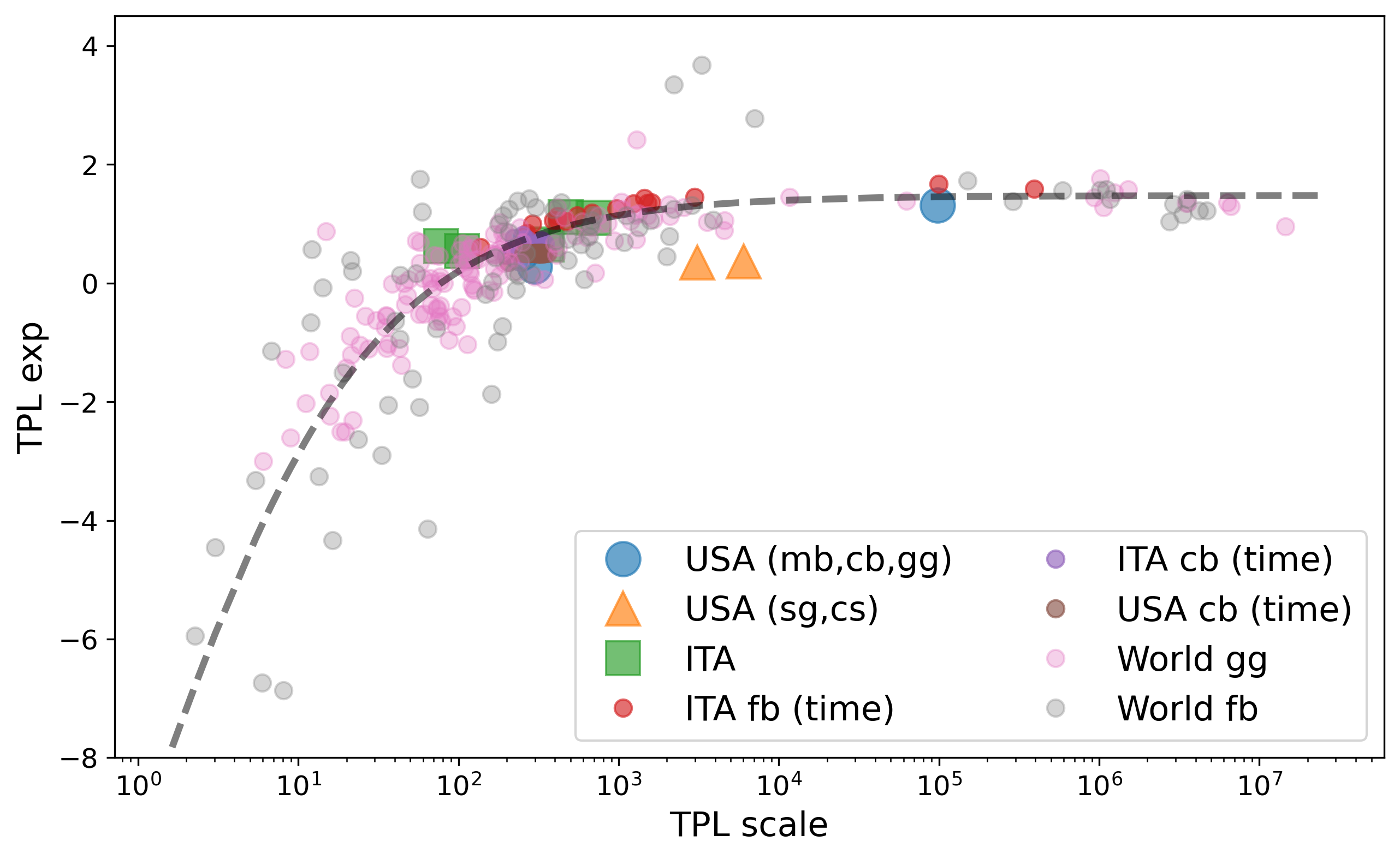} 
\end{center}
\caption{
  {\bf Comparison of the displacement distributions across datasets.} We estimate the fitting parameters of $P(L)$ (scale and exponent) for various countries using a truncated power law.  We observe a wide variance of these parameters. We include the results for the Figure 1 for USA --Cuebiq (cb), Mobile (mb)~\cite{gonzalez2008understanding}, Google (gg), Safegraph (sg), Census (cs) -- and Italy (as illustrated in Figure~\ref{fig1} it also includes Octotelematics and Facebook (fb) data), the values for the longitudinal weekly analysis of Cuebiq and Facebook  data, and the values estimated for the Google and Facebook baseline. We observe that all data points fall along a single curve, which suggests strong correlation between the free parameter and, ultimately, the existence of an underlying law having a single degree of freedom. This curve (dashed line) is estimated with a fit of the form $y=\frac{A}{1+x^{-1/B}}-C$ where $A \approx 22 \pm 1$, $B \approx 1.7 \pm 0.1$ and $C \approx 20 \pm 1$. We note how the USA-long data points (sg, cs) seem to deviate strongly from the average curve. This is because as they are capturing a behaviour totally different from the other sources of data.}
\label{fig3}
\end{figure}
%%%%%%%%%%%%%%%%%%%%%%%%%%%%%%

These CCDFs, together with data coming from other aforementioned datasets, are compared in panels Fig~\ref{fig1} ({\bf g,h}), where we observe very large deviation depending on the dataset and the processing methodology (see Supplementary information for more details) used for analyzing the mobility of the same country. Similar comparisons are made in Supplementary Fig.~\ref{figSI_other_countries_pl} for France, Spain and Portugal using Facebook and Google data along with mobile phone data coming from~\cite{Tizzoni:2014cl}), that also . Remarkably, the observed differences appear larger than the changes associated with the changes in mobility due to lockdowns in the US and Italy in 2020 (see Supplementary Fig.~\ref{figSI_time_dependency}), which represented, in some sense, a huge natural experiment and which we could safely consider as `major disruptions'. We fit the curves for $P(L)$ to compare Google and Facebook data across 58 countries and used different functional forms for the fit (see Methods and Supplementary Fig.~\ref{figSI_google_facebook_parameters} where we observe, again, large differences between the curves observed in different datasets). We will discuss here the case of the truncated power law (TPL) form with two parameters. In Fig.~\ref{fig3}, we display these fitting parameters for 143 Google curves, 88 Facebook curves,  the curves described in Fig~\ref{fig1}, and the time dependent curves of Supplementary Fig.~\ref{figSI_time_dependency}. We see that the parameters of the TPL (but also for the other fitting models, see Supplementary Fig.~\ref{fig3_rak_ln}) are distributed along a relatively narrow curve (that can be approximated by a sigmoidal function), suggesting the possibility of an underlying weakly universal behaviour, where the curve exponent is a function of the scale. These results demonstrate the very large variability of mobility networks over different countries and data providers, pointing out differences that might bias models which are known to be sensitive to mobility, such as epidemics spreading. Additionally, they point to limitations of the different data design and the need for enhancing data collaborations to ensure the reproducibility of scientific studies based on mobility networks data.

Furthermore, an international comparison based on Facebook and Google data across 61 countries reveals that the expected universality of human displacement is weak, as significantly different tail behaviours are observed in the smaller and larger countries (see Supplementary Figure~\ref{ccdf_comparison_google}). While we are still convinced that the underlying physical process driving individual displacements is universal~\cite{gallotti2016stochastic}, bounding it to the geographical constraint of national borders clearly influences the aggregated $P(L)$. This effect, added to the data biases coming from both the socio-economical characteristics of behaviour captured by the data, and the choices made in the processing of these data, renders the characterization of individual mobility from the distribution $P(L)$ of a single country not trivial.

%%%%%%%%%%%%%%%%%%%%%%%%%%%%%%%%%%%%%%%%%%%%%%%%%%%%%
\subsection*{The large sensitivity of spreading processes}

The long range mobility information captured in individual trajectories datasets are used for a variety of applications, such as seasonal changes in population distribution \cite{deville2014dynamic}, migration \cite{simini2012universal}, tourism \cite{lenormand2015human}, and international mobility \cite{hawelka2014geo}. As a representative case study, we focus our attention on a class of models where human mobility plays an important role: the spreading of an infectious disease. To better understand the consequences of the aforementioned variations observed on $P(L)$, we consider a standard metapopulation SIR model (see \cite{pastor2015epidemic} and the Methods section), where spatial patches plays the role of nodes and human flows connect them. The choice of this process is justified by the high societal impact it has, since human mobility is a standard target for non-pharmaceutical interventions -- such as curfews and lockdowns -- to mitigate the evolution of infectious disease epidemics~\cite{kraemer2020effect,schlosser2020covid,maier2020effective,davis2021cryptic}. Also, it is worth remarking that this class of models is emblematic and widely adopted for practical applications: in principle, other models could be considered as well, but they might depend on a larger number of parameters and hyperparameters, thus making more difficult the analysis of the impact of mobility data on the final results. The choice of a metapopulation SIR keeps the number of parameters small while still providing a valuable model used for realistic infectious diseases.

We start with the Facebook data for 71 countries where at least 10 nodes are represented in the network. We first observe (see Supplementary Fig.~\ref{small_world_pruning}a) that the average degree of the networks appear to be a (sub-linearly) growing function of the number of users, which means that more populated countries have denser networks. This is most likely due to a pruning procedure that removes edges with smaller flows for both privacy enhancing and disk space reduction. Networks with higher average degree have a larger fraction of shortcut nodes (as can be seen in Supplementary Fig.~\ref{small_world_pruning}b, where shortcuts are defined as links at least twice as long as the average distance between a node and its closest neighbour). This biased pruning heavily influences the flows, and imposes in particular strong cutoffs on the longer connections (see more discussions about this point in Supplementary Fig.~\ref{figSI_gravity_pruning}, where we use a toy model in order to analyse the relation between the density of edges and the displacement distribution). As a consequence of tampering the displacement tails, the spreading behavior on mobility networks ranges from lattice-like to a small-world behaviour, depending on the dataset processing details. Also, with less shortcuts, the spreading dynamics is heavily penalized in networks captured with smaller user-bases and  smaller average degrees as it is indeed observed in simulations (see again Supplementary Fig.~\ref{small_world_pruning}).

\begin{figure*}[ht!]
\begin{center}
\begin{tabular}{cc}
\raisebox{2.3cm}{(a)} \includegraphics[angle=0,width=0.45\textwidth]{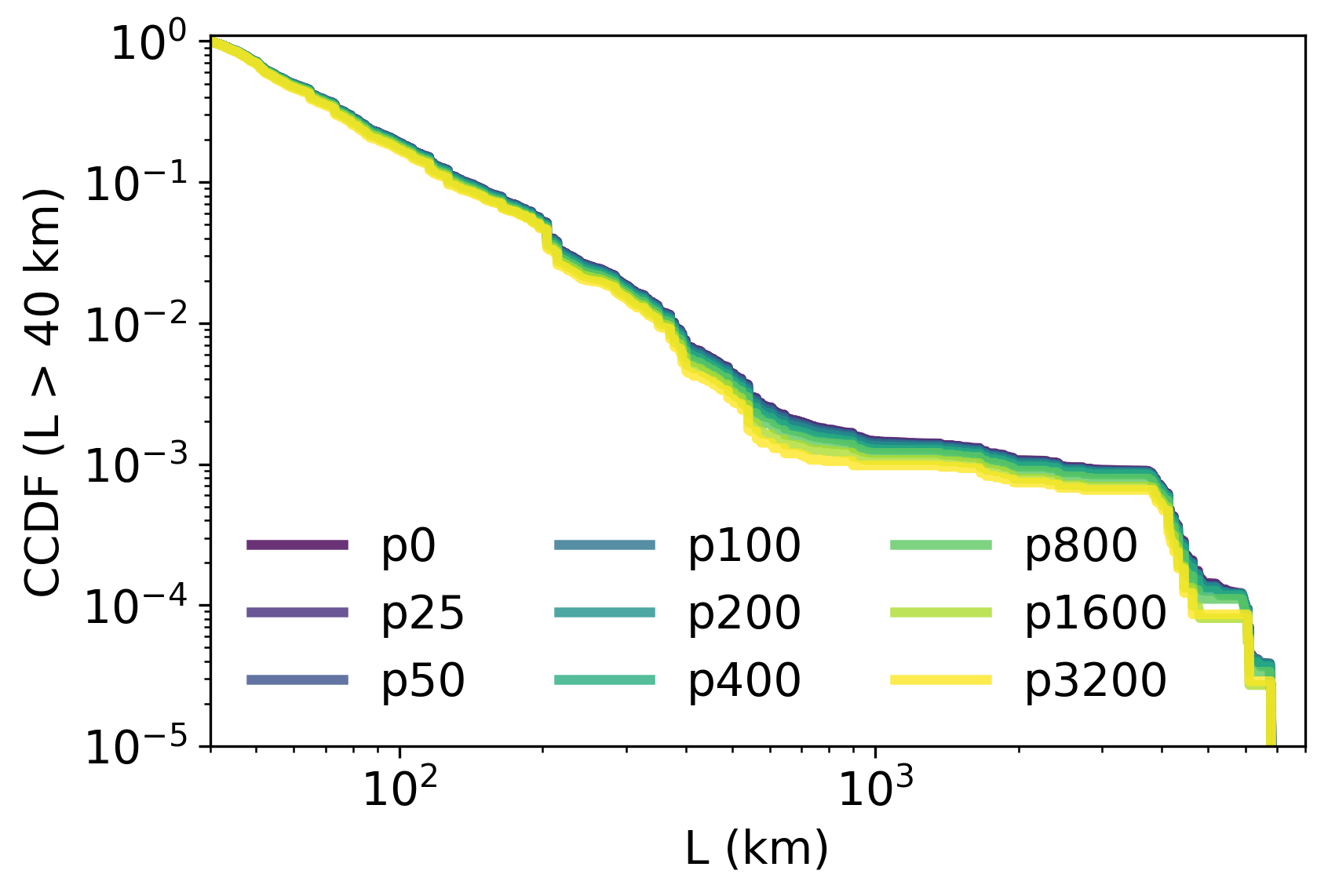} &
\raisebox{2.3cm}{(b)} \includegraphics[angle=0,width=0.45\textwidth]{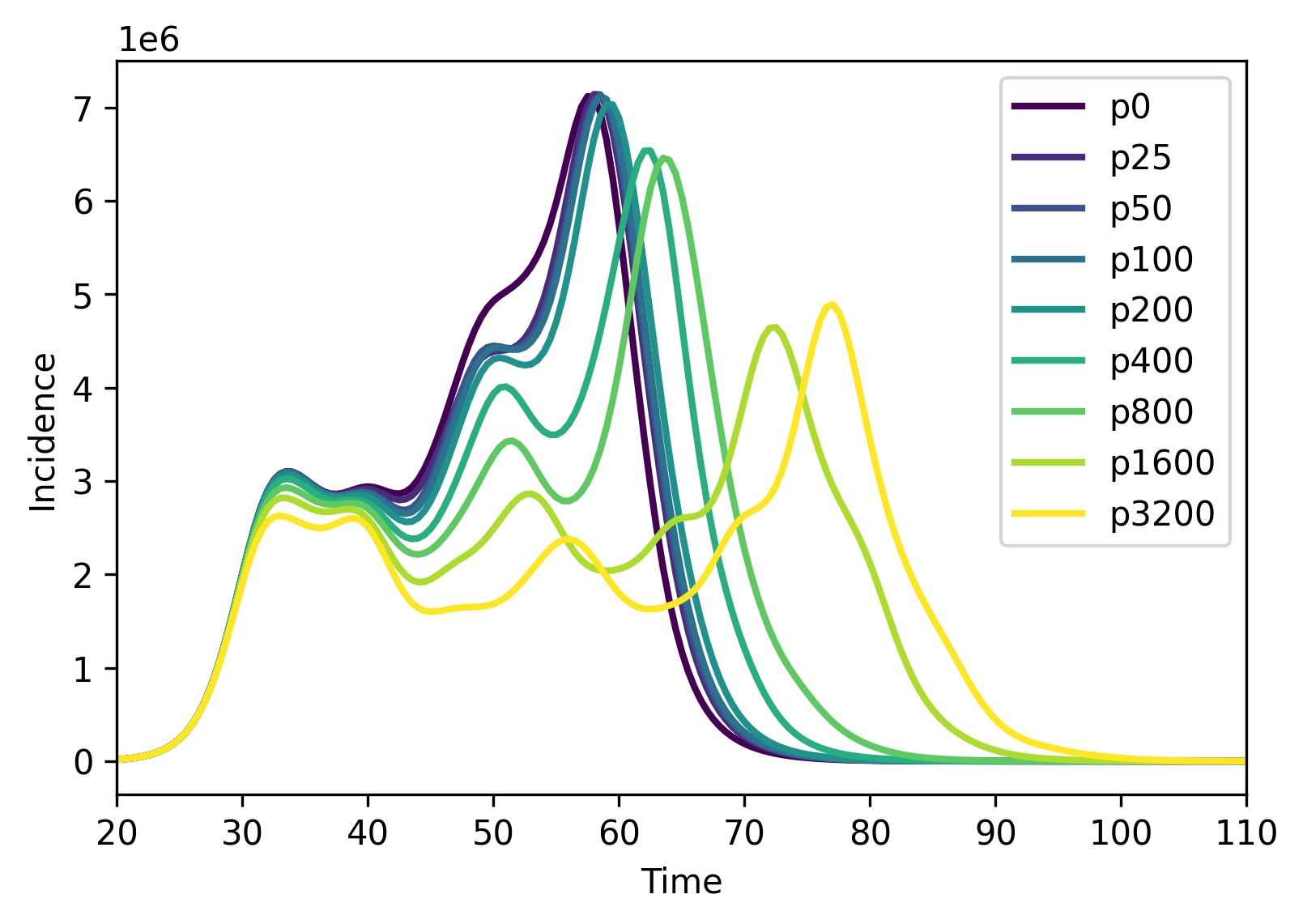} \\
\raisebox{2.3cm}{(c)} \includegraphics[angle=0,width=0.45\textwidth]{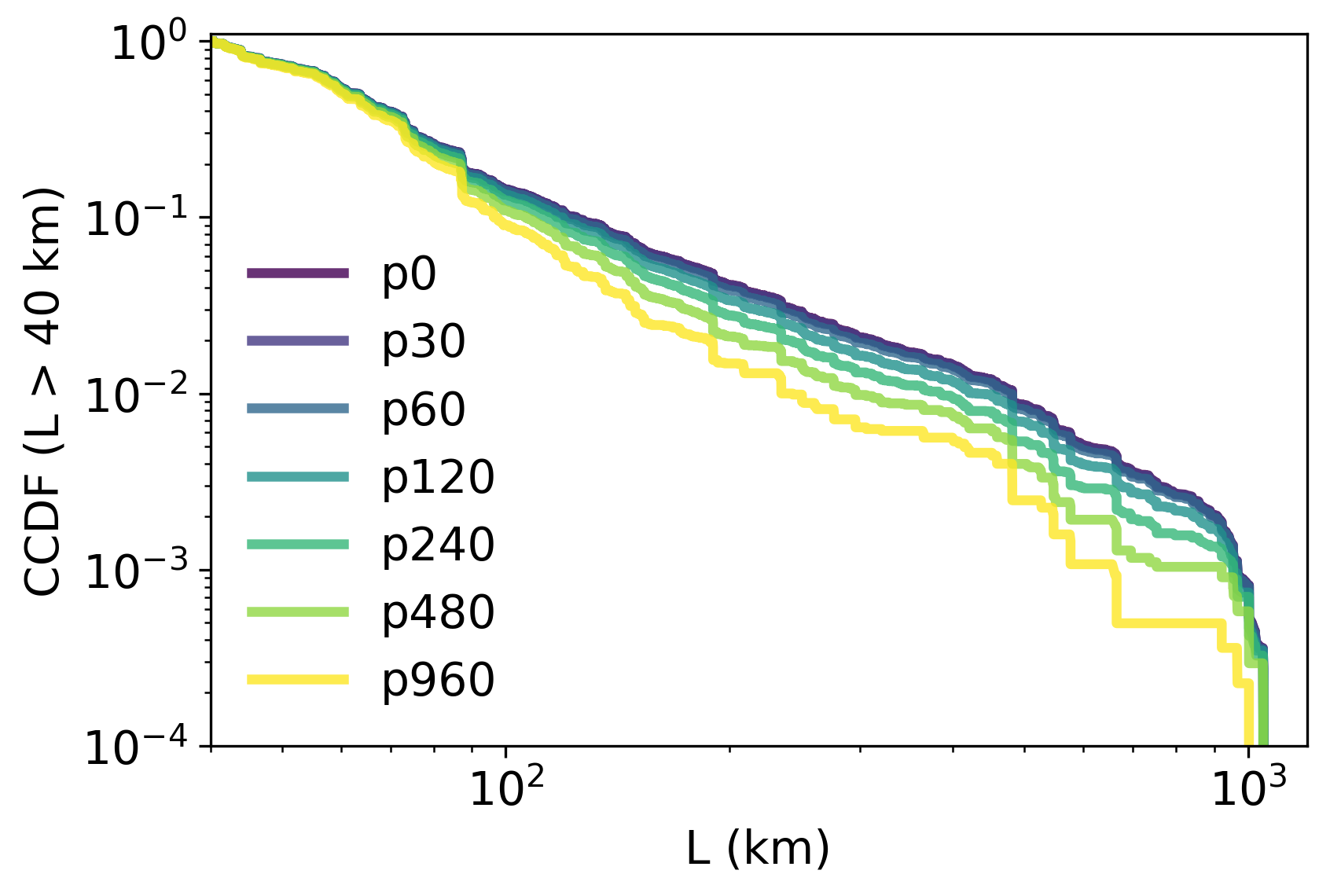} &
\raisebox{2.3cm}{(d)} \includegraphics[angle=0,width=0.45\textwidth]{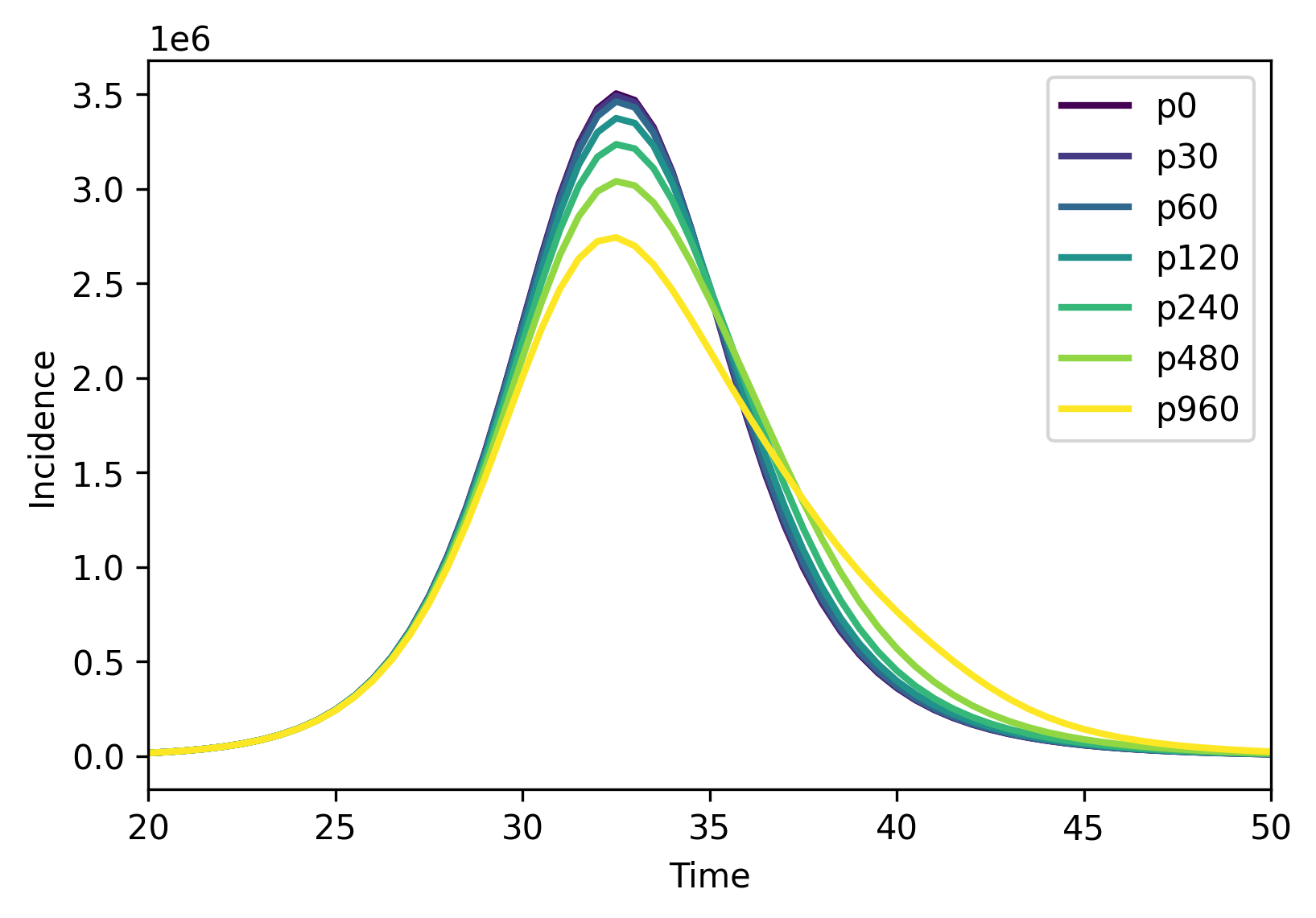} \\
\end{tabular}
\end{center}
\caption{{\bf Effect of flow pruning on the spreading dynamics.}  In these figures, the notation $p_x$ indicates the edge pruning with flow smaller than $x$. In (a) and (c) we show by examining the CCDF of the displacement distribution that the pruning of flows does not affect significantly $P(L)$. Despite the small effect on the displacement distribution, we observe very different dynamics for disease spread on the corresponding mobility networks, in particular in the US case (b), while in the Italian case (d), the impact on the dynamics is marginal, probably due to the smaller size of this country. 
}
\label{fig4}
\end{figure*}

The differences between the spreading dynamics in the US and Italy computed with different datasets is shown in Fig.~\ref{fig4}. We first observe that for the US Safegraph and Census data, the results are exactly the same (figures shown in Supplementary Fig.~\ref{fig_baseline_epi}). In contrast, simulations done with the Cuebiq data, being at a shorter scale, produce a more complex spreading behaviour with two peaks. In Supplementary Fig.~\ref{figSI_peak_map}, we show how this first peak is characterised by a localized behaviour reflecting a `lattice like' spreading~\cite{gross2020epidemic}. Italian data displays less differences, with the peak that appears not delayed but rather lowered for shorter scale Facebook and Census data.

To further investigate the differences observed between the US and Italy, we use the Cuebiq dataset and create a set of networks by pruning flows below an increasing threshold.  In Fig.~\ref{fig4}(a,c),  we show that the displacement distribution remains unchanged. However, despite the small effect on $P(L)$, the pruning has a very important impact on the spreading dynamics (Fig.~\ref{fig4}(b,d)) that appears to be progressively delayed as the pruning is increased~\cite{gross2020epidemic}. This is particularly true for the US case, while the Italian case - being at a smaller spatial scale - is not as sensitive to the pruning which induces a drop in the peak but not a delay. The differences observed in Fig.~\ref{fig4} as a consequence of the pruning are, again, smaller with respect to the differences observed when comparing across datasets and countries. This fact can be better appreciated in Supplementary Fig.~\ref{fig_epi_scale}, showing how the relative incidence at peak simulated with the SIR model are determined by the characteristics of the $P(L)$.

%%%%%%%%%%%%%%%%%%%%%%%%%%%%%%
%%%%%%%%%%%%%%%%%%%%%%%%%%%%%%

%%%%%%%%%%%%%%%%%%%%%%%%%%%%%%%%%%%%%%%%%%%%%%%%%%%%%
\section*{Discussion}\label{sec:discussion}

While aware that data gathering and processing techniques used for human mobility are different, rather unexpectedly we have found extremely wide differences between datasets of human movements which are commonly used interchangeably as a proxy for universal human behavior, and genuinely supposed to give comparable results. On the one hand, our empirical findings from the analysis of multiple datasets call for a more transparent access to data and its protocols. On the other hand, our findings demonstrate that many of the limitations that were only supposed to affect results do, in practice, significantly affect the insights about human behavior that can be obtained from mobility data.

Specifically, our study demonstrates that it is not sufficient to analyze human mobility and its multifaceted aspects, such as universal dynamics, by means of a single dataset: ensembles are needed to allow for a careful evaluation of potential biases in data gathering and processing protocols.
The first step is to become more aware than before of such a critical impact: in particular, we noticed a series of technical aspects about these various datasets, that are often overlooked by both their producers and by scientists using them (see the SI for more details). In this respect, our analysis provides an overview of the limitations inherent to empirical studies and point towards the urgent need for setting shared standards in the design and elaboration of human mobility datasets. The focus was recently on how to model complex socio-technical systems~\cite{citron2021comparing}, but their complexity also affects their empirical analysis, implying a special care is needed for their preparation due to their high sensitivity to specific processing.

Mobility datasets being passively collected from ICT data records, are necessarily influenced by the behaviour that produces the record itself~\cite{bonnel2015passive,olteanu2019social,escap2019handbook,de2021updating}. The method used is often undisclosed and results can be very sensitive to parametric choices. For instance, this is the case when aggregating data over any areal units which influences the observed results (see Supplementary Fig.~\ref{figSI_granularity}),
or when stop locations are defined according to radically different methodologies~\cite{gallotti2018tracking}. Also, pruning flows for anonymity reasons is a critical point and can have a dramatic large-scale impact as we saw for the Google or Facebook data. In particular, scale choice and pruning lead to different displacement distributions and affects particularly its tail behavior.  Spreading processes over wide areas is more than the sum of the spreading processes over its parts and the tail of the displacement distribution is critical here, as we have shown with our application to epidemic dynamics. 

The differences in methodologies might render meta-analysis and comparing different countries difficult, if not impossible. In the worst case, results obtained for a specific region might not be straightforwardly ported to another geographic area. Without verifying that such results are robust to changes in the protocols for data processing, the knowledge gained from one place -- under the implicit assumption of universal behavior, which we have only partially reproduced with our extensive analysis -- could not agree with the one gained from another place, jeopardizing global policies. 

Most of the problems we have outlined in this study could however be solved and we propose a series of good practices that largely align with what already largely discussed in the Mobile Phone Data Community~\cite{escap2019handbook,de2021updating}. A few contributions~\cite{chang2021mobility,moro2021mobility,schlosser2021biases,mercier2021effective} indeed already tested, or controlled for, some of the issues we illustrate in the Supplementary Materials, but this kind of testing is however often unfortunately neglected or impossible. Certainly, encouraging a direct access to device-level data, although naturally strictly regulated to the higher anonymization and privacy-preserving standards, as Cuebiq and Mobile phone providers did, will enforce a high level data quality. Also, more transparency in the data production and pre-elaborating baselines and metadata (as Facebook did) is needed. Building of communities as encouraged by Cuebiq and Safegraph, creating open platforms and a share framework for mobility data~\cite{kishore2020measuring} as proposed by initiatives such as DataCollaboratives~\cite{datacollaboratives} (especially associated to emergencies like COVID) are also measures that will improve the quality of data.  

On a more general perspective, it is always desirable that the study of complex systems relies on more than a single instance~\cite{parisi2002complex}. Even trivial operations on the data, like thresholding, can be very delicate~\cite{cantwell2020thresholding} and require a robustness analysis that, either for lack of generalized access to multiple datasets or other plausible reasons, too often is overlooked. In fact, results might be platform dependent~\cite{malik2016identifying} and data quantity cannot compensate for lack of data quality~\cite{bradley2021unrepresentative}. Therefore, a quantitative approach to social sciences aiming at the highest standards of reproducibility should go beyond single datasets, requiring by design to be tested with comprehensive meta-analyses~\cite{broido2019scale}.

\clearpage

%%%%%%%%%%%%%%%%%%%%%%%%%%%%%%%%%%%%%%%%%%%%%%%%%%%%%
\section*{\label{sec:methods}Methods}

\subsection*{Datasets}

\subsubsection*{Census data}

Census data provide statistical information about the urban and inter-urban commuters' flows using a widely statistical representative sample. The data can in principle be collected at the street address level of granularity, but is often aggregated at different administrative levels (district, municipality, provinces, counties, \dots). Census data  typically covers the whole adult population in a country. However, the quality of these data as proxies for mobility may be limited by the fact that the home (or work) location in the national registers needs to be regularly updated as individuals move to new homes and are still registered at their original homeplace. This mismatch may vary strongly between countries as it depends on how enforced or rewarded is the update of the public register. Moreover, these data fail to capture other types of mobility besides home-work commuting of adults, such as the movements of tourists seasonal or freight traffic. At a urban level, census survey data are reasonably similar to the estimates coming from Mobile Phone or Twitter data~\cite{lenormand2014cross} and allow for the opportunity of studying epidemic spreading considering the movements of different age groups~\cite{Dalziel:2013cf,Arenas:2020eg}, although the use at a national scale~\cite{Arenas:2020eg,Gatto:2020cb} can be influenced by the aforementioned limitations, as discussed in this paper. Here, we collected and analyzed official 2015 census data describing aggregated commuting patterns for the US~\cite{uscensus}, at the county level, and the official 2011 Census data for Italy~\cite{itcensus}, aggregated at the province level.

\subsubsection*{Mobile phone data}

Call detail records (CDR) from mobile phone data were at the root of the first ICT data-informed attempts at
modelling human mobility~\cite{gonzalez2008understanding,song2010modelling}. The information provided by the CDR are records of when and where a user exchanges information (audio, text or data) with an antenna. Unless some triangulation methods are applied, the spatial distancing between the antennas represent the characteristic granularity of the mobile phones trajectory data. In the earlier days of mobile communication, when it was limited to calls and SMS, the data temporal sampling was uneven and sparse, which would cause biases in how the trajectories were reconstructed~\cite{gallotti2018tracking}, with an over-representation of long trips. This is most likely not the case with modern data, since smart-phones are continuously connected to the mobile network for data exchange. Thanks to its wide  coverage and the fact that it includes a wide range of mobility behaviors, mobile phone data represent one of the more important assets that can be used for studying human interaction~\cite{Sobolevsky:fv} and as proxies of human mobility for epidemic modelling~\cite{Tizzoni:2014cl,Panigutti:2017it} with applications in the study of the spread of a wide range of diseases~\cite{Wesolowski:2015fm,Ruktanonchai:2016ju,Kramer:2016jj,bosetti2020heterogeneity,schlosser2020covid,pullano2020evaluating,Valdano:2021fb}. The access to these data is usually limited to a small number of research groups that build tight partnerships with the mobile industry (with the notable exception of D4D challenges~\cite{blondel2012data,de2014d4d}), but the active collaboration between industry and academia in principle grants that the data processing can be custom built for the scientific use and the underlying assumption shared for reproducibility purposes. Here, we will reconstruct the displacement distribution function for the US on the basis of published fitting parameters~\cite{gonzalez2008understanding}, and those for Portugal, Spain and France, from the data published for reproducibility purposes~\cite{Tizzoni:2014cl}.

\subsubsection*{GPS data}

Small blackboxes equipped with GPS trackers and accelerometers are installed in private vehicles for insurance reasons and record trajectory data. This type of data is naturally limited to a single transportation mode, but is able to capture locations with a great spatial accuracy (typically of order $10m$), and the dynamics of the trip with stops and velocity, which allows the reconstruction with high accuracy of a driver's displacements. The scientific use of a dataset describing the vehicular movements in Italy, provided by Octotelematics, has given the opportunity for exploring different facts of human mobility behaviour~\cite{gallotti2012towards,pappalardo2013understanding,gallotti2013entropic,gallotti2015understanding,pappalardo2015returners,gallotti2016stochastic}. However, with the exception of a small dataset provided for a data challenge~\cite{barlacchi2015multi}, their access has been limited to a limited number of groups who had granted access to the raw data by an industrial partner. Here, we analyze the displacement statistics of Italian drivers in various cities as described in~\cite{gallotti2016stochastic}.

\subsubsection*{Google datasets}

Google data are elaborated from the opt-in service `Google location history' and the trajectory data, capturing the movement of Android smartphone users, has been segmented using a machine learning technique~\cite{kirmse2011extracting} into a series of displacements between pairs of stop locations. The displacement have been then aggregated over a grid and on a weekly bases. The grid size is been described as a $5 km \times 5 km$ grid in~\cite{Kraemer:2020hy} and as a $5 km^2$ cell grid in other publications on similar datasets. To ensure the users' privacy, a network pruning has been applied where flows between cells is smaller than $100$ users/week. The strength of this dataset is its wide international coverage. However,  since longer connections have smaller flows, the pruning procedure systematically cuts inter-urban mobility and is a clear limitation of these datasets. In particular, this leads us to think that some results derived from Google data using this procedure are biased towards an over-representation of short movements. While the analysis at a urban scale~\cite{bassolas2019hierarchical,aguilar2020impact,barbosa2021uncovering,Hazarie:2021jc} is probably marginally affected by this problem, the relative weight of long range connections is certainly biased. This bias might be one of the factors leading to the observation~\cite{Kraemer:2020hy} of a steeper decline in the displacement distribution in less developed countries where, coincidentally, smartphones experience also a smaller market penetration and thus the effect of the threshold-based pruning of small long range flows can be naturally stronger. As we will see (see text and Fig.~\ref{fig4}), this pruning effect also influences all analysis describing the disease spreading process over the mobility network at a country scale, such as those discussed in~\cite{rader2020crowding,Ruktanonchai2021}. The Google data analysed here has been shared through a Data4Good programme. We do not however have access to the data and a reproducibility request associated to two published papers~\cite{Kraemer:2020hy, rader2020crowding}, was made in October 2020 is still pending at the time of the submission of this paper. We will therefore use the displacement statistics released by the authors of~\cite{Kraemer:2020hy} and which have been computed using bins of $1 km$. The data covers almost all countries for any displacement larger than $1km$ but we restricted our analysis to displacement larger than $5 km$ taking into account the cut-off induced by the underlying cell dimension.

\subsubsection*{Facebook}

Facebook data is obtained by tracking the movements of the mobile phone application users that opted-in to the Location History and Background Location collection services~\cite{iyer2020large}. The access is granted through a Data4Good platform which is accompanied by a limited description of the dataset and very limited methodological description.
These datasets provided by the Facebook Data4Good can describe, in different forms, the mobility at different scales in correspondence to natural disasters (disaster maps) or the COVID-19 epidemics. The COVID-19 DiseaseMaps dataset included three descriptors of human mobility that have been used for research: i) small scale mobility flows between small tiles of size approx $1km \times 1km$ (the dimension of the tiles varies between datasets~\cite{kishore2021lockdowns}) and internal to administrative areas~\cite{kissler2020reductions,mena2021socioeconomic,kishore2021lockdowns}; ii) long range mobility flows between administrative areas~\cite{bonaccorsi2020economic,galeazzi2021human}; iii) co-location matrices computed at an administrative level~\cite{iyer2020large,kishore2020measuring,chang2021variation}. Co-location matrices being intrinsically different from OD matrices, are not considered here.
Similarly to what is happening to Google datasets, Facebook performs the pruning of small flows and long distances, thus introducing an important bias and might severely affect results at a large scale (and less at a urban level), and leading to national origin-destination matrices that are visibly sparse as observed in~\cite{bonaccorsi2020economic}, Figure 1e and Supplementary Fig.~\ref{figSI_maps_same_density}. (This effect of pruning appears as reduced in co-location matrices.)
Also, we note that Facebook also provides `baseline' values that represent a typical flow prior to the event (disaster or disease) characterising the dataset. The strength of the Facebook data is its wide international coverage, coupled with the possibility of providing metadata such as user gender. Limitations here come from the data processing under the form of small flow pruning and the fragmentation of the network into smaller sub-networks. For instance, U.S.A. flows for the COVID-19 disease maps are provided only at state level, with no information about the movements between states. Here, we will use the mobility network where the nodes represent administratives areas and the edges the long range mobility flows between these areas. The flows are aggregated every 8 hours and we obtain temporal networks. If the flow is below a certain threshold (that we estimate it to be of about 10 users), it is not reported in the data. For these temporal networks, we aggregate flows on a weekly basis. Similarly, we also define an aggregated baseline network combining all baseline flows reported for different 8 hours bins and day of the week. Baseline flows are present however only for origin-destination pairs and time bins where it the flow reported is above the filtering threshold in at least one of the weeks comprised in the dataset, thus some information is necessarily lost also in this case. For epidemic simulations, we also used, as a proxy for population, the baseline values describing the number of facebook active users recorded in a given node. Similarly to flows, active users are reported in bins of 8 hours and for each day of the week and have been accordingly averaged over a whole week. 

\subsubsection*{Cuebiq}

Cuebiq collect mobility data of anonymized mobile app users who opted-in to a large number of different location based services in different countries including U.S.A., U.K., Italy, Spain, France and Germany. Access to this data has been provided under data governance frameworks to address the COVID-19 emergency via a Data4Good program that constitute a clear attempt at building a collaborative environment around the data. Cuebiq provided to the members of the program access to device-level privacy-enhanced data, where noise is added to home and work locations at the census block group level, and stops associated with privacy-sensitive locations are removed entirely from the dataset, in order to preserve privacy. This data has been used in the early days of the pandemic to produce (non peer reviewed) mobility reports~\cite{santana2020analysis,pepe2020covid,klein2020reshaping}, often accompanied by interactive visualisation of the mobility reduction patterns. A clear advantage of these datasets is the access to the device-level trajectory data at GPS precision, from which it has been possible to study the behavioural changes associated to lockdowns~\cite{gauvin2020socioeconomic,hunter2021effect,lucchini2021living}. The limitation coming with this abundance of data is the need for building a strong preprocessing pipeline in order to analyse this data, but nevertheless allows for a tailored and transparent design of the segmentation algorithm. Here, we use Cuebiq-HDR mobility data for Italy derived from the device-level data and aggregated weekly at the province scale by Pepe and collaborators~\cite{pepe2020covid}. We also use flows describing the movements across the U.S.A., computed directly by Cuebiq and provided within the framework of the Data4Good program. The flows are here also aggregated weekly and at the county scale, based on a proprietary segmentation method. We computed the baseline flows aggregating those of the weeks preceding the beginning of lockdowns in Italy and U.S.A respectively, rescaled the flows to the total country population and used the associated census population for epidemic modelling.

\subsubsection*{Safegraph}

Safegraph collects statistics about visiting patterns of different points of interests (PoI) by aggregating anonymised location data from mobile applications. The data covers the US only, where the home location is identified for each user at the level of census block group.  Safegraph data activated a Data4Good program that gives free access to Academics for non-commercial work. Safegraph actively attempted at building a community around their data by organising regular seminars and a platform where results and issues can be exchanged. The limitation of this dataset is that, since it is focusing on PoIs, mobility flows are not directly available from the data~\cite{chang2021mobility}, as the flows recorded are between home locations and visited PoI. This means that two subsequent visits to two location A and B would be recorded not as a movement between A and B but as two movements from Home to A and from Home to B. Deriving the mobility network at the urban level requires a rather complex procedure~\cite{chang2021mobility}, which is likely not suitable for being extended at a country scale. Nevertheless, these mobility flows have been released as OD matrices to describe population flows during the COVID-19 pandemic~\cite{kang2020multiscale} and used to reconstruct mobility at long range~\cite{hou2021intracounty}. The advantages of this data is that there is a great abundance of detailed information about the users activity at PoI level, including social distancing estimates that have been also subject to scientific use~\cite{charoenwong2020social,weill2020social}. Here, we will use the aggregated flows published in Nature Scientific Data~\cite{kang2020multiscale}, aggregated weekly and at county level, and used a two month period before the U.S.A. lockdown do define the baseline flows.

\subsubsection*{Network features}

We recap in Table~\ref{table_networks}, the main features of the baseline networks obtained from the various datasets used for Italy and the US. 
\begin{table}[ht!] 
\renewcommand{\tablename}{Supplementary Table}
\caption{{\bf Baseline Networks characteristic numbers.} We display here the dimensions of the networks shown in Fig.~\ref{fig1}. We compute the number of nodes $N$, the number of directed Edges $E$, the average (in- or out-) degree $\langle k \rangle = E/N$ and the edge density $D=E/(N(N-1))$. The Italian Cuebiq (ITA Cuebiq) network has been aggregated over different provinces which contains different numbers of nodes. Having neglected self-loops, the USA Cuebiq dataset has $31$ disconnected nodes.}
\begin{tabular}{lcrrr}
\hline
Dataset & $N$ & $E$ & $\langle k \rangle$  & $D$ \\
\hline
ITA Cuebiq      &   107/107      &   3,735      & 34.9      & 0.329 \\
ITA Facebook    &   110/110      &   926        & 8.4       & 0.077 \\
ITA Census      &   110/110      &   4,019      & 36.5      & 0.335 \\
USA Cuebiq      &   3077/3108    &   60,181     & 19.6      & 0.006 \\
USA Safegraph   &   3108/3108    &   1,574,604  & 506.6     & 0.163 \\
USA Census      &   3108/3108    &   131,391    & 42.3      & 0.014 \\
\hline
\end{tabular}
\label{table_networks}
\end{table}

\subsection*{Curve fitting as analysis method}

In this paper we use the distribution of displacements $P(L)$ as one of the analysis tools to inspect at the same time the flows and the distance covered in a flow network/origin destination matrix. In the past years, there has been an open discussion about the functional form of this distribution, that we know to be governed by the multi-scale characteristics of human mobility~\cite{gallotti2016stochastic,alessandretti2020scales}. As in many similar cases, we use curve fitting as an analytic tool for extracting information from the data collected. Curve fitting is clearly a method limited by the functional form chosen. A large number of functional forms have been used for mobility datasets, and the most common ones are:
\begin{enumerate}
\item a truncated power law (TPL) with three~\cite{gonzalez2008understanding} or two~\cite{kraemer2020effect} free parameters;
\item a lognormal (LN)~\cite{alessandretti2017multi};
\item model-driven generalized gamma functions~\cite{gallotti2016stochastic}.
\end{enumerate} 

An exponential form is also sometimes considered a ta shorter scale, but can be captured by the exponential cutoff using TPL and model-driven generalized gamma functions.
In our analysis, we attempted fits using 2 and 3 parametric TPL, LN and three different model-driven forms: the Random Uncorrelated Accelerations (RUA, 1 parameter), the Random Acceleration Kicks (RAK, 2 parameters) and the Weibull (WB, 2 parameters). 

The RUA follows the functional form $P(L) \propto L^{-\frac{3}{4}} \exp(-(L/L_s)^\frac{1}{2})$, and is what expected when a traveller speed follows a brownian motion~\cite{gallotti2016stochastic}, the RAK is in practice a generalisation of RUA, as follows the similar form $P(L) \propto L^{-\gamma} \exp(-(L/L_s)^\frac{1}{2})$, where $\gamma$ is a free parameter, and represents the saddle points approximation for a model where a traveller accelerates and decelerates at fixed increments following a poisson process~\cite{gallotti2016stochastic}.

The plots were technically carried out by fitting directly the CCDF function, without any data binning, using the python {\emph lmfit} library, which performs a (non-linear) least square fit. This procedure clearly weights more the short distances in the distribution, where data is more dense, and basically ignores the tail. Since the detailed characterisation of tail for the curves at hand was essential, we compensated by fitting the logarithm of the y-axis of the empirical curves. The results are satisfactory, as illustrated in Supplementary Fig.~\ref{figSI_fits_are_good} and in the collection of figures attached to the Supplementary Materials.

\subsection*{Metapopulation SIR model}

In order to understand the geographical diffusion of diseases, one has to combine the microscopic contagion processes with the long-range disease propagation due to human mobility across different spatial scales. In order to tackle this problem, epidemic modeling has relied on
reaction-diffusion dynamics in metapopulations \cite{brockmann2013hidden}. Metapopulations can be thought as nodes of a complex network of spatial patches, where links encode human flows from one place to another and are responsible for between-patch transmission \cite{hagenaars2004spatial}.

We denote by $M$ the number of patches, $N$  the total number of agents and $N_i$ the population of the i-th patch. At any time, we have $\sum_i N_i = N$ and, if the system is closed (i.e. there are no births and deaths), this number $N$ is conserved. The mobility of the agents between the patches is ruled by a weighted adjacency matrix $\mathbf{W}$, whose entry $W_{ij}$ is the flux from patch $i$ to patch $j$. The probability $P_{ij}$ that an agent placed in $i$ moves to $j$ must be proportional to the flux $W_{ij}$ and reads (as  in \cite{brockmann2013hidden})
\begin{equation}
    P_{ij} = \frac{W_{ij}}{\sum_{j=1}^M W_{ij}}
\end{equation}
At this point one has to introduce the reaction dynamics, that takes place independently within the patches. A very wide-used and simple model for this is the so called SIR-model: agents belongs either to the susceptible (S), the infected (I) or the recovered (R) compartment; therefore, in each patch $i$ we have that $S_i + I_i + R_i = N_i$.
The allowed reactions are the following: one agent that is in patch $i$ can move from the $S_i$ to the $I_i$ state by getting the infection from another infected agent with infection rate $\beta$ \cite{brockmann2013hidden}
\begin{equation}
    S_i+I_i\xrightarrow{\beta} I_i + I_i
\end{equation}
and one agent (in patch $i$) can move from the $I_i$ to the $R_i$ state by healing from the infection with recovery rate $\mu$ \cite{brockmann2013hidden}
\begin{equation}
    I_i\xrightarrow{\mu} R_i
\end{equation}
therefore, the continuous-time equations for the infection dynamic are
\begin{equation}\label{eqn: infections}
    \begin{cases}
        \frac{dS_i}{dt} = -\beta I_i\frac{S_i}{N_i} \\
        \frac{dI_i}{dt} = \beta I_i\frac{S_i}{N_i} - \mu I_i \\
        \frac{dR_i}{dt} = \mu I_i
    \end{cases}
\end{equation}
Now, if we work under the assumption that the state of an agent does not affect its diffusive behavior, the mobility for all the agents is described by a unique mobility matrix $\mathbf{P}$, and the continuous-time equation relative to the mobility for a generic compartment $X$ is
\begin{equation}\label{eqn: mobility}
    \frac{dX_i}{dt} = \sum_{j=1}^M\mathcal{P}_{ji}X_j-\sum_{j=1}^M\mathcal{P}_{ij}X_i
\end{equation}
Therefore by summing up Eqs.(\ref{eqn: infections},\ref{eqn: mobility}) one obtains the system of $3M$ differential equations of the model (that basically is the same of \cite{castioni2021critical} with $\epsilon=1$)
\begin{equation}\label{eqn: system}
    \begin{cases}
        \frac{dS_i}{dt} = -\beta I_i\frac{S_i}{N_i} + \sum\limits_{j=1}^M\mathcal{P}_{ji}S_j-\sum\limits_{j=1}^M\mathcal{P}_{ij}S_i \\
        \frac{dI_i}{dt} = \beta I_i\frac{S_i}{N_i} - \mu I_i + \sum\limits_{j=1}^M\mathcal{P}_{ji}I_j-\sum\limits_{j=1}^M\mathcal{P}_{ij}I_i\\
        \frac{dR_i}{dt} = \mu I_i + \sum\limits_{j=1}^M\mathcal{P}_{ji}R_j-\sum\limits_{j=1}^M\mathcal{P}_{ij}R_i
    \end{cases}
  \end{equation}
This system of equations cannot be solved analytically but only numerically; notice that for the basic reproductive ratio we simply have $R_0=\beta/\mu$ (see for example \cite{castioni2021critical} for details).

\subsection*{Simulation details}
All the simulations were done by choosing the free parameters $\beta$ and $\mu$ in order to have $R_0 = 2.6$, building the mobility matrix from the data and initializing the population of the patches with the data. The local population was set as proportional to the reported baseline userbase for Facebook data in countries different from Italy, while for Facebook Italy, Cuebiq, Safegraph, and Census data the resident population has been used. The other free parameter to be chosen (over the total duration of the simulation, that was selected to be $t=200$ days just to be sure to reach the end of the epidemic) is the time step, that was chosen as $dt=0.5$ days. The effect of this parametric choice has been tested by performing simulations with $dt=0.1$ days, which yielded exactly the same numerical results.
Given the initial conditions and the parameters, the code for the simulations solves Eq.(\ref{eqn: system}) by using the built-in \texttt{ode} function of the software \texttt{R}, providing the time evolution for $S$,$I$,$R$ individuals in all the patches at each time step starting from $t=0$. All the simulations have at $t=0$ a fully susceptible population but one infected (the seed), whose location was fixed, in each network, in the node with the larger population.
From the time evolution, the code trivially calculates the total incidence (namely the total number of new infected) at each time step, the number of recovered at the end of the epidemic in each patch and the attack rate in each patch. Notice that by the moment that in some networks unconnected components are present, in order to calculate the attack rate the recovered were not divided by the total population of the network, but only by the population of the component of the network in which the initial seed was placed.

%%%%%%%%%%%%%%%%%%%%%%%%%%%%%%%%%%%%%%%%%%%%%%%%%%%%%

\bibliographystyle{naturemag} 

\bibliography{biblio}% Produces the bibliography via BibTeX.

\section*{Acknowledgments}
We thank Cuebiq and Facebook for providing us free access to their Flow Network Data through their Data for Good programme.\\

\clearpage

%%%%%%%%%%%%%%%%%%%%%%%%%%%%%%%%%%%%%%%%%%%%%%%%%%%%%
\section*{\label{sec:SM}Supplementary Text}

\subsection*{Technical limitations affecting human mobility datasets}

In this manuscript, we observe great differences in how mobility datasets coming from different providers are able to describe mobility flows at a country level. In this section of the Supplementary Materials, we briefly discuss here a series of possible causes for the differences. This list does not however represent a comprehensive text illustrating all biases associated with mobility data collection and elaboration, for which we suggest further readings~\cite{bonnel2015passive,olteanu2019social,escap2019handbook,de2021updating}.

\subsubsection*{Behaviour captured}

Most datasets discussed are being passively collected from ICT data records. They are therefore necessarily influenced by the type of behaviour that produces the records. Issues associated with the behaviour captured cannot in general be solved, but awareness of the consequences of passive collection of movement behaviour can limit the errors introduces in the analysis.

Mobile Phone Call Detail records  used to be associated with phone calls and SMS messages, but as the use of the ``data'' connection is becoming widespread, the data captured are now less coupled with the communication behaviour, but can still be limited by the closure of the data connection by users. 

Providers aggregating data from several mobile apps, such as Cuebiq or Safegraph, and on a broader sense also Google and Facebook, capture data when the device GPS connection is used by one of the location based applications associated with the providers, which may require the user to be actively using that app or not, depending on the specifics of the app. 

Other GPS data such those of Octotelematics coming from devices installed to cars for insurance reason, but also data from Tomtom data coming from single apps specifically gathering the movements of bikers or runners, are necessarily limited to movements associated with a specific mode of transport.

Finally, the accuracy of census data can be hindered by the tendency of users to not update their records in case of relocation. This tendency may depend upon the incentives  set by each country for this update (for example, in the case of Italy, citizens are highly incentivised to update their public records to have access to the Public Health System).

\subsubsection*{Segmentation and sampling}

Mobility data is generally gathered in form of raw trajectories formed by a sequence of position where a user has been observed at a certain time. For understanding the user movement behaviour, these trajectories are usually segmented into a sequence of movements in space and stops where the user stays in a particular place for a certain duration in time. There are several alternative methods for this segmentation. Many data provider designed their own optimised for their data which details are often undisclosed private. The observed segmented results can be very sensitive to algorithmic and parametric choices, as discussed in~\cite{gallotti2018tracking}. The impact of this issue can be reduced for academic use by allowing to operate directly on the raw trajectories using open source methods with explicit parametric choice.

\subsubsection*{Spatial Granularity} 

Segmented trajectories are, in most cases, further aggregated over (administrative) areal units to obtain the so-called Origin-Destination (OD) matrices. The choice of those areal units clearly influences the observed results, a problem known in literature as ``Modifiable areal unit problem''. Many datasets circulating are influenced by this problem. In Supplementary Figure~\ref{figSI_granularity}, we illustrate this with an example using it on our main indicator, the displacement distribution $P(L)$. In panel (a) we show how the distances estimated between the centroids of the unit areas may differ from the average distance of users moving between these areas, while in panel (b) we show how different spatial granularity yields to different observed $P(L)$. While this problem is intrinsic with the choice of aggregating data for obtaining a flow matrix, allowing researchers to use the granularity most adapt to the analysis performed instead to adapting to a standard one chosen by the data provider can without any doubt limit the biases introduced.

\subsubsection*{Upscaling}

All data gathered, with the exception of Census, describe only a fraction of the total population. Often, for ICT-based datasets, the representativity of this sample is also biased towards high-income, more urban and mostly male populations. To estimate the real mobility flows of the whole population between different areas, that is the information captured in the OD matrices, it is necessary to ``upscale'' the observed flows.  Several methods can be used for upscaling, in most case they are based on a proportionality with respect to an estimate of the local observed population. The knowledge of the active populations in each areal unit, which could also be a function of time, becomes therefore a crucial piece of information to rebalance the observed flows to what is expected to be the mobility of the total population. Lack of knowledge about active population can conversely limit the analysis due to the need of having more imprecise upscaling. This issue can be easily solved by the company, by providing explicit information about the user-base population as a function of time (as Facebook and Cuebiq do in the data analysed in this paper), when possible with data disaggregated by income and gender in order to allow for alleviating the potential biases in the data using post-stratification methods.

\subsubsection*{Self-loops}

For many applications, knowing how large is the fraction of people who exited an areal unit with respect to those who circulated only locally is very relevant. This information is captured, in an OD matrix, in the diagonal elements that describe, from a network perspective, self-loops of the mobility flows network. This self loops are in general present in most aggregated datasets. However, it is often not very clear what they do represent. It can encompass at the same time information on the number users observed in the dataset, i.e. those activate the capturing device but never left their home location. Or alternatively users who performed a local movement (as identified by the segmentation algorithm), but never arriving at leaving the areal unit. Unless specified, it is in principle not clear if users who belong to the self loop are also counted it outwards flows. All this info information can be very relevant for mobility analysis and epidemic modelling and we advise, as a minimum, to describe clearly what is represented by self loops and to further include any other ``non movement'' information available among the information accompanying the OD matrices.

\subsubsection*{Pruning}

In all datasets, small flows between two areas can be unobserved because of the local user-base population is too small or because the behaviour captured or the segmentation method has a limited efficiency at identify the users' movement. This makes the observed OD matrices systematically sparser than the real ones. This effect is further worsened by an active ``pruning'' of the mobility network done to limit the size of the datasets shared and to ensure K-Anonymity. As discussed in the paper, in the Facebook data we estimated that the pruning happens for flows below (10 users)/(8 hours), while Google declare cuts below 100 users/week, and this pruning heavily influences our possibility of using the data collected for studying long-range mobility. This major problem can be solved, again, by providing to researcher a way of elaborating directly raw data without applying any heavily lossy pre-elaboration such as pruning.

\subsubsection*{Fragmentation}

Again for limiting the size of data archives, the datasets provided are often divided into large administrative ares (countries, US states, ...). Similarly to pruning, also this process forcefully cuts  long range mobility. This problem can be illustrated by the data provided by Facebook for USA, where they used state-level fragmentation, not allowing for any analysis about the spreading behaviour across the continental USA. Spreading processes over wide areas is more than the sum of the spreading processes over its parts. The issue of fragmentation is unfortunately not easy to solve, as in many cases the data providers only has data associated to a single country. This makes the problem of studying international flow one of the hardest we face. Only a few sources have international coverage, but aviation data are limited to a single mode of transport that can underestimate mobility between contiguous countries. Google data, as we saw, is heavily influenced by pruning cutting long range flows and Facebook offers only fragmented data. We believe that this great limitation in the opportunity of analysing international flows, information nowadays so vital for epidemic containment, represent a paradigmatically example illustrating the need of a shared platform where to share, elaborate and analyse mobility data for academic purposes.

\clearpage

%%%%%%%%%%%%%%%%%%%%%%%%%%%%%%%%%%%%%%%%%%%%%%%%%%%%%

\section*{\label{sec:SM}Supplementary Figures}

\setcounter{figure}{0} %reset figure counter

%MAPS SAME DENSITY
\begin{figure*}[ht!]
\renewcommand{\figurename}{Supplementary Fig.}

\begin{center}
\begin{tabular}{ccc}
\raisebox{2.3cm}{(a)} \includegraphics[angle=0,width=0.3\textwidth]{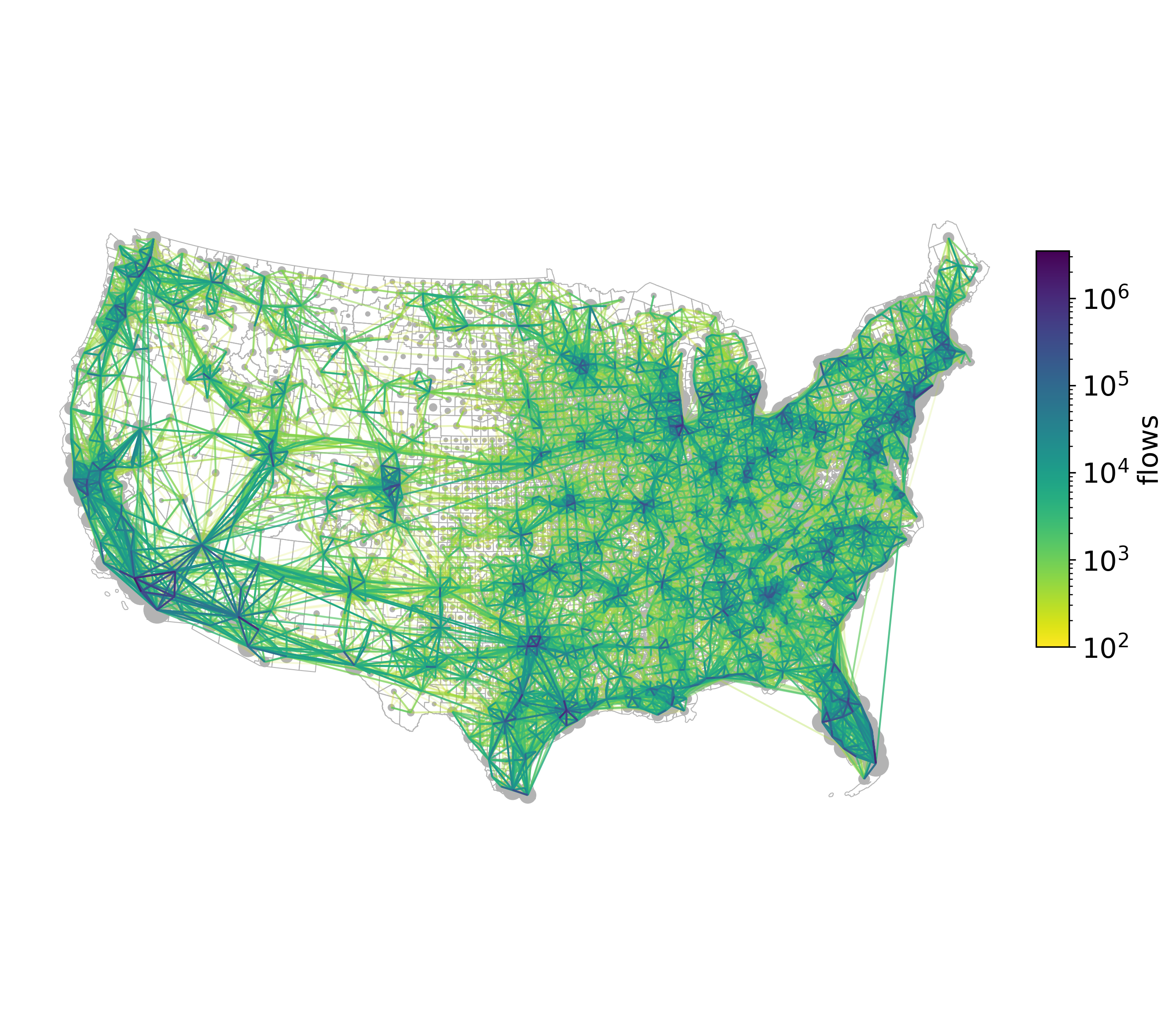} &
\raisebox{2.3cm}{(b)} \includegraphics[angle=0,width=0.3\textwidth]{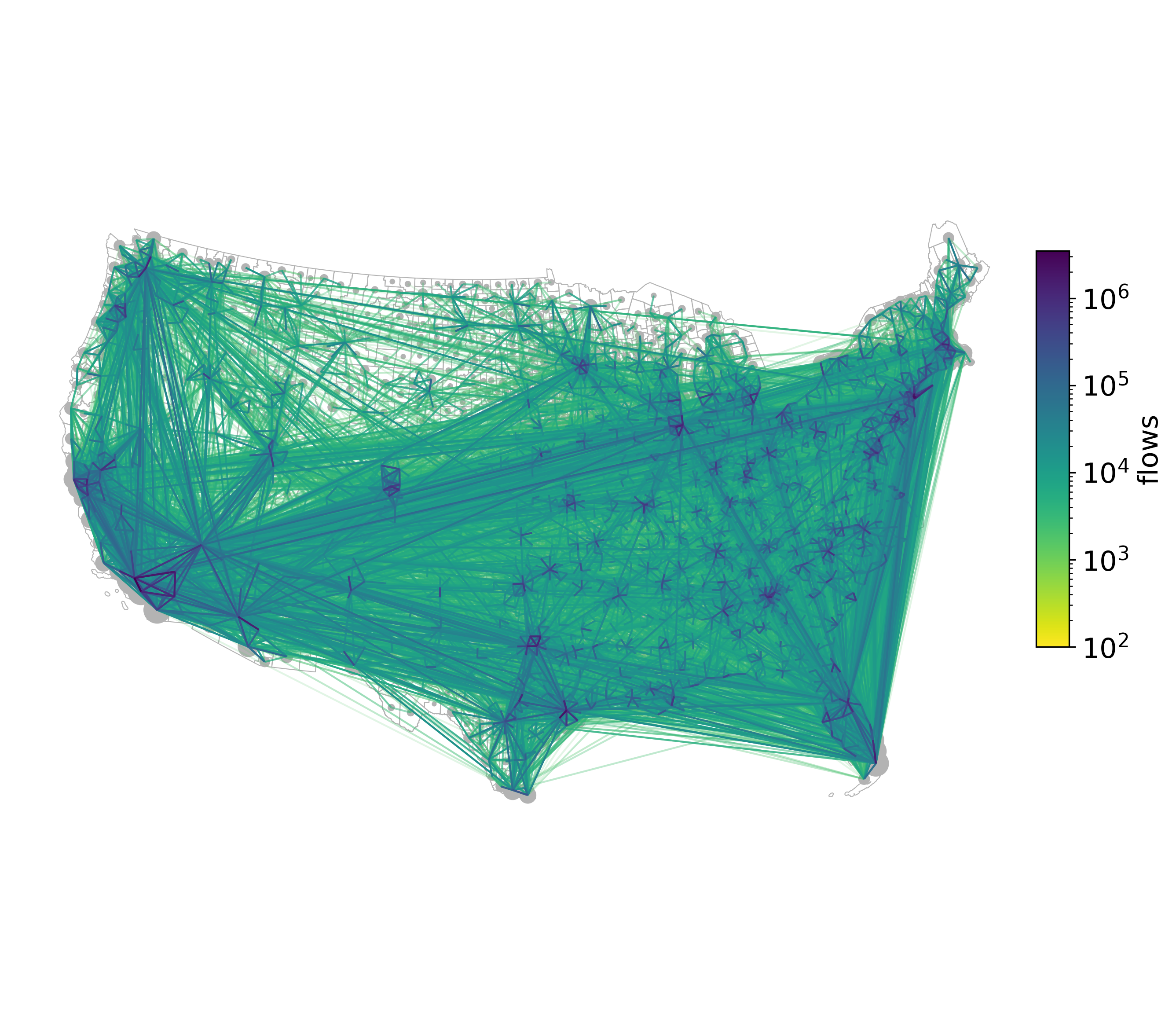} &
\raisebox{2.3cm}{(c)} \includegraphics[angle=0,width=0.3\textwidth]{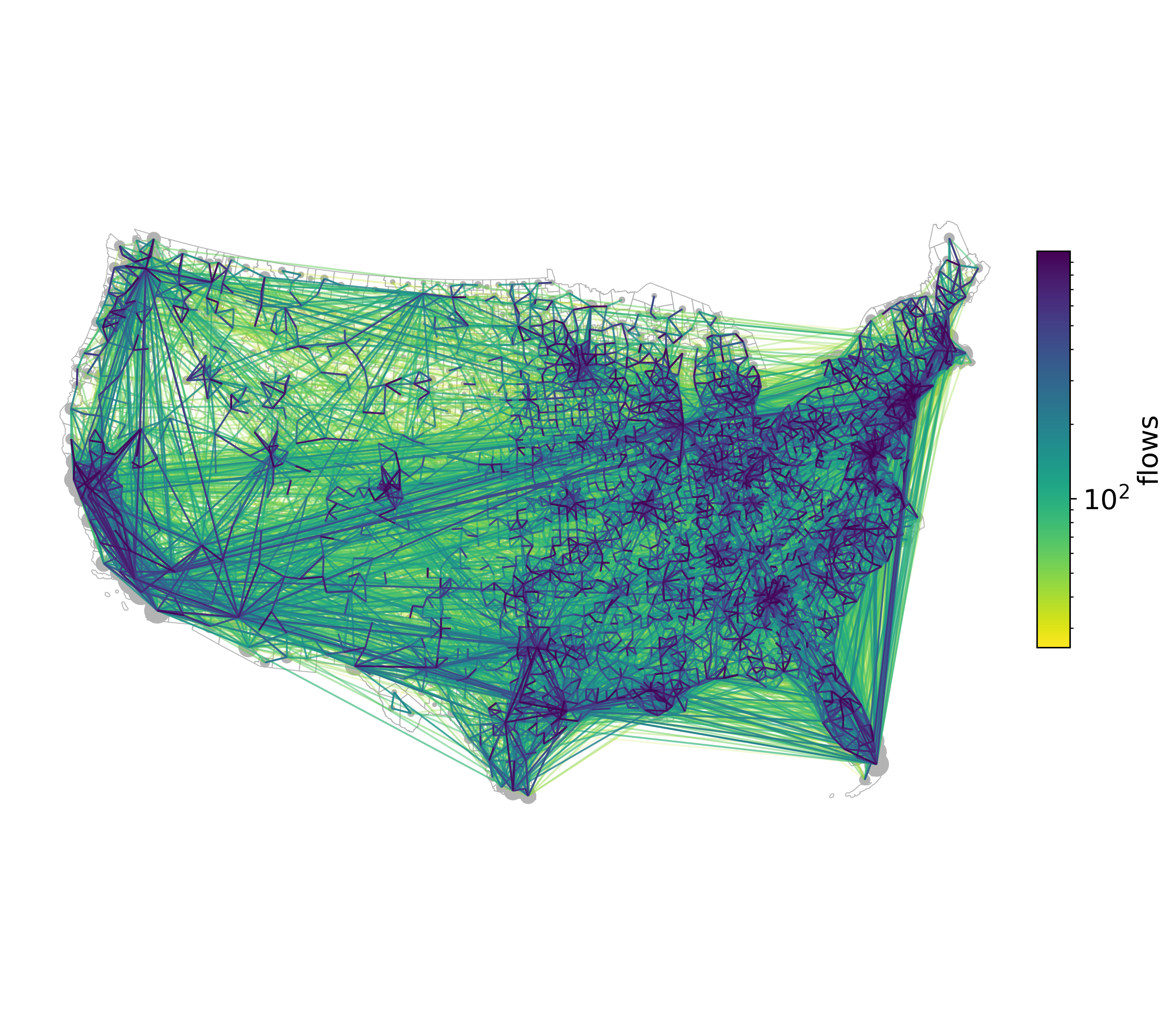} \\
\raisebox{2.3cm}{(d)} \includegraphics[angle=0,width=0.3\textwidth]{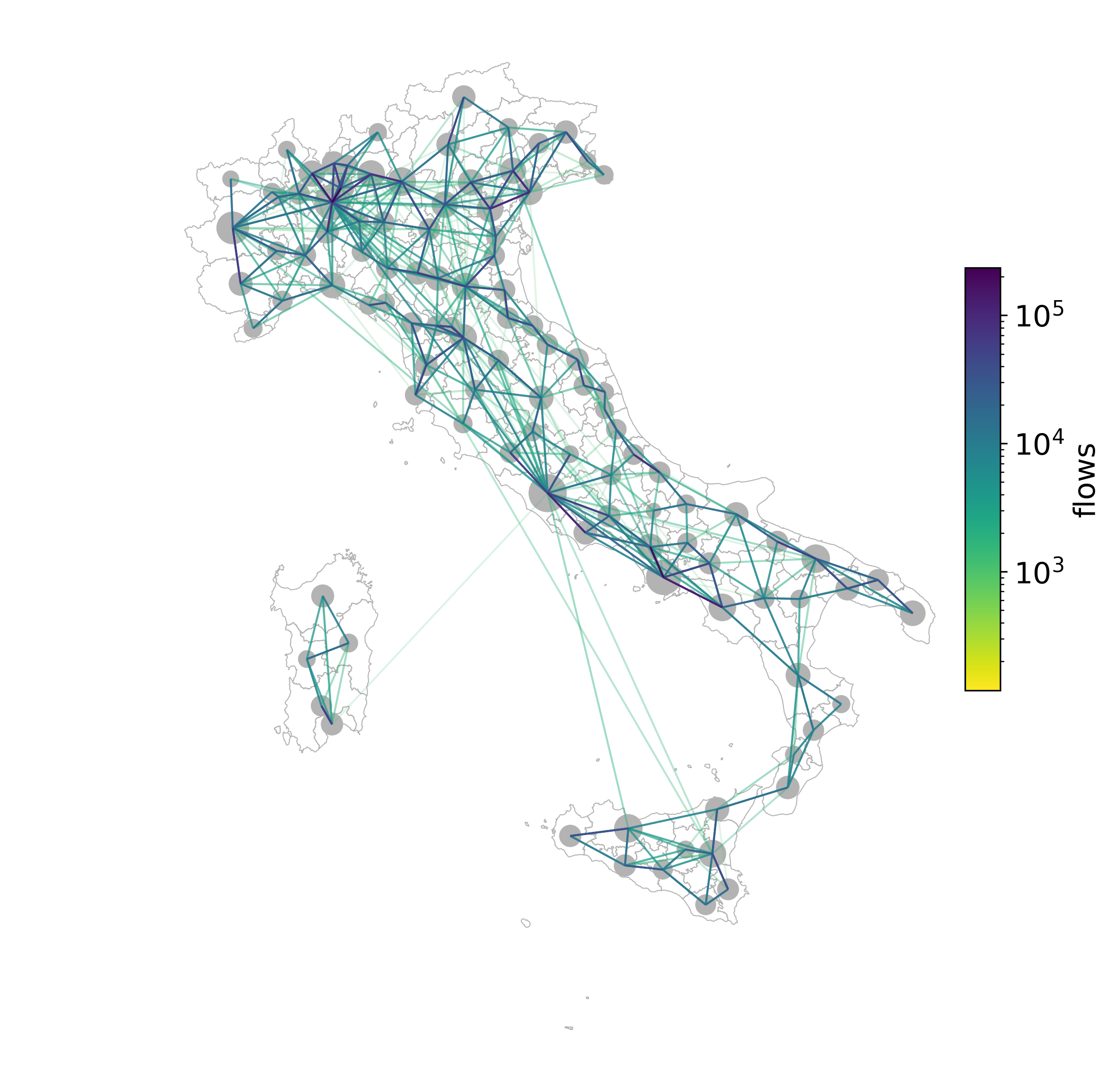} &
\raisebox{2.3cm}{(e)} \includegraphics[angle=0,width=0.3\textwidth]{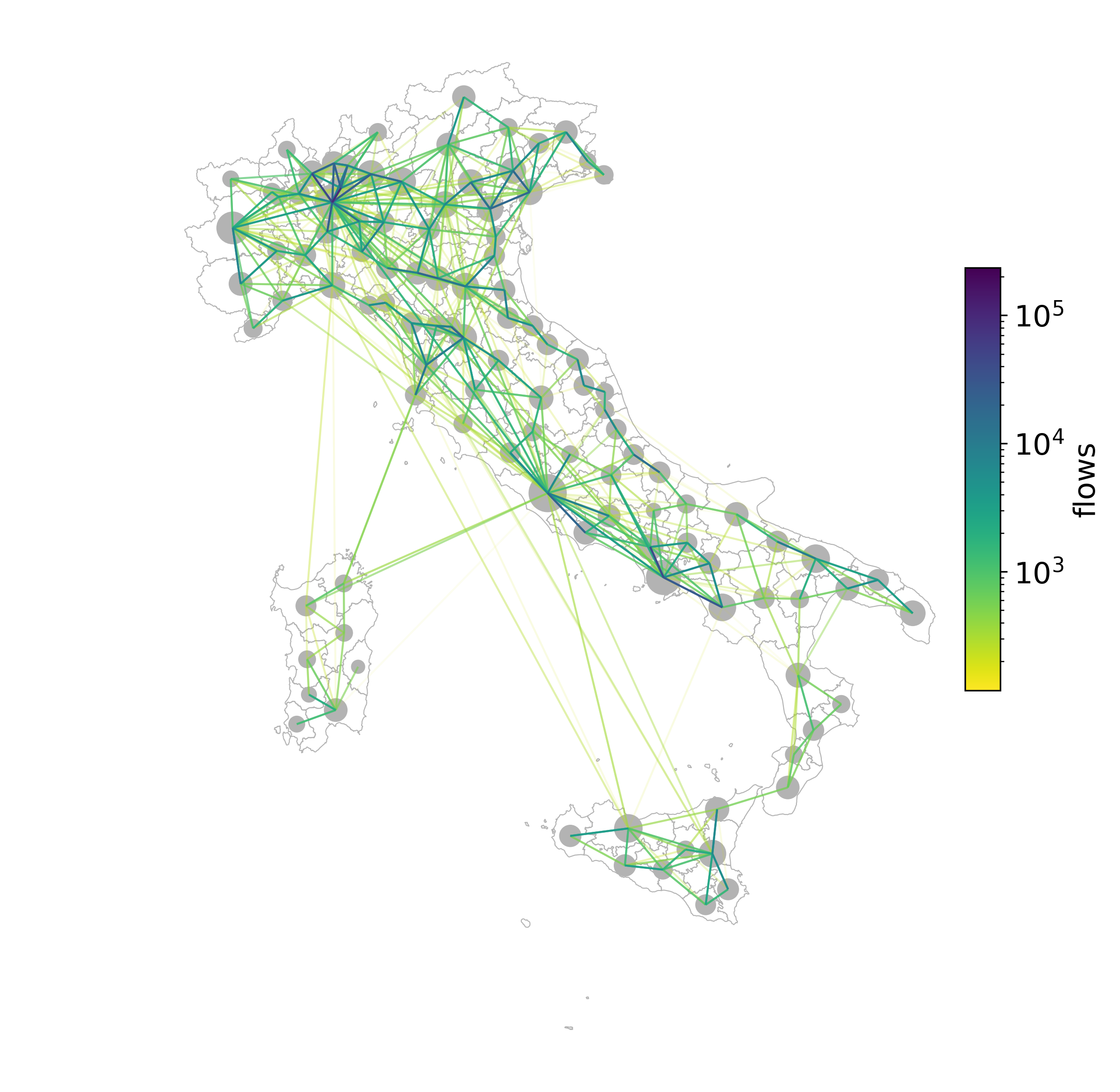} &
\raisebox{2.3cm}{(f)} \includegraphics[angle=0,width=0.3\textwidth]{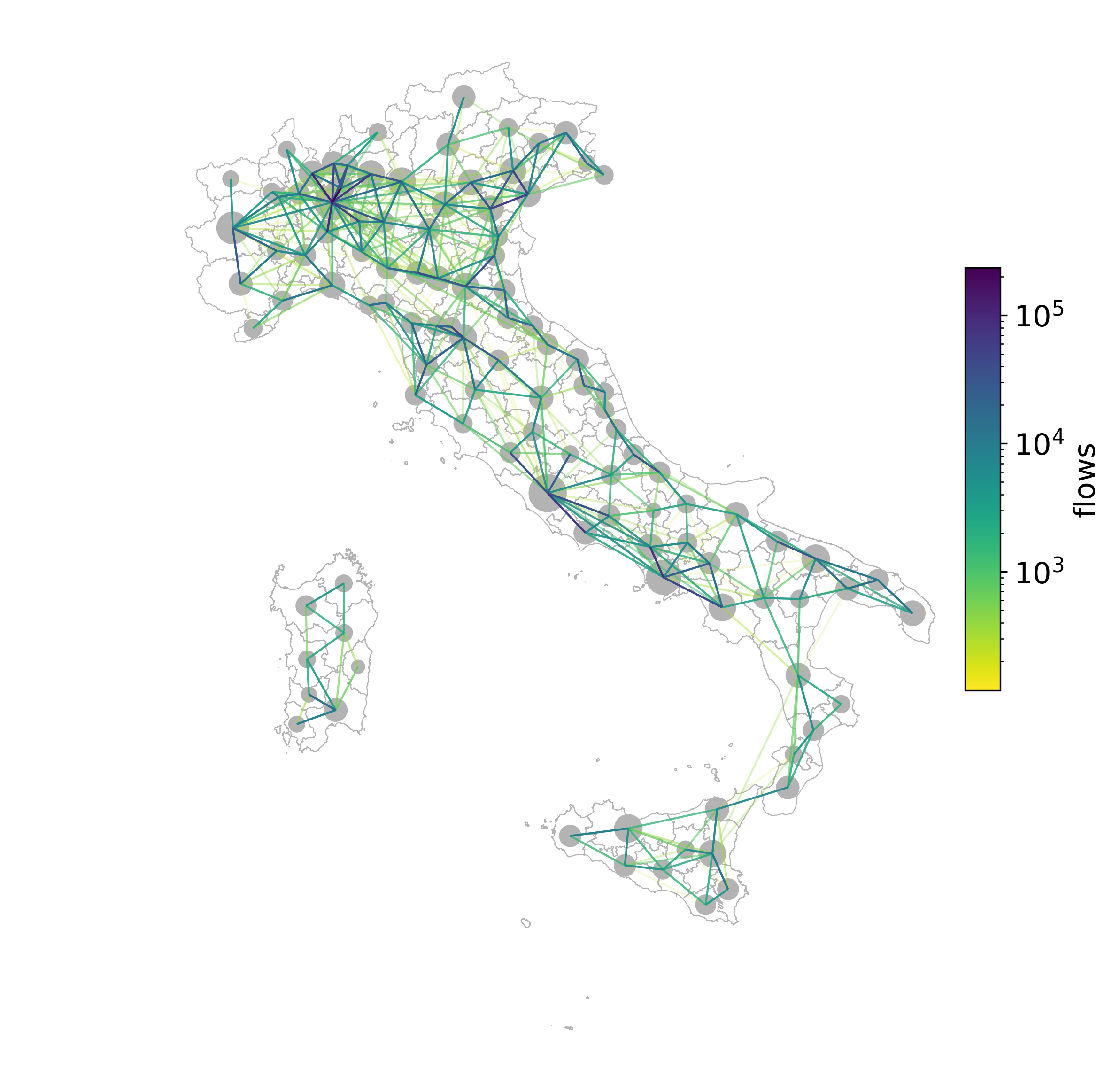} \\
\end{tabular}
\end{center}
\caption{
{\bf Visual comparison of baseline flow maps selecting a fixed number of high flow edges.} Here we display the same networks displayed in Fig.1, but limiting each network to having at max the number of edges of the sparser network of the same country by removing edges with smaller flows. For U.S.A. we have Cuebiq {\bf (a)} which is the sparser, Safegraph {\bf (b)} and Census {\bf (c)}. Here networks are visibly different even if one compensate for the different edge density. For Italy we have Cuebiq {\bf (d)}, Facebook {\bf (e)} and Census {\bf (f)}. In this case, the differences between networks are less pronounced, with only notable difference the lack of connectivity between the peninsula and the islands of Sardigna and Sicily for Census data (panel c)
}
\label{figSI_maps_same_density}
\end{figure*}

%OTHER COUNTRIES
\begin{figure*}[ht!]
\renewcommand{\figurename}{Supplementary Fig.}

\begin{center}
\begin{tabular}{cc}
\raisebox{2.3cm}{(a)} \includegraphics[angle=0,width=0.45\textwidth]{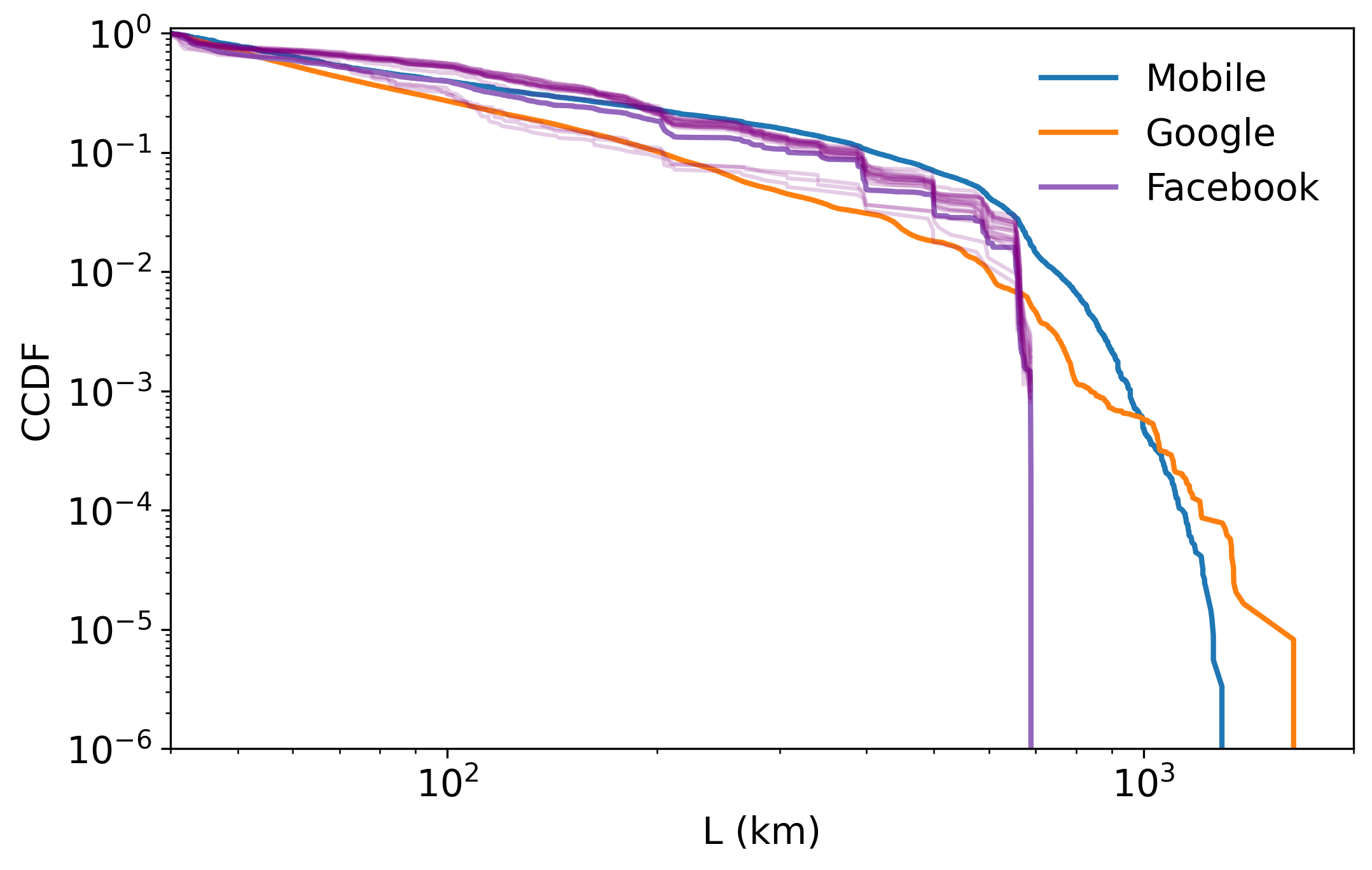} &
\raisebox{2.3cm}{(b)} \includegraphics[angle=0,width=0.45\textwidth]{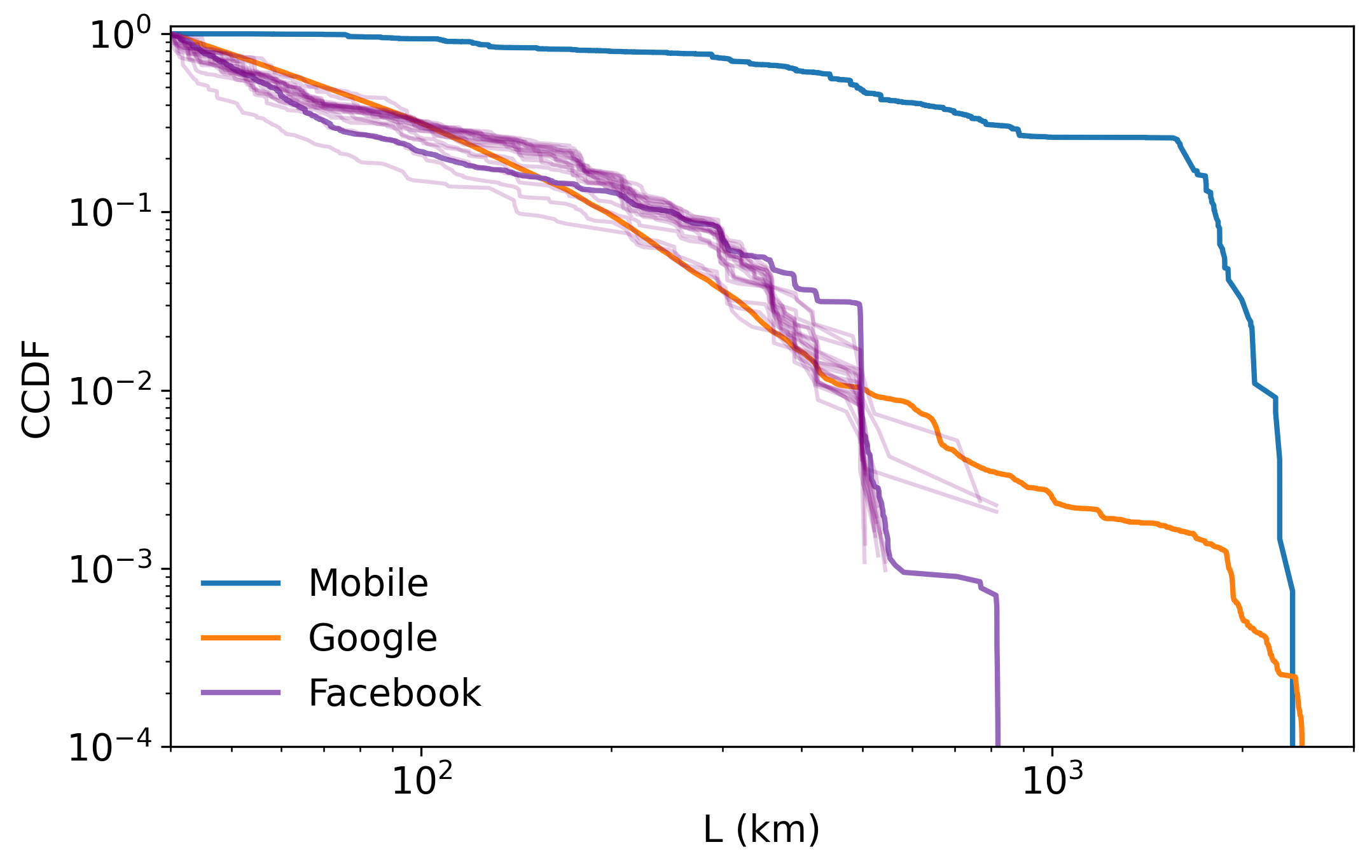} \\
\end{tabular}
\begin{tabular}{c}
\raisebox{2.3cm}{(c)} \includegraphics[angle=0,width=0.45\textwidth]{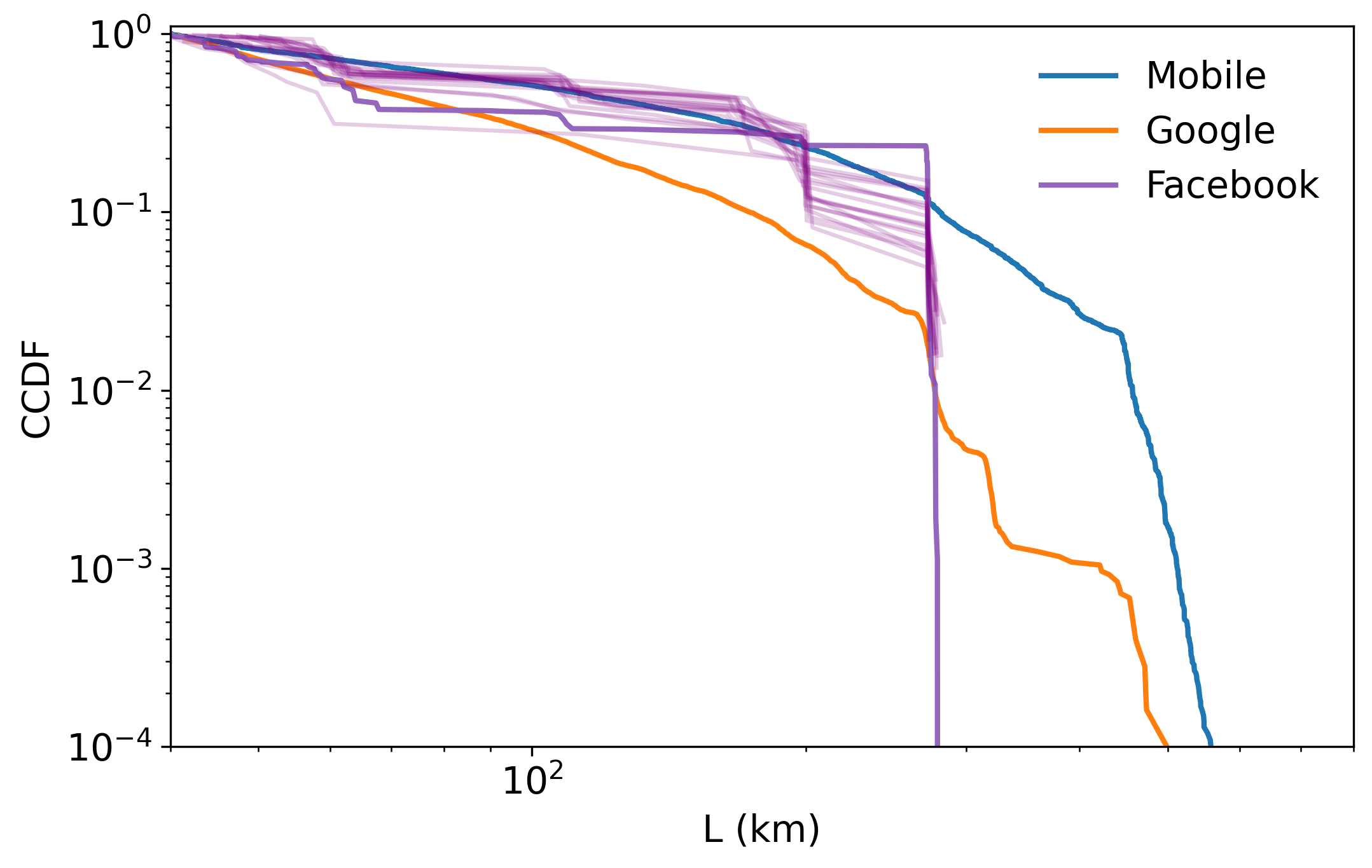}
\end{tabular}

\end{center}
\caption{
{\bf The Displacement Distributions of France (a), Spain (b), Portugal (c).} Here, we rescaled the mobile data coming from~\cite{Tizzoni:2014cl} to become representative of the underlying population similarly to what done with Cuebiq and Facebook data. The light purple lines represent the evolution in time of the Facebook dataset, while the thick line represents the baseline. The differences associated with the COVID-19 lockdown, one of the more impactful events in world history, is smaller than the observed difference between datasets. 
}
\label{figSI_other_countries_pl}
\end{figure*}

% DIFFERENCE BETWEEN DATASETS & FITTING P(L)
\begin{figure*}[ht!]
\renewcommand{\figurename}{Supplementary Fig.}

\begin{center}
\begin{tabular}{ccc}
\raisebox{2.3cm}{(a)} \includegraphics[angle=0,width=0.3\textwidth]{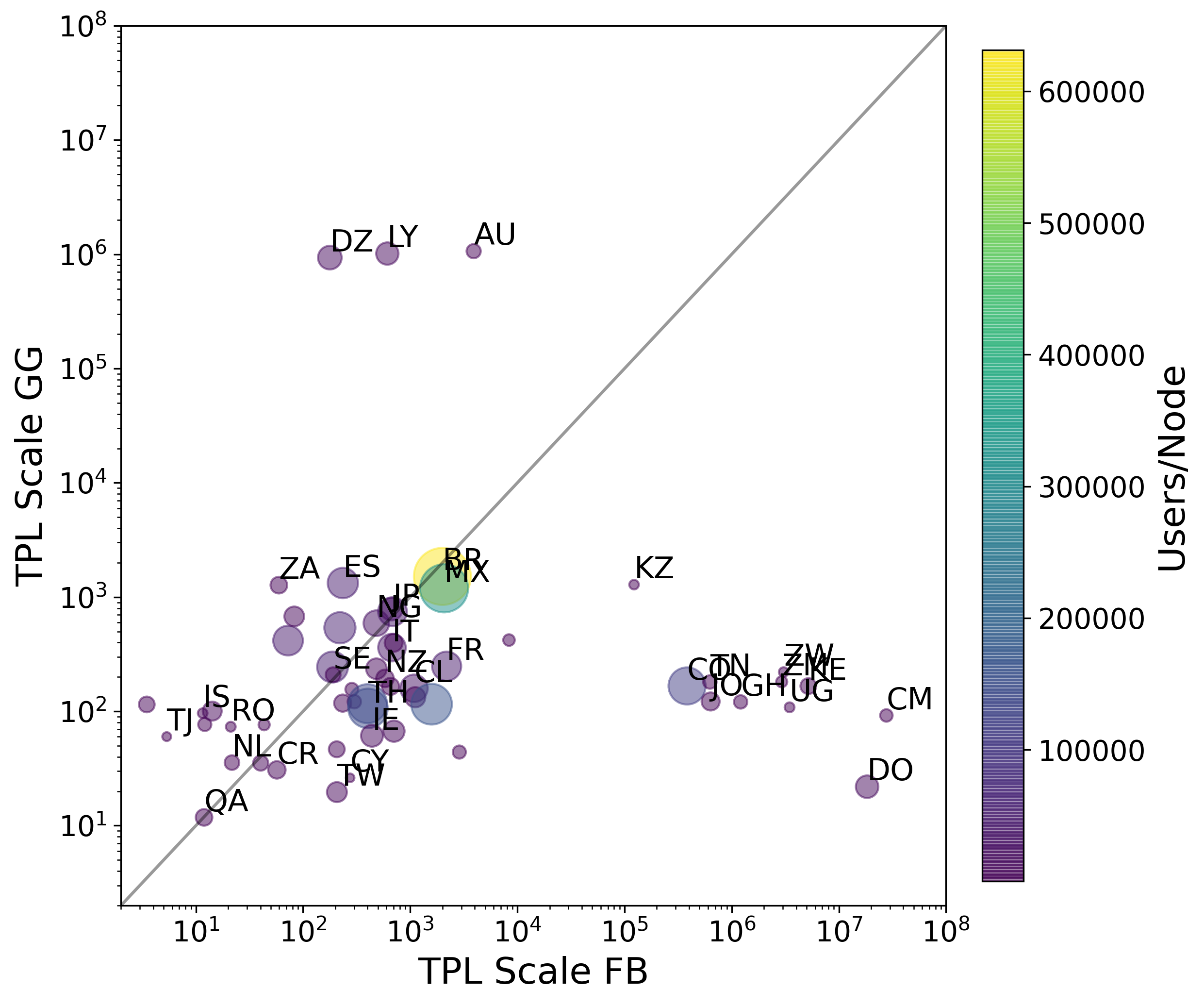} &
\raisebox{2.3cm}{(b)} \includegraphics[angle=0,width=0.3\textwidth]{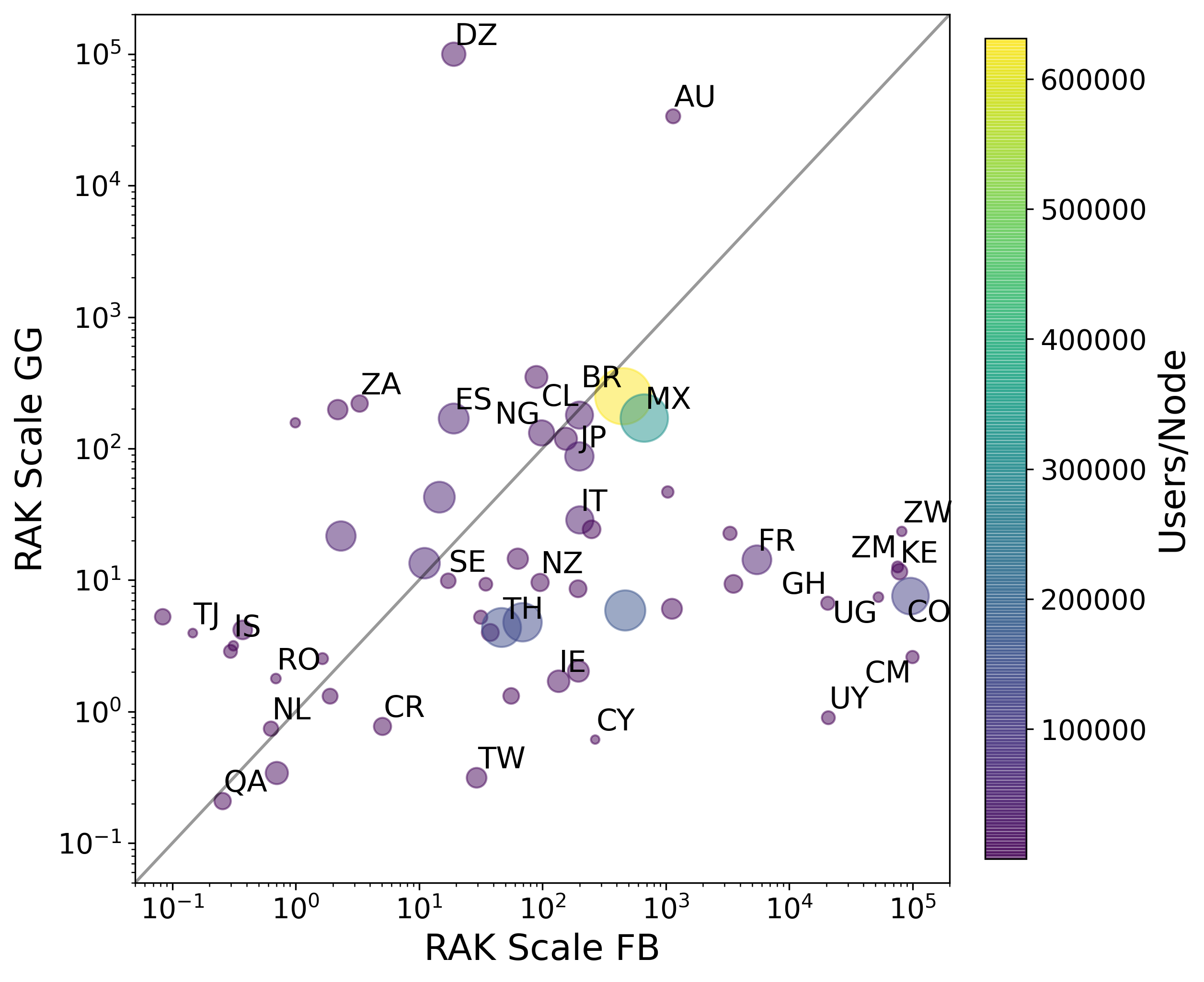} &
\raisebox{2.3cm}{(c)} \includegraphics[angle=0,width=0.3\textwidth]{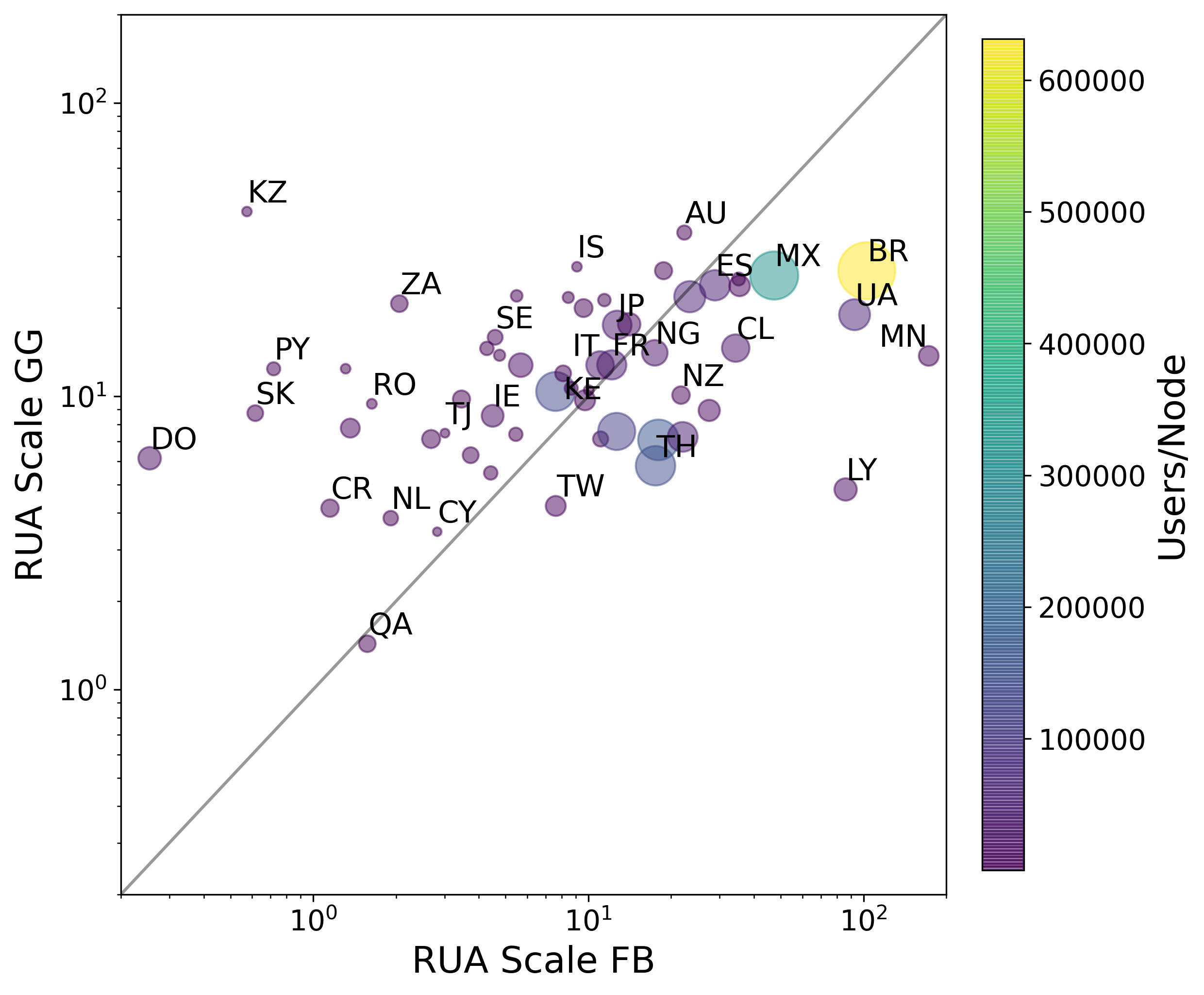} \\
\end{tabular}
\end{center}
\caption{
{\bf Comparison of the scale parameters fitted for Facebook and Google datas with different models.}  We illustrate here the difference between the fitted parameters for a set of 58 countries that are present in both the Facebook and the Google datasets. In panels a) and b) we represent the scale parameter of two bi-dimensional fits (truncated power law $P(L) \propto L^{-\alpha} e^\frac{L}{L_{s}}$ and the saddle point approximation of the random acceleration kicks model $P(L) \propto L^{-\alpha} e^{\left( \frac{L}{L_{s}}\right)^{0.5}}$ ). Due to the very sensible degree of freedom associated with the scaling parameter, the differences observed in the distance scale $L_s$ are way larger than what observed in c), where the scale is estimated using the monoparametric random uncorrelated acceleration model ($P(L) \propto L^{-0.75} e^{\left( \frac{L}{L_{s}}\right)^{0.5}}$ )
}
\label{figSI_google_facebook_parameters}
\end{figure*}

% TIME DEPENDENCY
\begin{figure*}[ht!]
\renewcommand{\figurename}{Supplementary Fig.}

\begin{center}
\begin{tabular}{cc}
\raisebox{2.3cm}{(a)} \includegraphics[angle=0,width=0.45\textwidth]{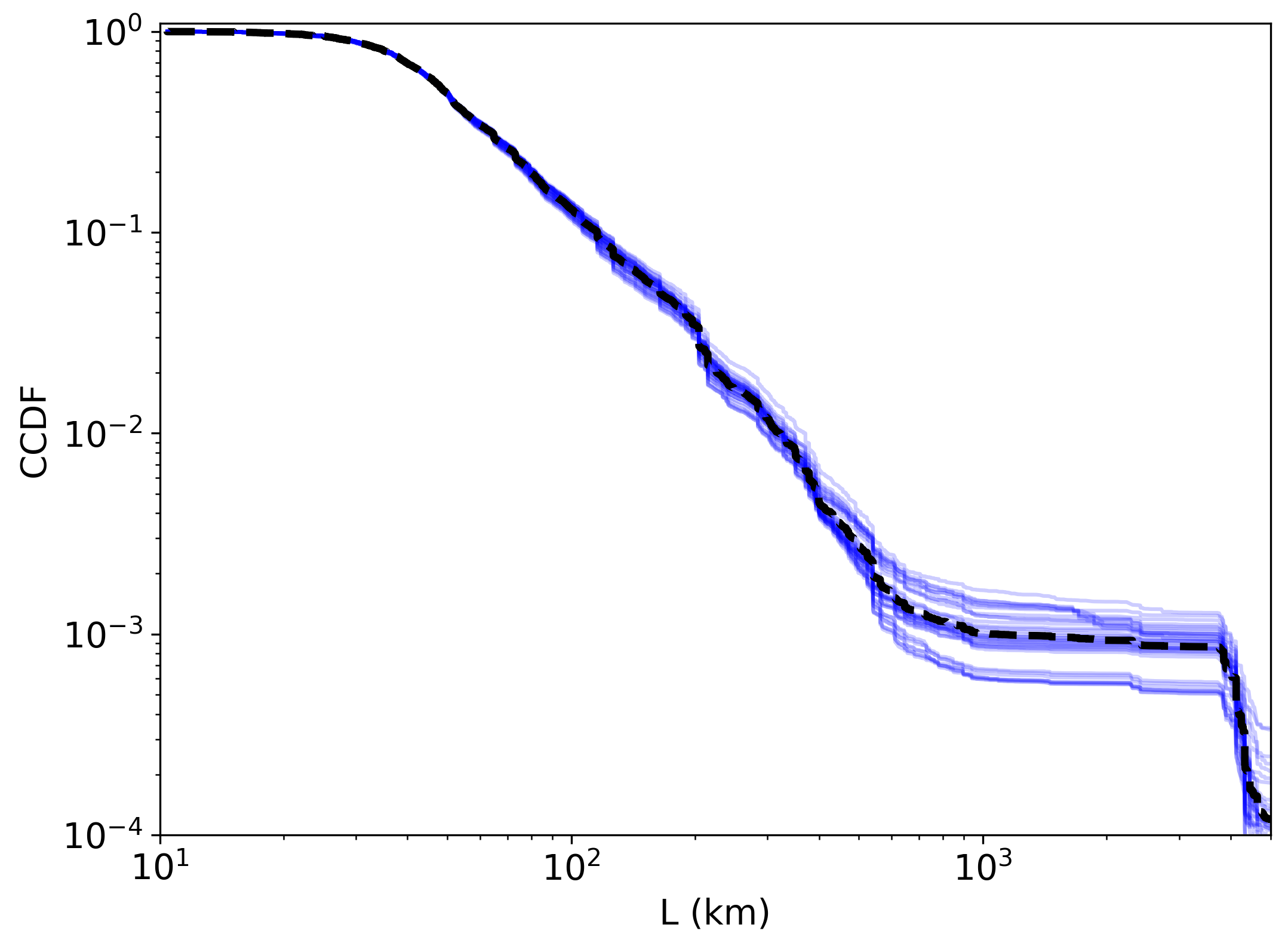} &
\raisebox{2.3cm}{(b)} \includegraphics[angle=0,width=0.45\textwidth]{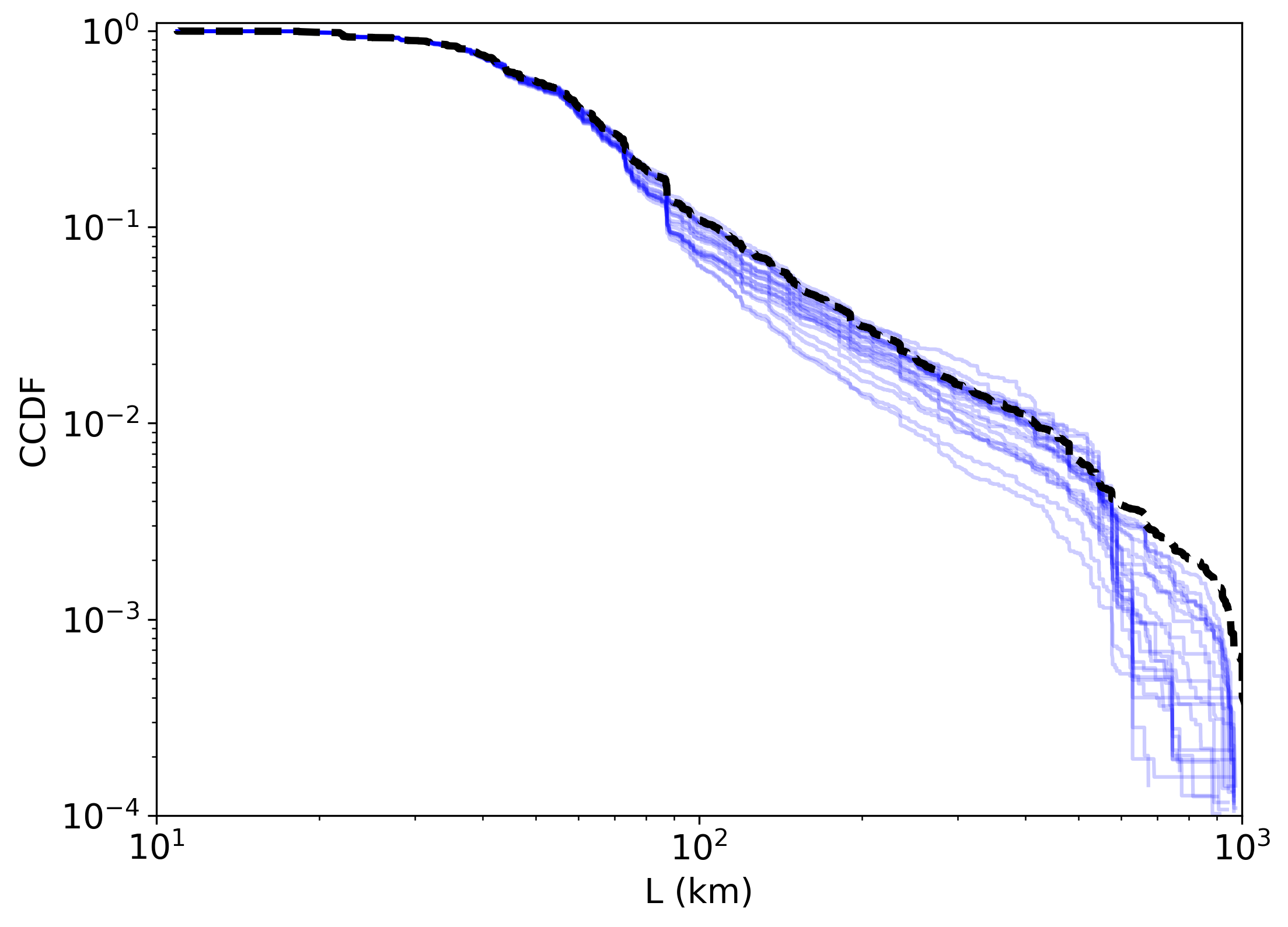} \\
\end{tabular}
\end{center}
\caption{
{\bf The evolution in time of the Cuebiq data in the USA (a) and Italy (b)} Again, the differences associated with the COVID-19 lockdown is smaller than the observed difference between datasets. There the CCDF is analyzed for $L > 10$ and not limited to the continental U.S.A. The black dashed line represents the baseline curve.
}
\label{figSI_time_dependency}
\end{figure*}

%%%% FIG 3 %%%%
\begin{figure*}[ht!]
\renewcommand{\figurename}{Supplementary Fig.}

\begin{center}
\begin{tabular}{cc}
\raisebox{2.3cm}{(a)} \includegraphics[angle=0,width=0.45\textwidth]{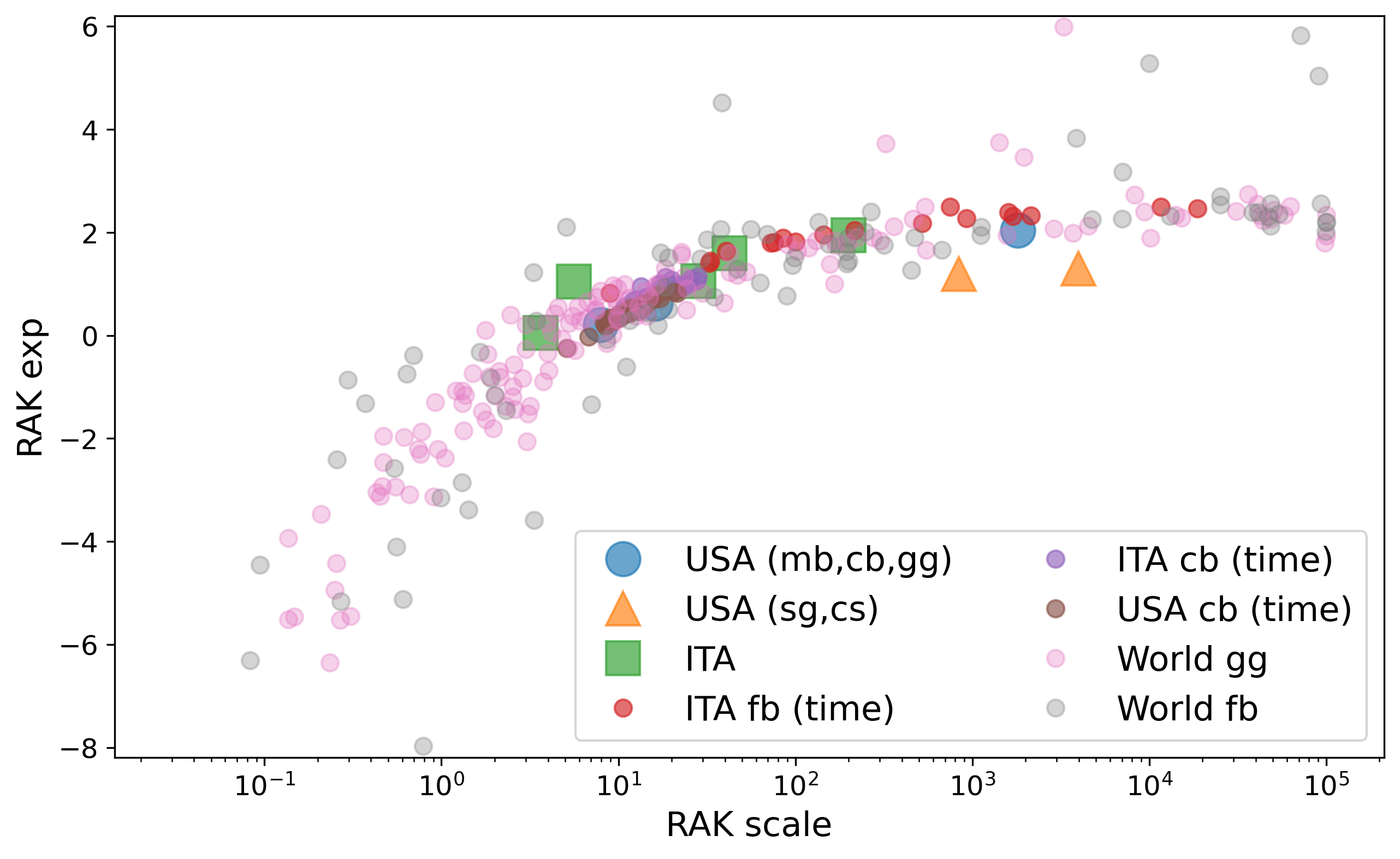} &
\raisebox{2.3cm}{(b)} \includegraphics[angle=0,width=0.45\textwidth]{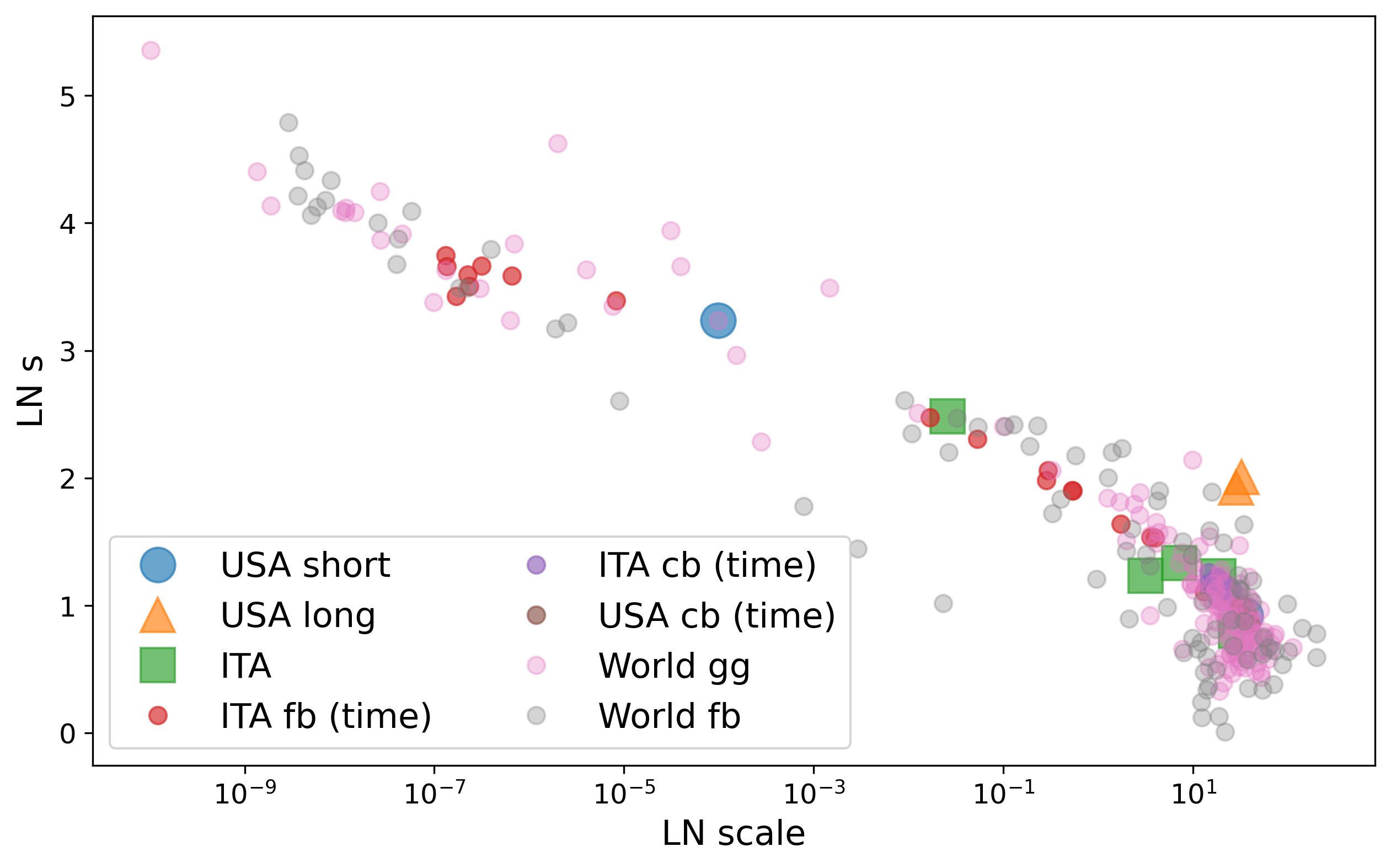} \\
\end{tabular}
\begin{tabular}{c}
\raisebox{2.3cm}{(c)} \includegraphics[angle=0,width=0.45\textwidth]{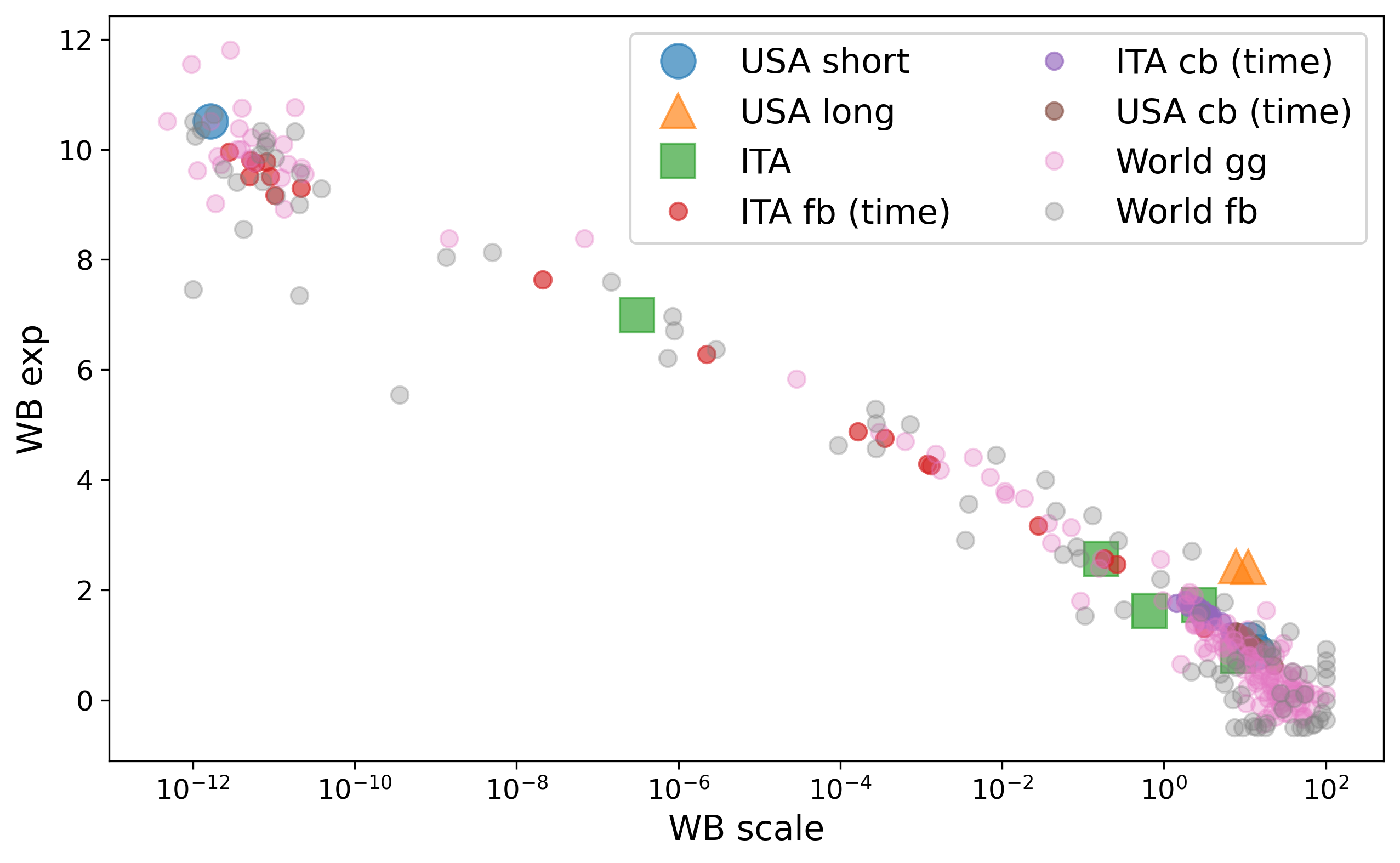}
\end{tabular}
\end{center}
\caption{
{\bf Comparison of the displacement distributions across the globe.} We observe a wide variance in the fit parameters estimated for the $P(L)$. In panels (a), (b), and (c) we represent the the scale and exponent parameter for the Random Acceleration Kicks models, a Lognormal fit, and a Weibull fit respectively. As in Fig.~\ref{fig3}, we include the curves of Figure 1 for USA --Cuebiq (cb), Mobile (mb)~\cite{gonzalez2008understanding}, Google (gg), Safegraph (sg), Census (cs) -- and Italy (that, as illustrated in Figure~\ref{fig1} includes also Octotelematics and Facebook (fb) data). The values for the longitudinal weekly analysis of Cuebiq and Facebook  data, and the values estimated for the Google and Facebook baseline. We observe how all data points align around a curve, which suggests strong correlation between the free parameter and, ultimately, the existence of an underlying law having a single degree of freedom.}
\label{fig3_rak_ln}
\end{figure*}

%%%% FIG 3 %%%%
\begin{figure*}[ht!]
\renewcommand{\figurename}{Supplementary Fig.}

\begin{center}
\begin{tabular}{cc}
\raisebox{2.3cm}{(a)} \includegraphics[angle=0,width=0.45\textwidth]{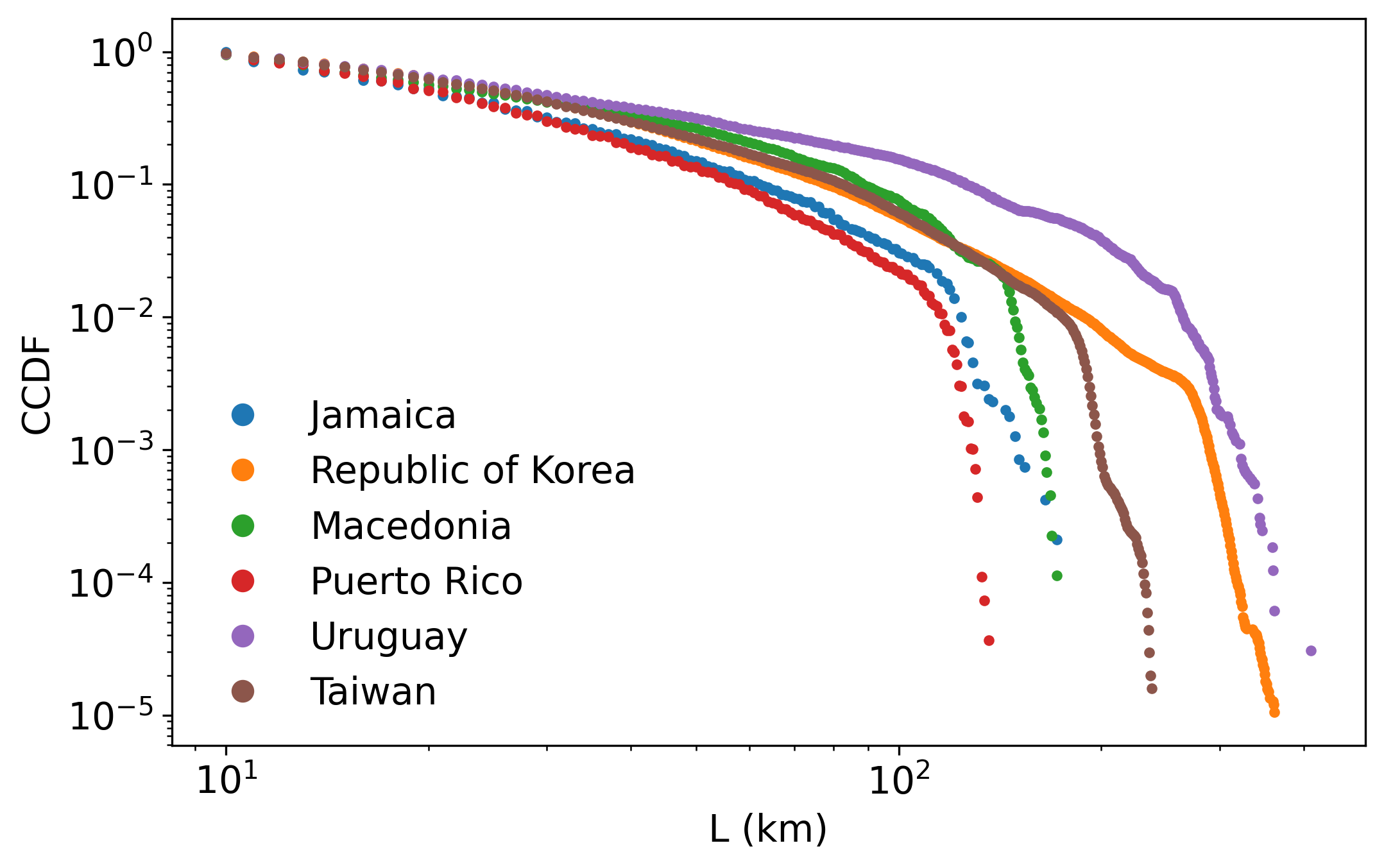} &
\raisebox{2.3cm}{(b)} \includegraphics[angle=0,width=0.45\textwidth]{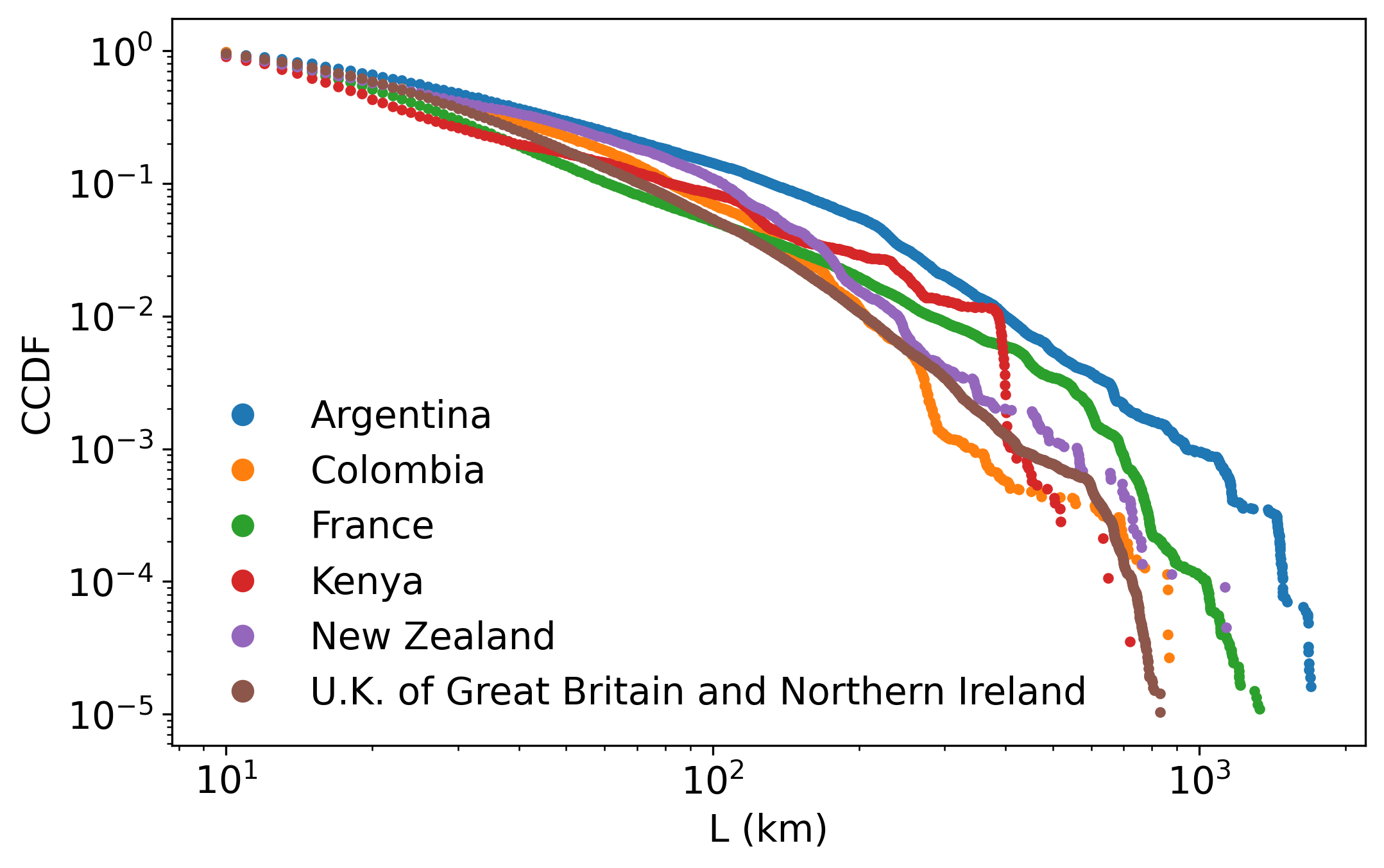} \\
\end{tabular}
\begin{tabular}{c}
\raisebox{2.3cm}{(c)} \includegraphics[angle=0,width=0.45\textwidth]{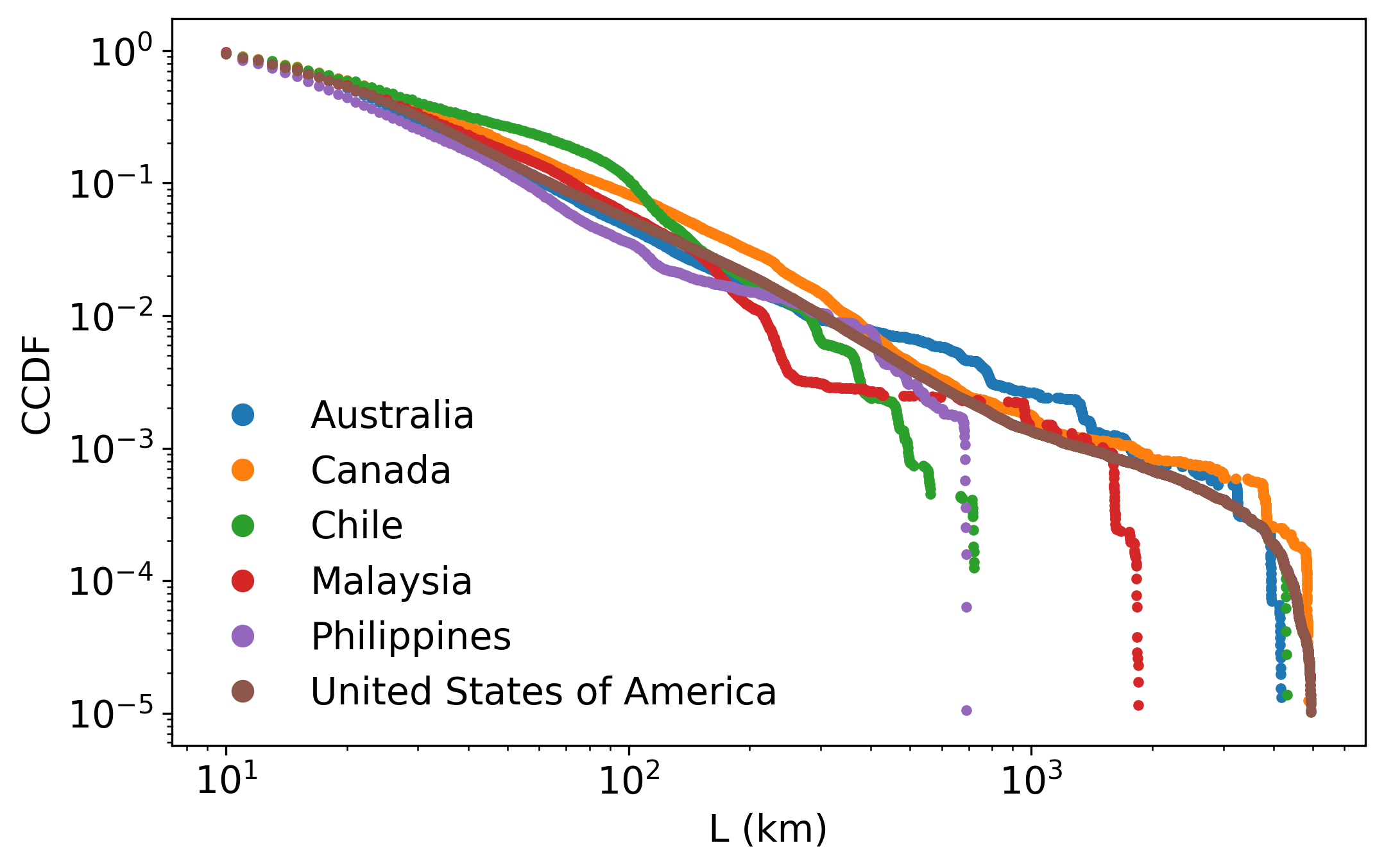} 
\end{tabular}
\end{center}
\caption{
{\bf Comparison of the displacement distributions across the globe.} Using the Google dataset, we observe how the tail behaviour of $P(L)$ is radically different if one consider small or insular countries (a), medium size countries (b),or large countries or archipelagos (c). These subsets of countries have been identified using the RAK fits as large countries have an exponent $>2$, the medium size countries have been selected between those with exponent between 0 and 1 and the small countries have exponent $<-3$. In the smaller countries we observe an exponential-like cutoff, while in the larger countries the curves are more stretched and suggest the existence of multiple scales.
This difference leads to a wide variance in the fit parameters estimated for the $P(L)$ in panels (d) and (e), where we represent the the scale and exponent parameter for the Truncated Power Law and Random Acceleration Kicks models respectively. We include the curves of Figure 1 for USA --Cuebiq (cb), Mobile (mb)~\cite{gonzalez2008understanding}, Google (gg), Safegraph (sg), Census (cs) -- and Italy (that, as illustrated in Figure~\ref{fig1} includes also Octotelematics and Facebook (fb) data). The values for the longitudinal weekly analysis of Cuebiq  and Facebook  data, and the values estimated for the Google and Facebook baseline. We observe how all data points align around a curve, which suggests strong correlation between the free parameter and, ultimately, the existence of an underlying law having a single degree of freedom. 
 }
\label{ccdf_comparison_google}
\end{figure*}

\begin{figure*}[ht!]
\renewcommand{\figurename}{Supplementary Fig.}

\begin{center}
\begin{tabular}{cc}
\raisebox{2.3cm}{(a)} \includegraphics[angle=0,width=0.45\textwidth]{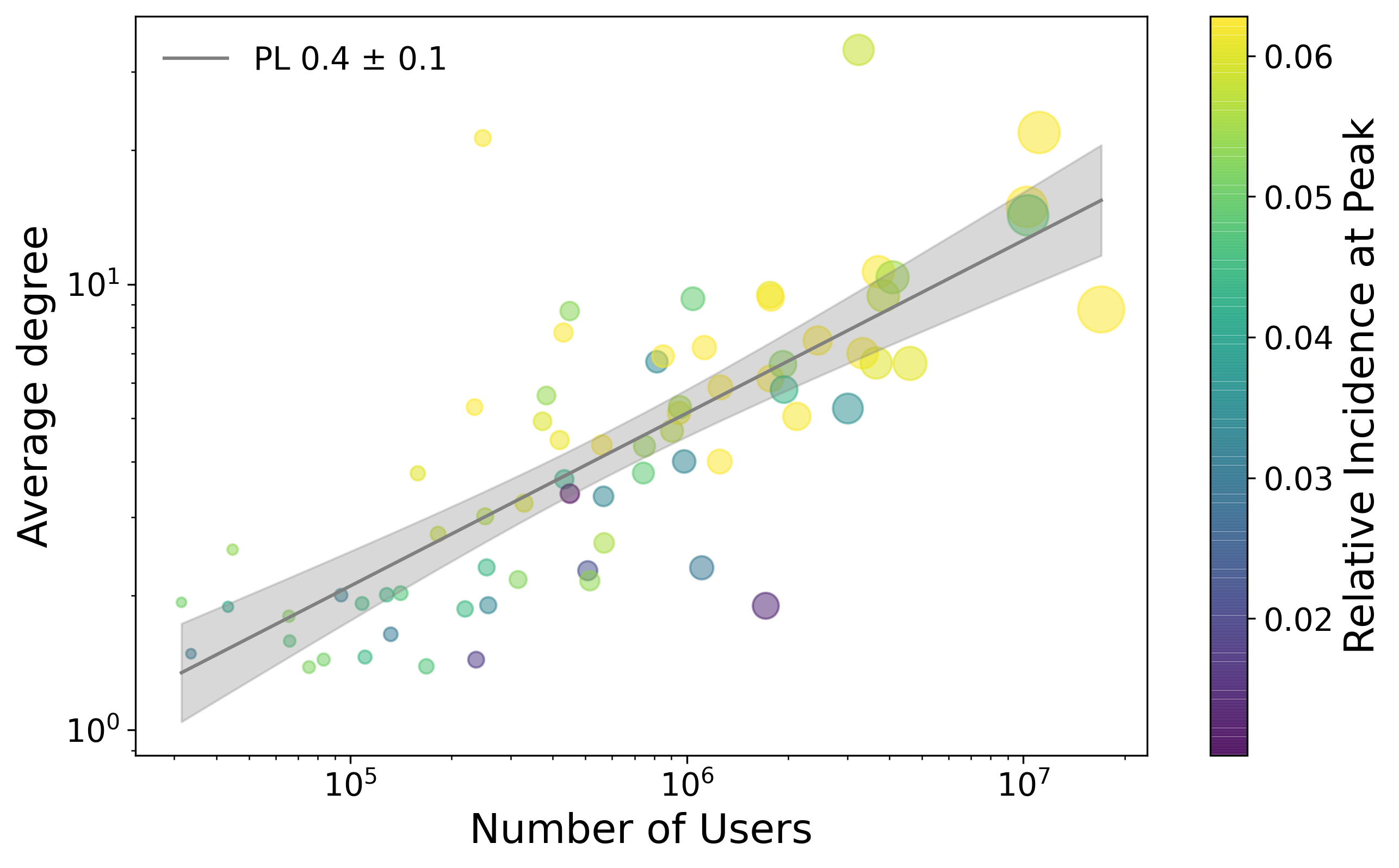} &
\raisebox{2.3cm}{(b)} \includegraphics[angle=0,width=0.45\textwidth]{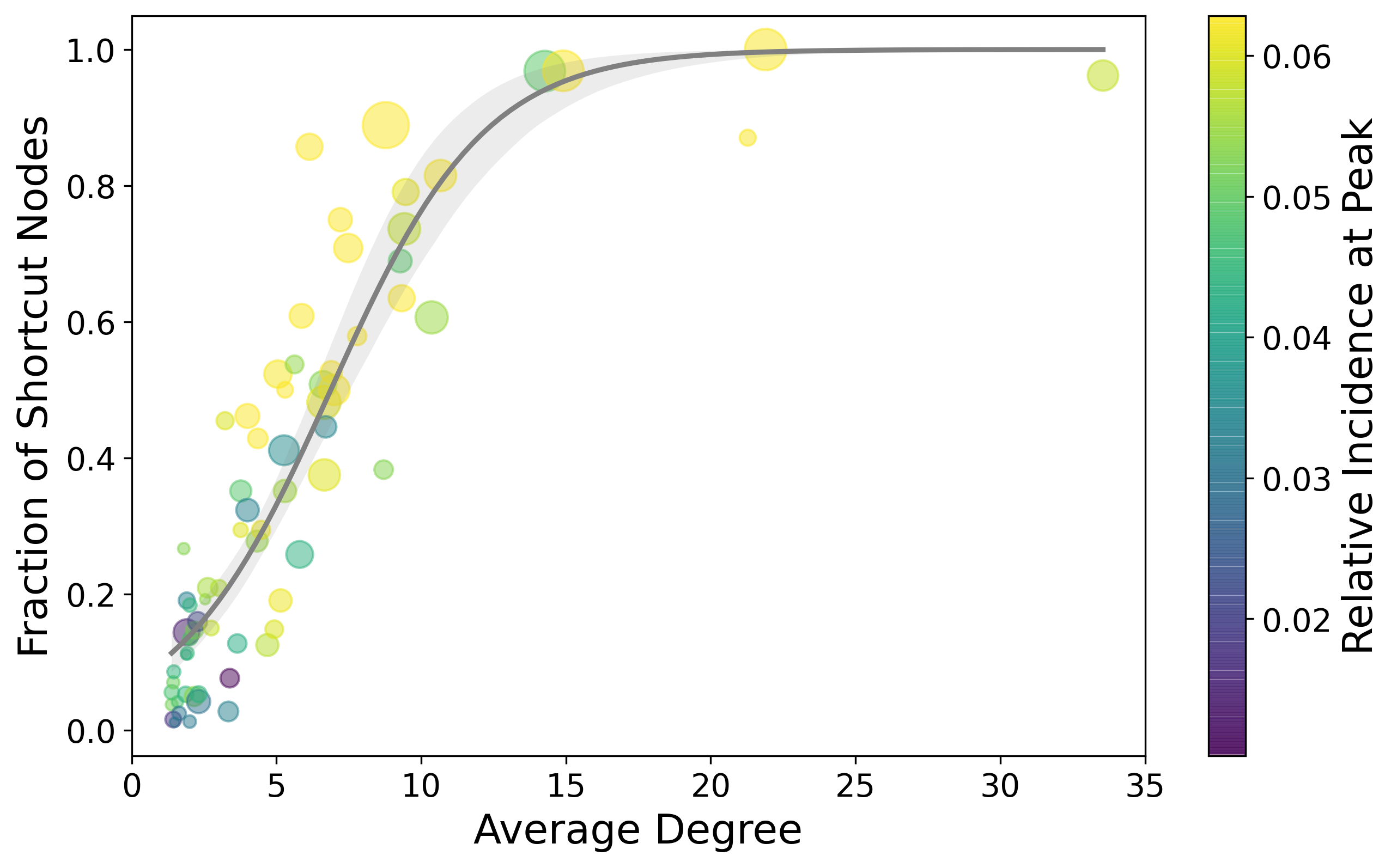} \\
\end{tabular}
\begin{tabular}{c}
\raisebox{2.3cm}{(c)} \includegraphics[angle=0,width=0.45\textwidth]{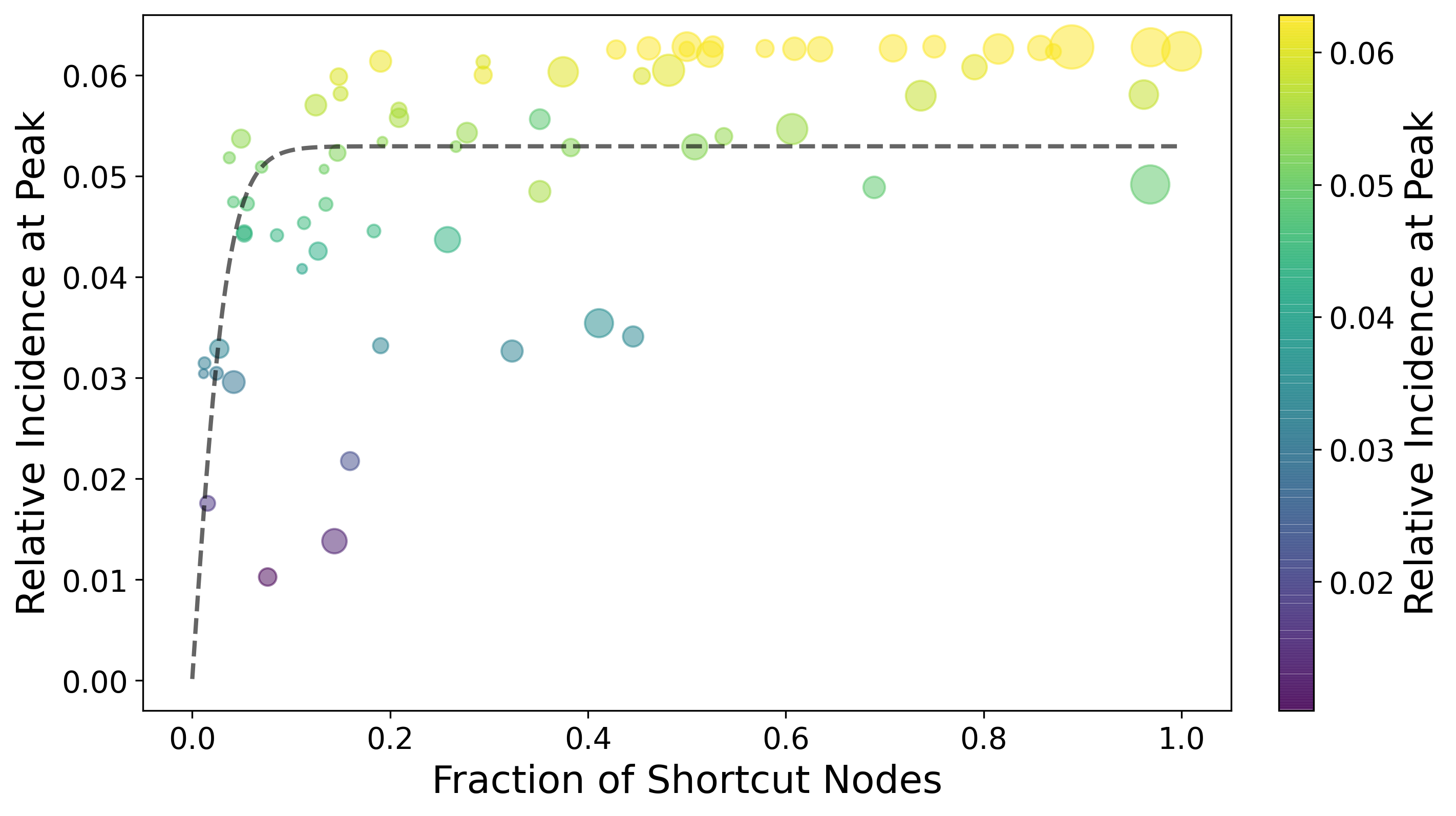}
\end{tabular}
\end{center}
\caption{
{\bf Small world effect in epidemic diffusion.} (a) Focusing on a single dataset, the Facebook baseline mobility, we observe how the average number of edges per node grows as the number of users increases (the fit suggest a scaling sub-linear behaviour). This is most likely due to a pruning procedure that removes edges with smaller flows for both privacy enhancing and disk space reduction. However, introducing a threshold to minimal flow (we estimate is about 10 users/8 hours for these data) systematically cuts longer edges where movement costs are higher, and in particular between marginal areas. 
(b) This reflects into a great change into the shape of the network. Networks with higher average degree has a larger fraction of shortcut nodes (where shortcuts are defined as links at least 2 times as long as the average distance between a node and the closest neighbour). The fit here suggests a logistic behaviour.
(c) The fraction of shortcut nodes, in turns, determines the behaviour of spreading processes on the network, here again we illustrate it over the Facebook baseline mobility networks. As we did also in Fig.~\ref{fig4} and Supplementary Fig.~\ref{fig_epi_scale}, we simulate the spreading of epidemics using a SIR model starting from a single seed in the node with the larger userbase. We see how when the fraction of shortcut nodes passes a relatively small threshold, we have consistently a faster spreading and thus a higher incidence at peak. 
}
\label{small_world_pruning}
\end{figure*}

% Gravity Pruning
\begin{figure*}[ht!]
\renewcommand{\figurename}{Supplementary Fig.}

\begin{center}
\begin{tabular}{c}
\includegraphics[angle=0,width=0.75\textwidth]{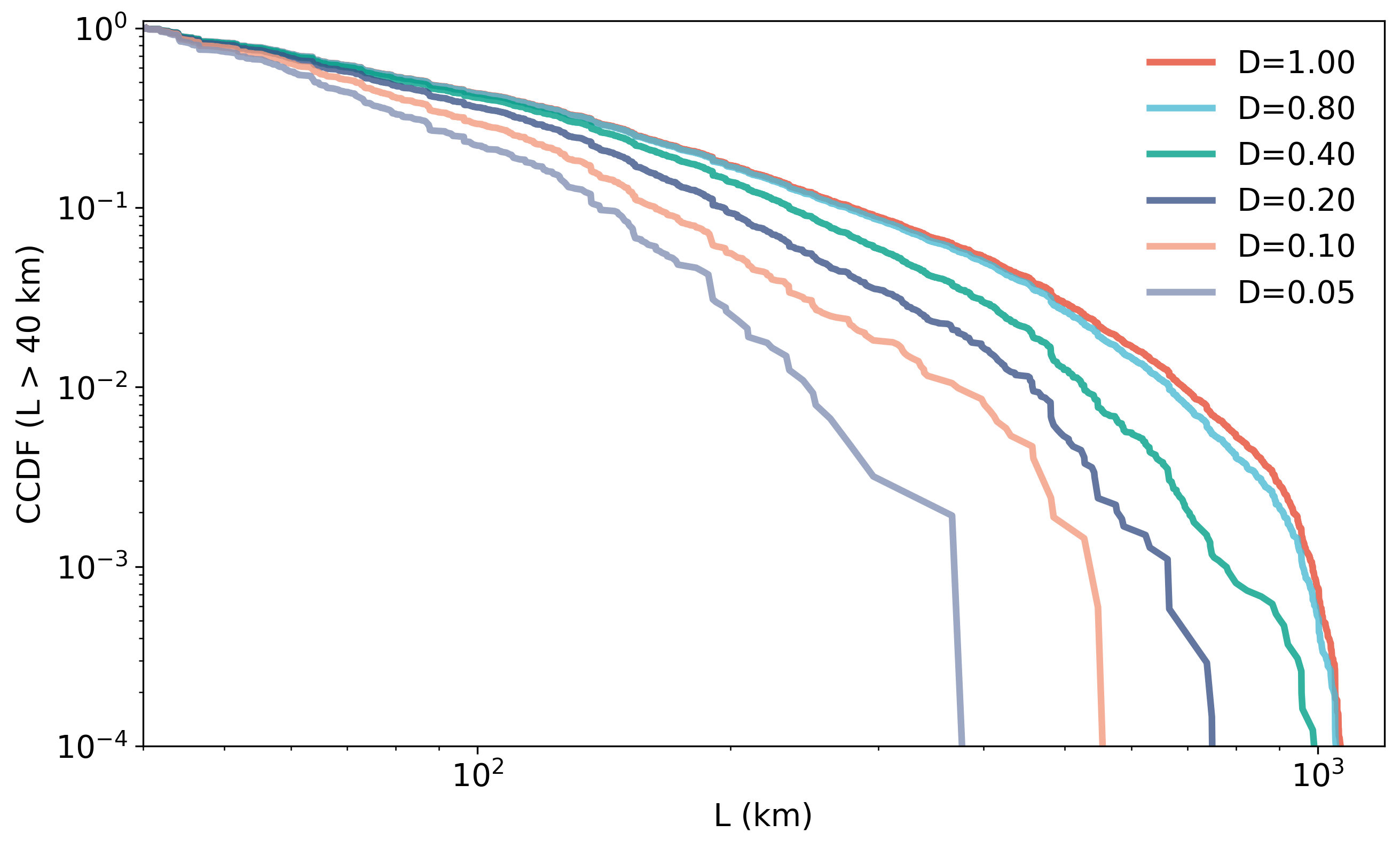} \\
\end{tabular}
\end{center}
\caption{
{\bf Displacement distribution of a pruned gravity-generated network.} 
Starting with resident population of Italy at province level, we generated the flow with a toy Gravity model as $\phi_{i,j} \propto P_i P_j L_{i,j}^{-2}$ where $P_i$ is the population at node $i$ and $L_{i,j}$ the distance between nodes $i$ and $j$. Then, we progressively removed edges having the smallest flows until reaching an edge density $D$ indicated in the figure legend. This process strongly influences the observed $P(L)$ when $D<0.8$. Now, as seen in Table~\ref{table_networks}, empirical edge densities are lower than $0.35$ in all three Italian datasets, suggesting that a significant part of long flows might be missing in the data captured. 
}
\label{figSI_gravity_pruning}
\end{figure*}

% SMALL WORLD
\begin{figure*}[ht!]
\renewcommand{\figurename}{Supplementary Fig.}

\begin{center}
\begin{tabular}{cc}
\raisebox{2.3cm}{(a)} \includegraphics[angle=0,width=0.45\textwidth]{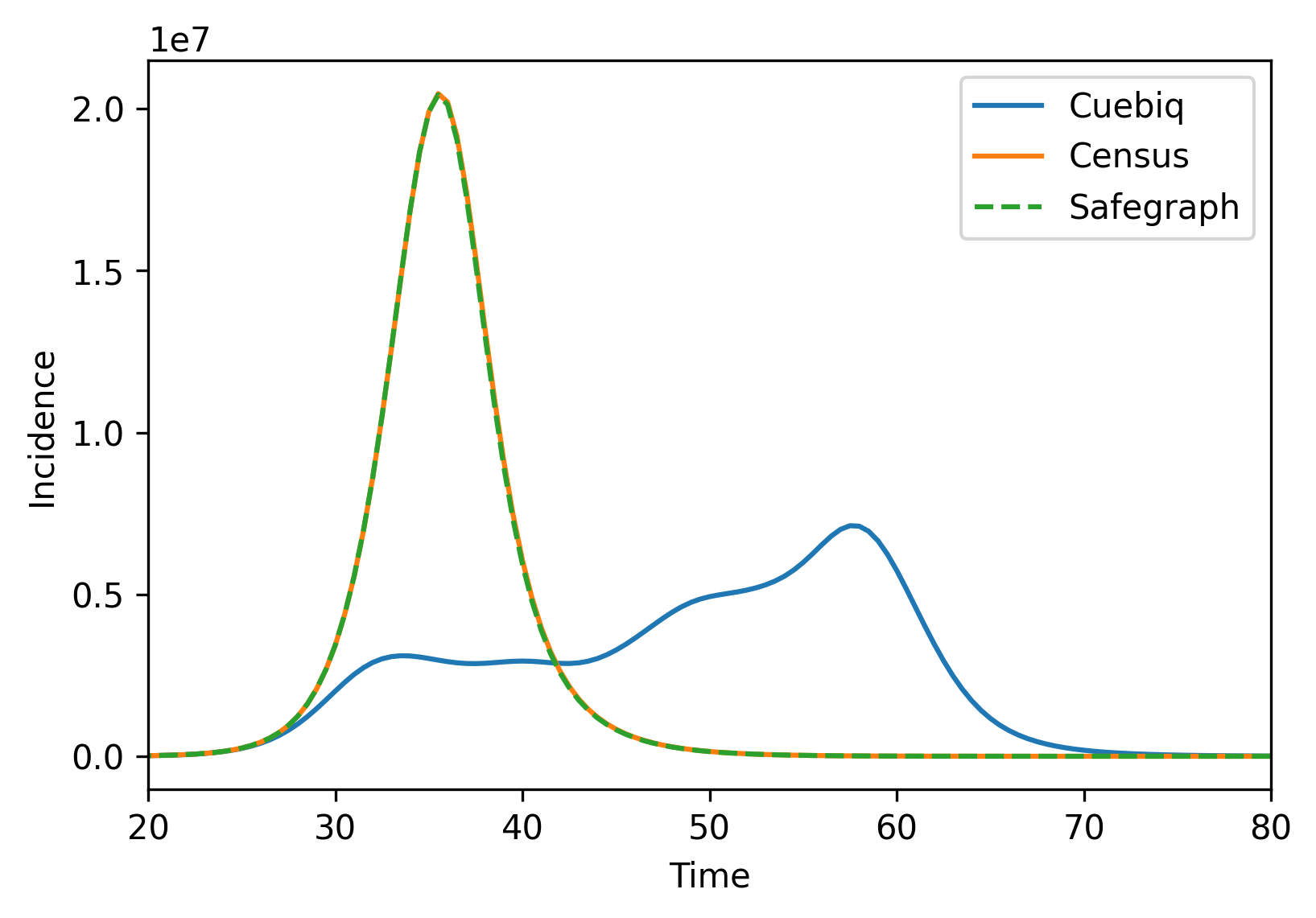} &
\raisebox{2.3cm}{(b)} \includegraphics[angle=0,width=0.45\textwidth]{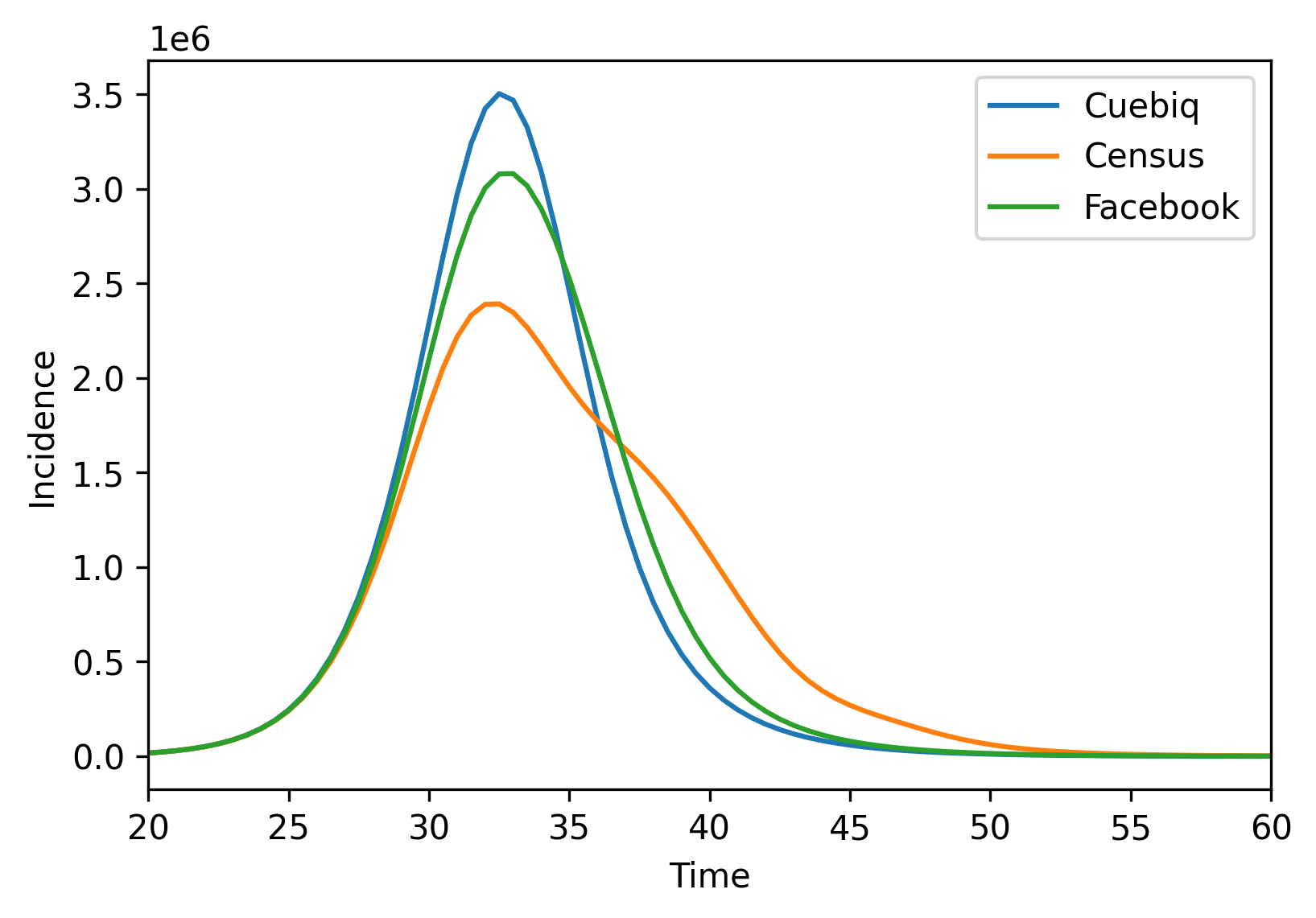} \\
\end{tabular}
\end{center}
\caption{
{\bf Effect of different datasets on spreading dynamics.} Here we represent the differences between the spreading dynamics in U.S.A. and Italy using the SIR model described in the paper. We present the evolution over time of the incidence, that accounts for the new infected in that unit time. In panel (a), we see how for U.S.A. Safegraph and Census (solid orange and dashed green lines), that as we discussed are characterized by an significantly larger weight of long range flows, yield exactly the same result that likely is equivalent to homogeneous mixing. Since Cuebiq data does not display this over-abundance of long range connections, the epidemic diffusion is mostly local (see Supplementary Fig.~\ref{figSI_peak_map}) and produces a more complex behaviour where the incidence curve has two peaks. In panel (b), Italian data display instead smaller differences, probably because of the smaller spatial scale of the Italian peninsula with respect to the spatial scales at hand. Here the peak appear not delayed but rather lowered for Facebook and Census data, that are in this case characterized by an under-representation of long-range connections.
}
\label{fig_baseline_epi}
\end{figure*}

% Spreading USA
\begin{figure*}[ht!]
\renewcommand{\figurename}{Supplementary Fig.}

\begin{center}
\begin{tabular}{c}
\includegraphics[angle=0,width=0.9\textwidth]{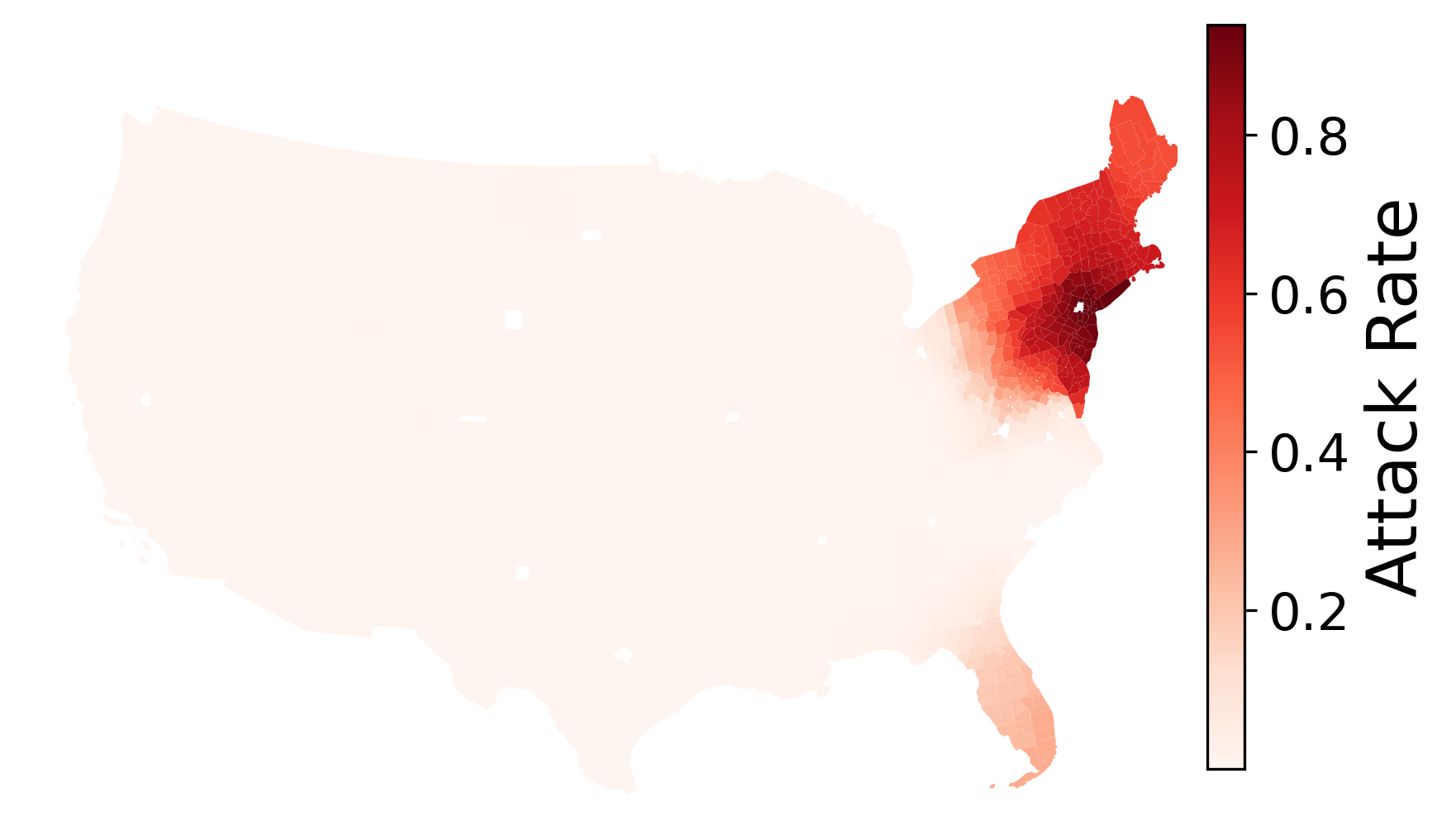} \\
\end{tabular}
\end{center}
\caption{
{\bf Mapping of the attack rate across U.S.A. at the first epidemic peak simulated with Cuebiq data.} The map represents the epidemic state in correspondence of the first peak in the U.S.A. Cuebiq simulation (day 40). The epidemics, starting in Manhattan (NY), appear geographically contained and spreading only locally. On the other hand, the equivalent Safegraph and Census plot would show a map of the U.S.A. homogeneously red. 
}
\label{figSI_peak_map}
\end{figure*}

\begin{figure*}[ht!]
\renewcommand{\figurename}{Supplementary Fig.}

\begin{center}
\begin{tabular}{cc}
\includegraphics[angle=0,width=0.8\textwidth]{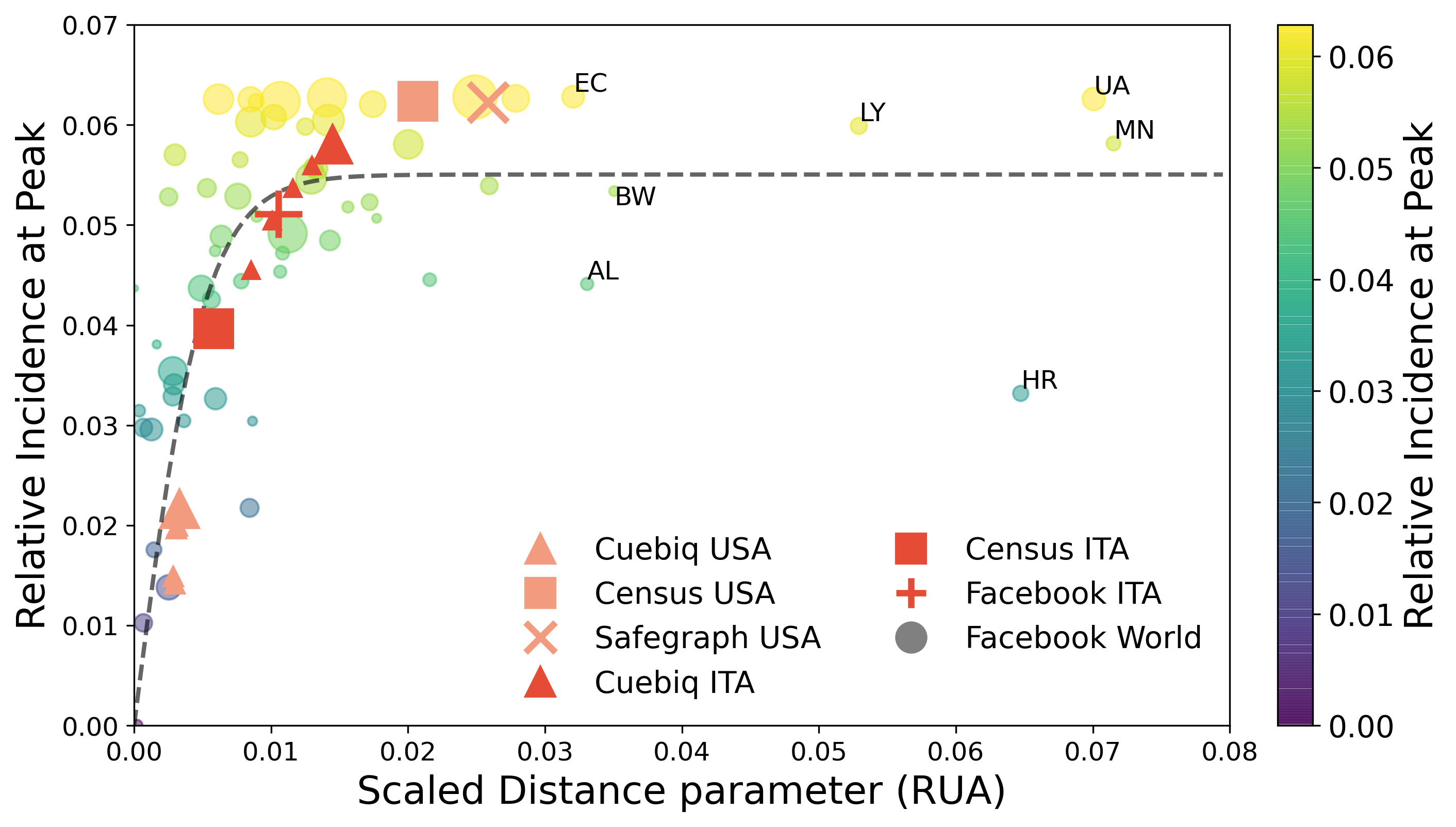} \\
\end{tabular}
\end{center}
\caption{
{\bf Understanding the small world effect in epidemic diffusion by fitting the $P(L)$} The relative incidence at peak, that is the largest fraction of population that become infected in a single day during the epidemic simulation, can be reconnected to the property of the $P(L)$.
We use here the scale parameter of the $P(L)$ estimated using the RUA model, that being characterized by a single parameters yields the more stable estimate for the spatial scale (see Supplementary Fig.~\ref{figSI_google_facebook_parameters}). This relative scale parameter is, for each country, obtained by dividing the RUA scale parameter by the maximal length of the continental landmass, estimated as the longest side of a minimum rotated rectangle shape of the continental outline of the country). We analyse in this plot not only the Facebook baseline, but also the networks illustrated in Figure 1 and observe clearly how to larger relative values of the relative RUA scale, that we can associated to a larger tendency to small world behavior and, similarly to what observed in Supplementary Fig.~\ref{small_world_pruning} (c), we have that larger scale mobility dictates significantly larger epidemic effects. At the same time, as observed in Fig.~\ref{fig4}, we also observe great variation in the incidence at peak in the same country depending on the different source of data used.
}
\label{fig_epi_scale}
\end{figure*}

% GRANULARITY
\begin{figure*}[ht!]
\renewcommand{\figurename}{Supplementary Fig.}

\begin{center}
\begin{tabular}{cc}
\raisebox{2.3cm}{(a)} \includegraphics[angle=0,width=0.45\textwidth]{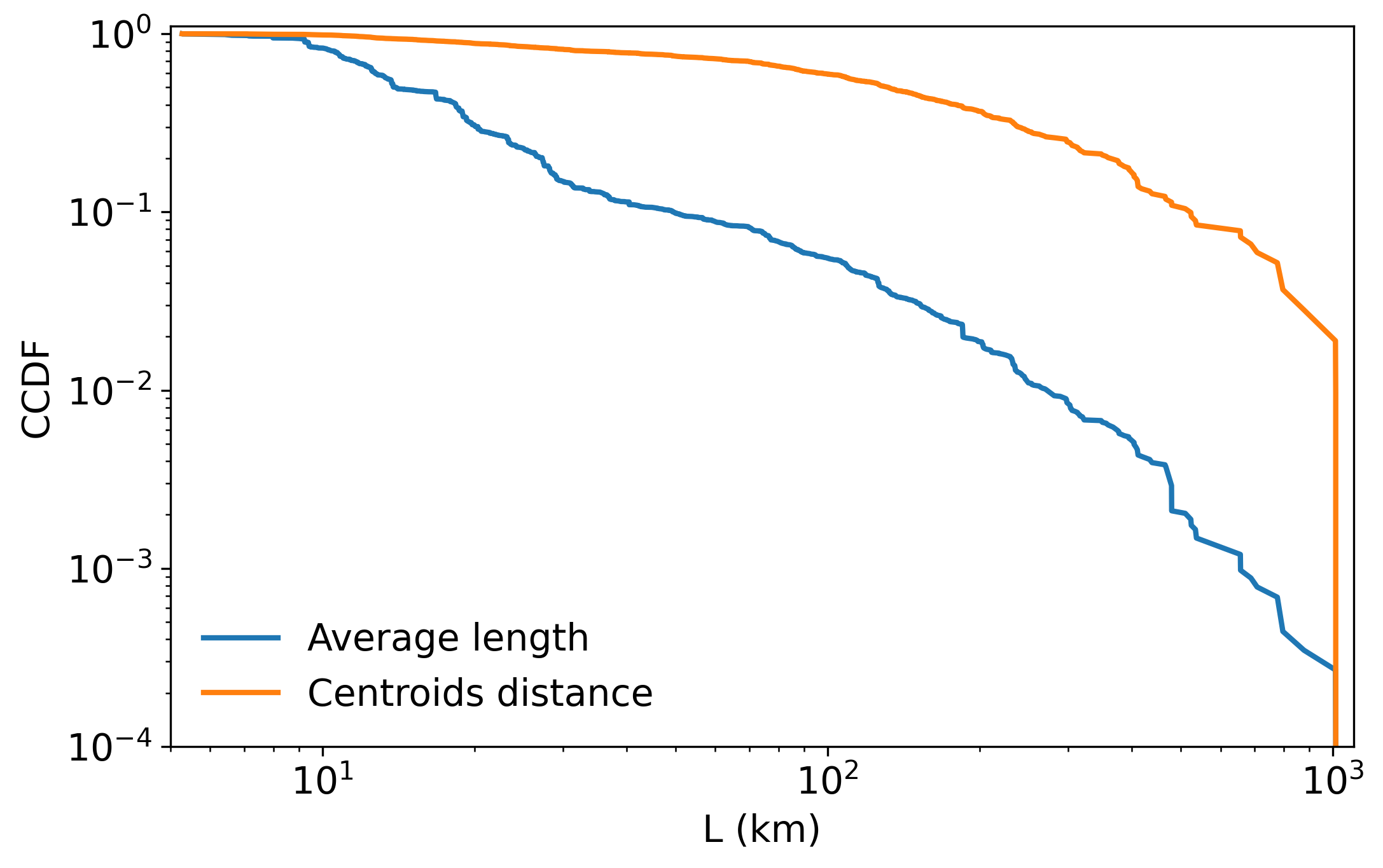} &
\raisebox{2.3cm}{(b)} \includegraphics[angle=0,width=0.45\textwidth]{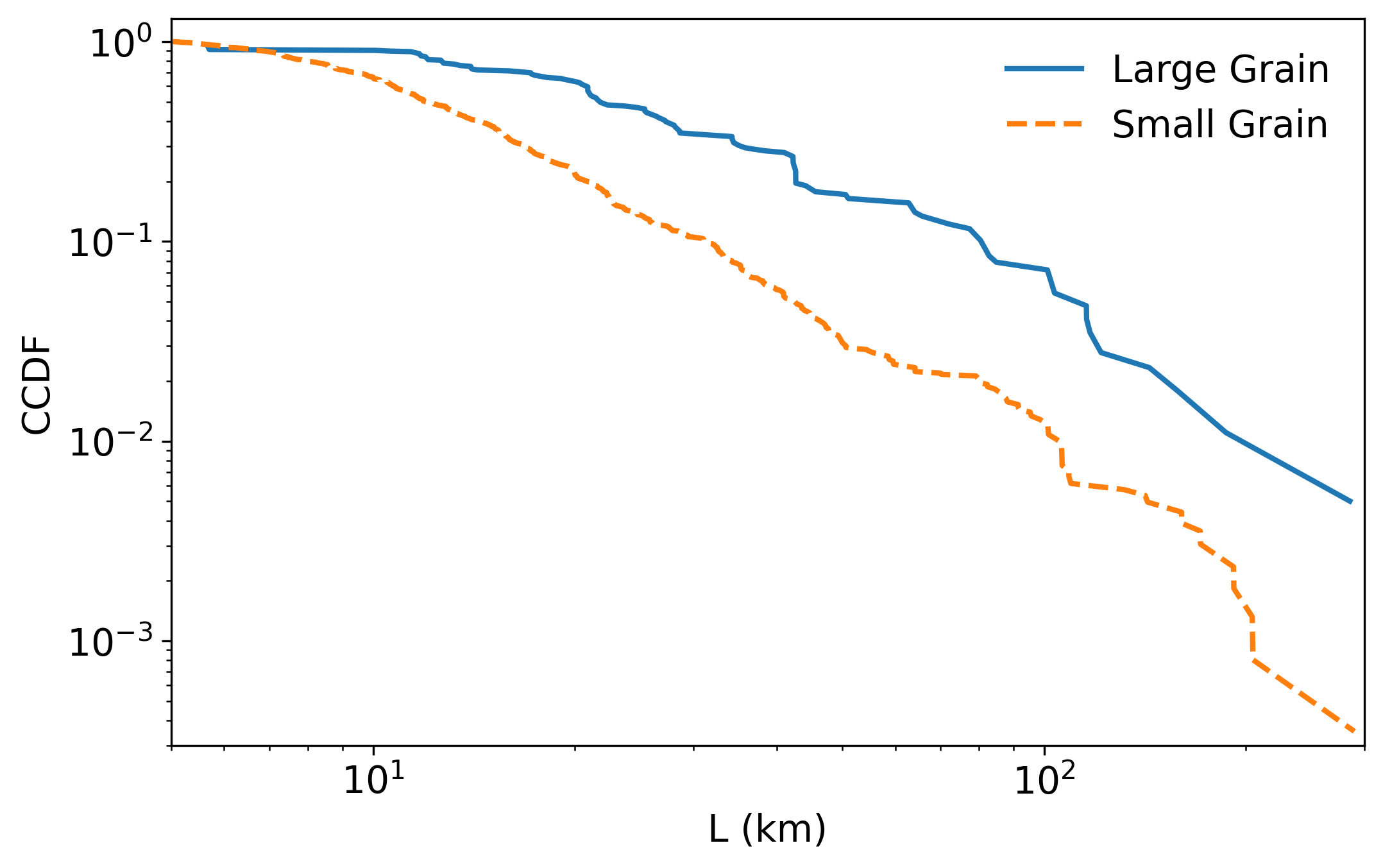} \\
\end{tabular}
\end{center}
\caption{
{\bf Analysis of the effects of granularity on Facebook baseline data.} {\bf a)} the differences in the $P(L)$ distribution function in Italy changes if one consider as displacement lengths the distance between the centroids of the administrative areas or the average length indicated by Facebook. From this graph, we can deduce that the real displacements are largerly smaller than what estimated using the centroid distance.
{\bf b)} the differences in the $P(L)$ distribution function in Hungary changes if one consider two different areal units lengths. The Hungarian datasets comes replicated with two different granularities, one consistent with the other countries (here indicated as Large grain), having 20 nodes, another with a smaller areal unit, having 821 nodes. We observe a wide difference between the displacement distribution estimated using the average lengths over these two datasets, with the larger grain dataset yielding larger distances estimates over the same dataset.
}
\label{figSI_granularity}
\end{figure*}

% FITS ARE GOOD
\begin{figure*}[ht!]
\renewcommand{\figurename}{Supplementary Fig.}

\begin{center}
\begin{tabular}{ccc}
\raisebox{2.3cm}{(a)} \includegraphics[angle=0,width=0.3\textwidth]{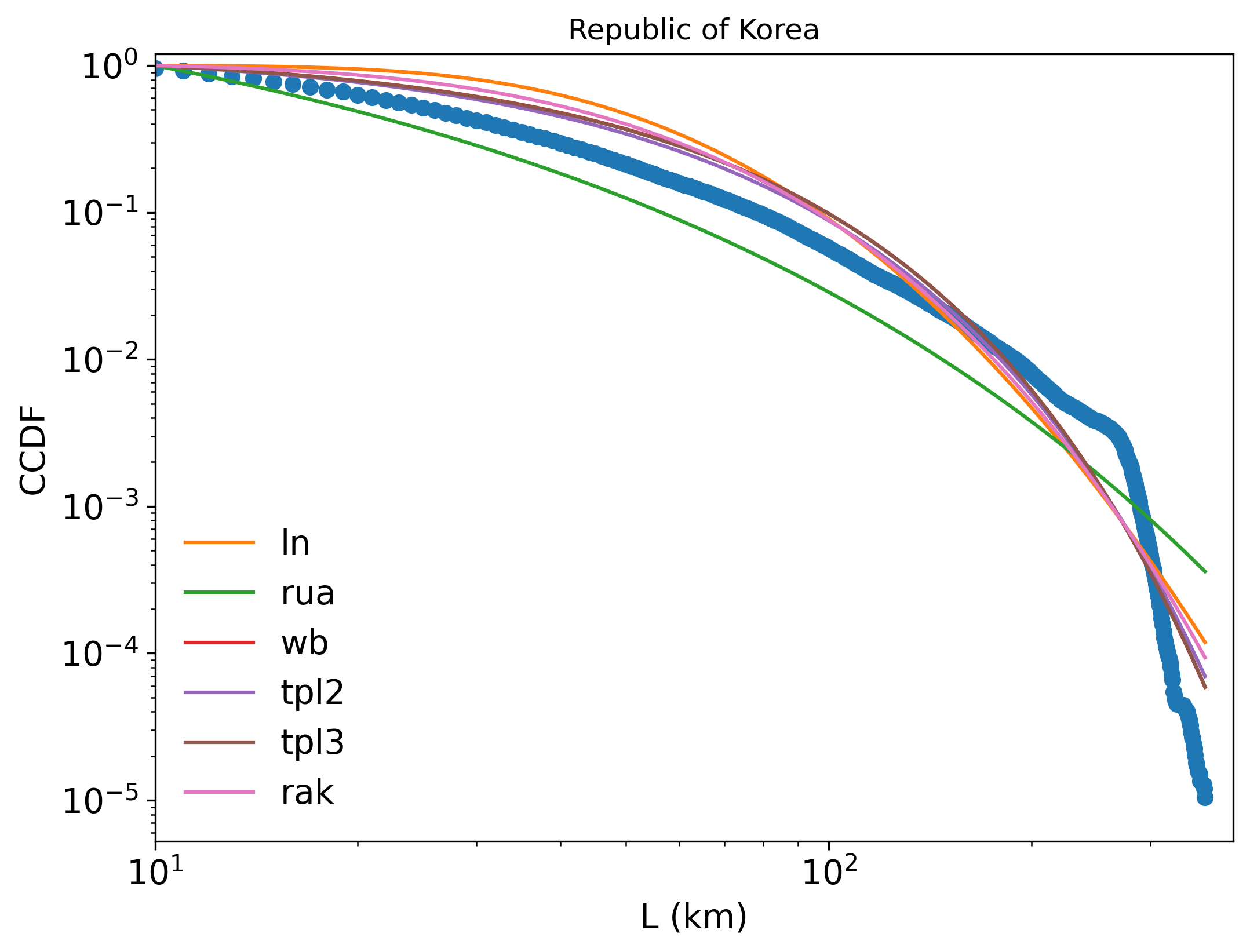} &
\raisebox{2.3cm}{(b)} \includegraphics[angle=0,width=0.3\textwidth]{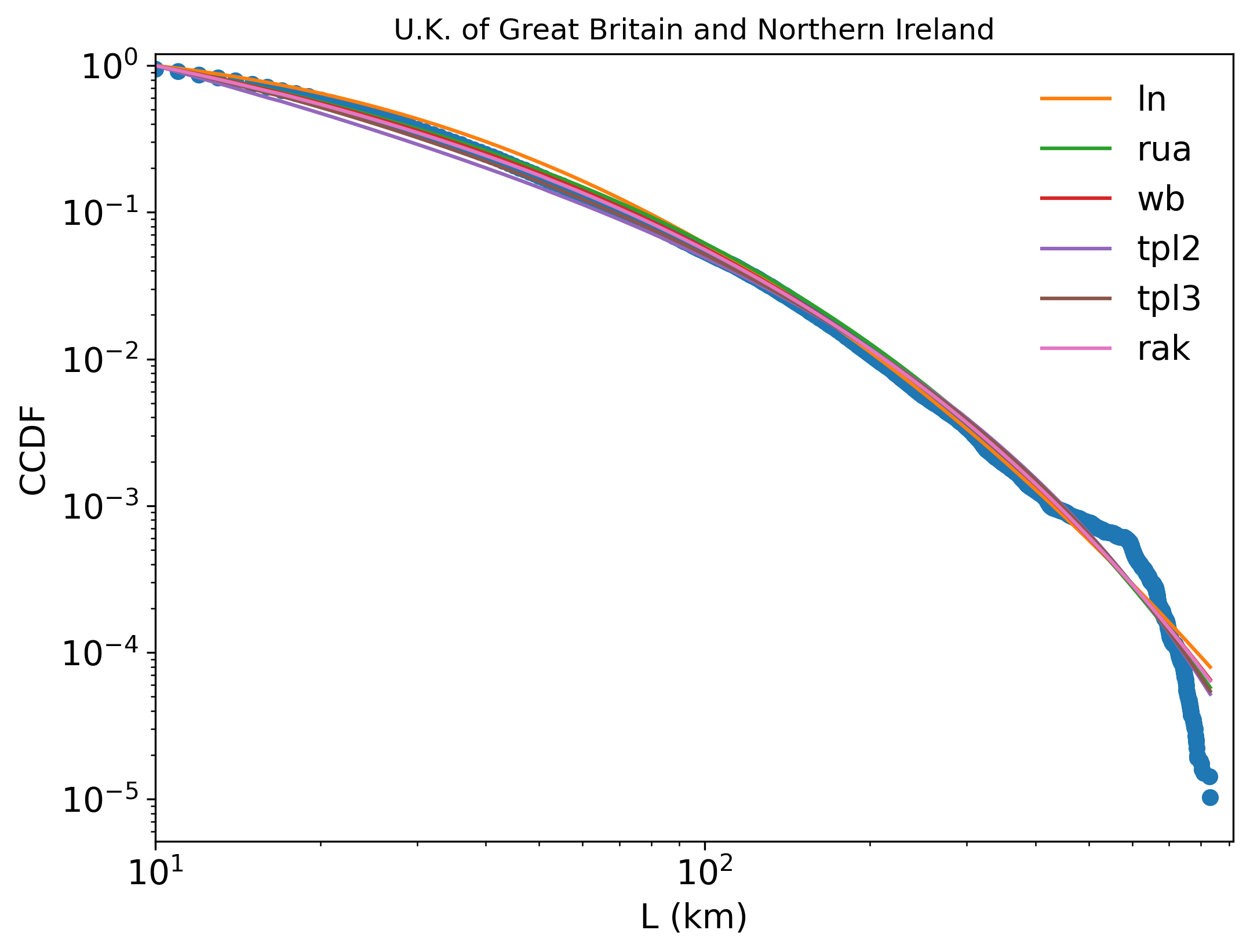}&
\raisebox{2.3cm}{(c)} \includegraphics[angle=0,width=0.3\textwidth]{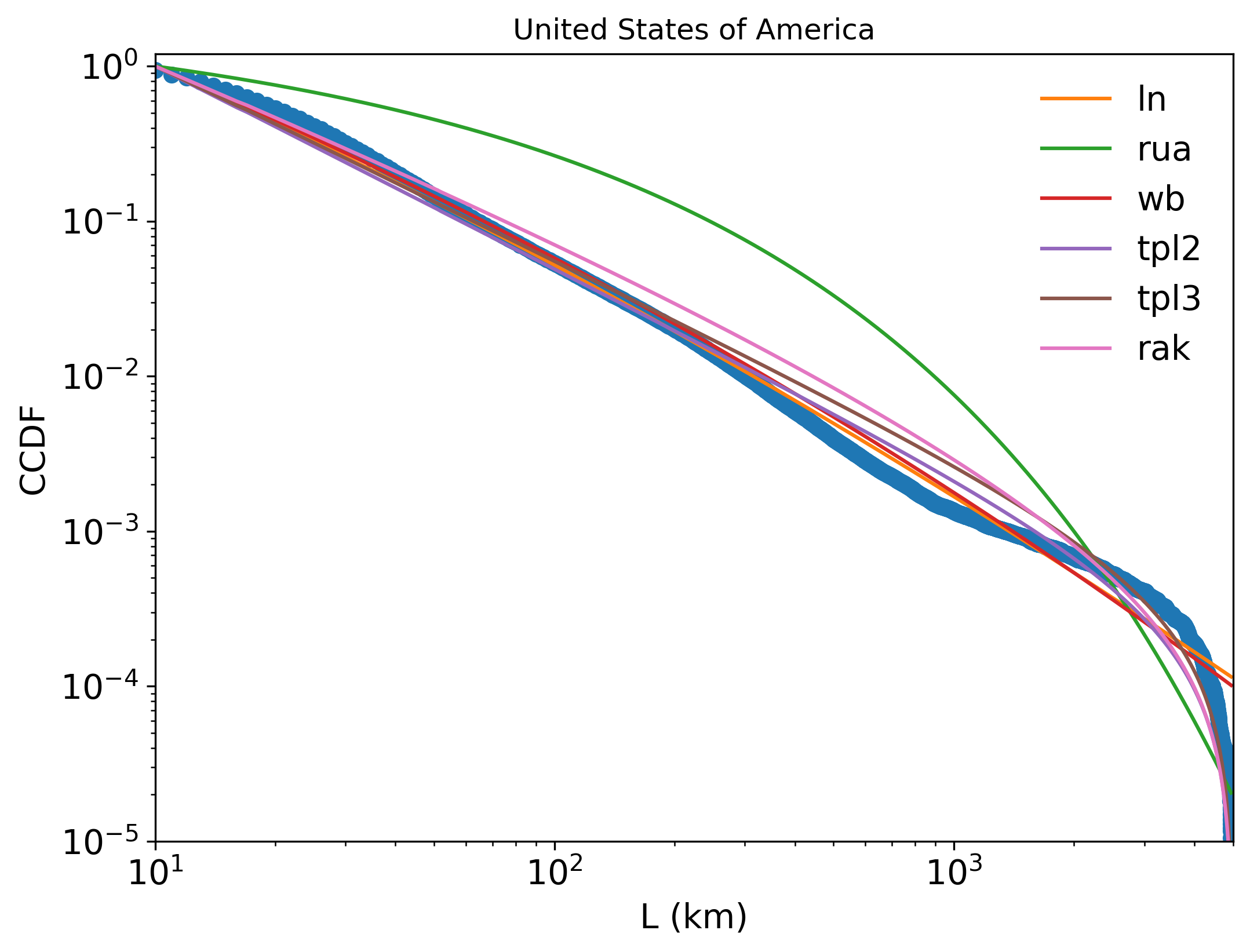}\\
\raisebox{2.3cm}{(d)} \includegraphics[angle=0,width=0.3\textwidth]{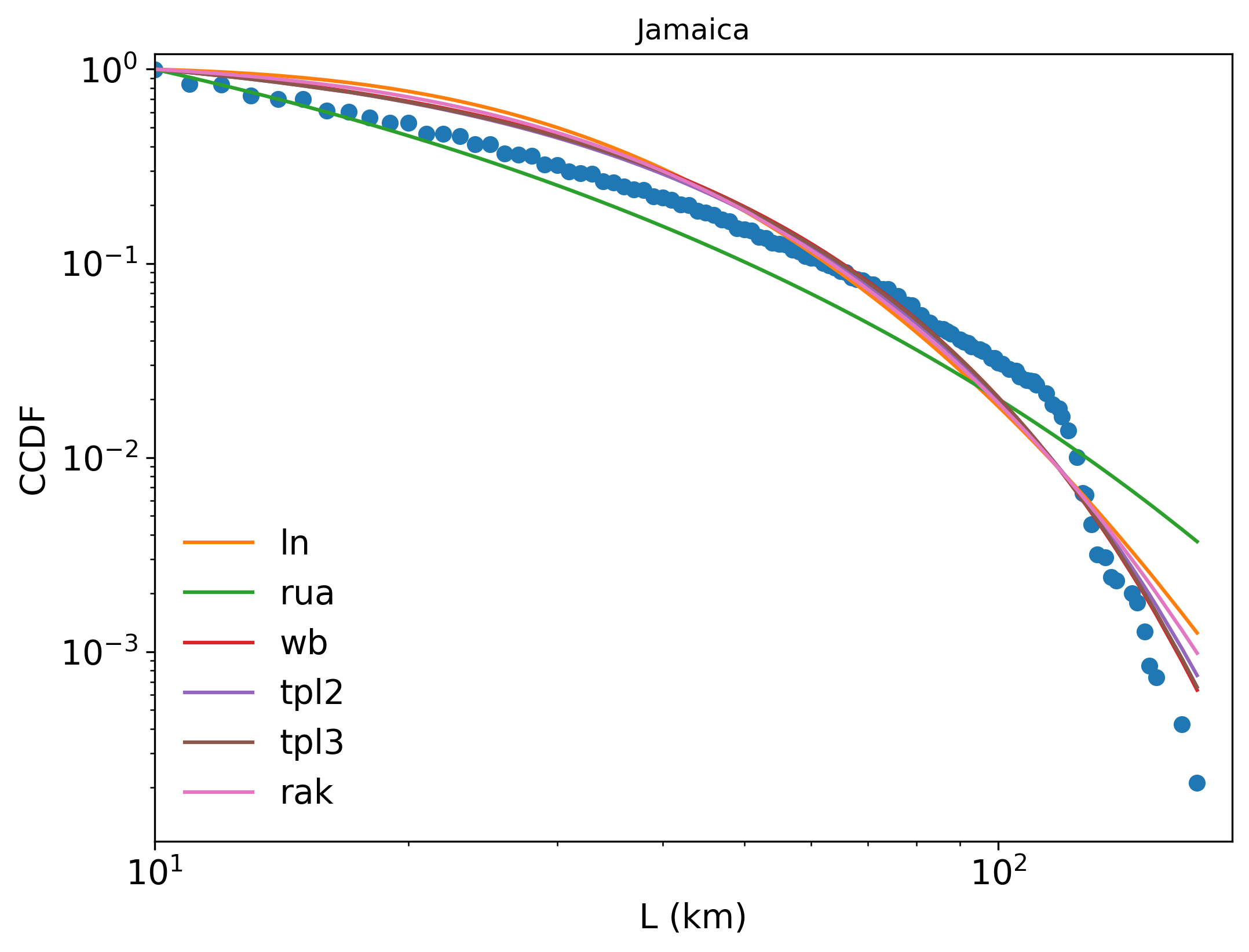}&
\raisebox{2.3cm}{(e)} \includegraphics[angle=0,width=0.3\textwidth]{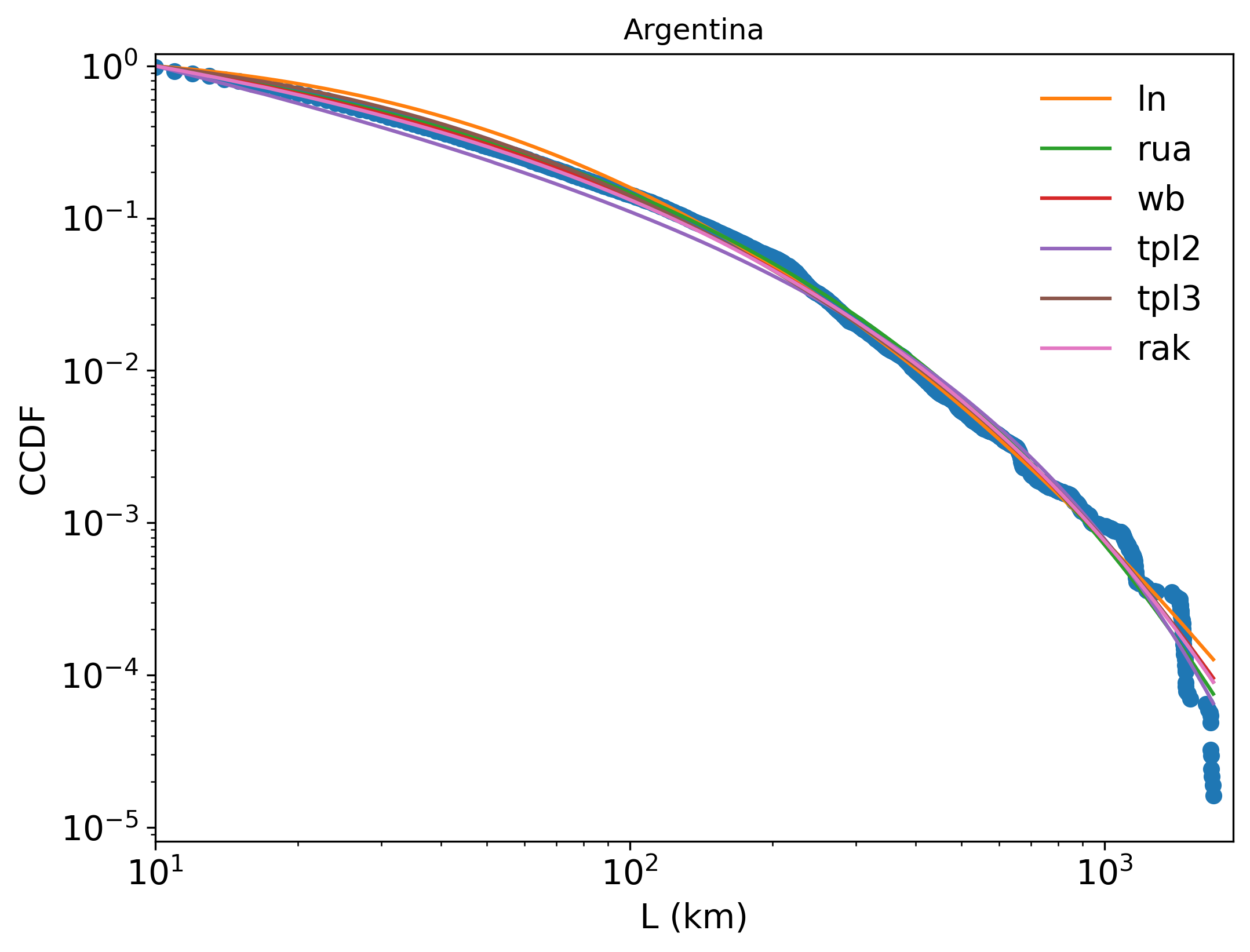}&
\raisebox{2.3cm}{(f)} \includegraphics[angle=0,width=0.3\textwidth]{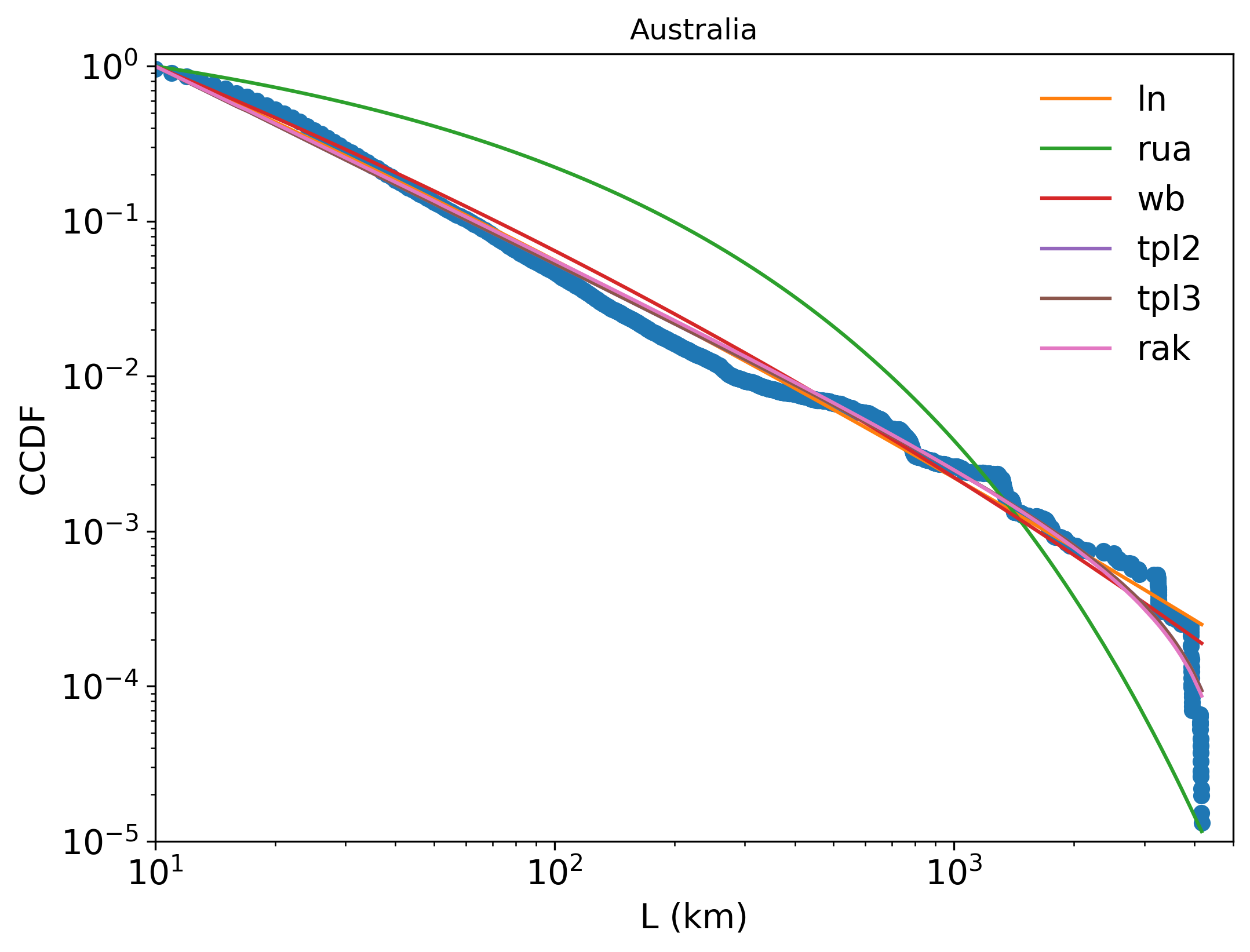}\\
\end{tabular}
\end{center}
\caption{
{\bf Visual comparison of baseline flow maps selecting a fixed number of high flow edges.} We illustrate here the alternative fits executed on Google data using the models described in the Methods. We can appreciate how the fits appear to be well converged. For mid-size countries (UK and Argentina), the alternative fit functions give very similar results. For smaller (Republic of Korea, Jamaica) or larger (U.S.A. or Austraila) countries, the observed behavior diverges from what it can be fit with the models proposed, especially for the monoparametric RUA fit.
}
\label{figSI_fits_are_good}
\end{figure*}

\clearpage

\begin{center}
\renewcommand{\tablename}{Supplementary Table}

\begin{table}
\tiny{
\begin{tabular}{c|ccccccccccc}
Provider 							
&	Source				& 	Area				    &	Elaboration		&  Precision		&	Stop time 		&	NW Aggregation	 	    &	Self-loops		&	Pruning		&	Availability	\\
\hline
Tel\'efonica\cite{schlosser2020covid}	
&	Mobile Phone			&	Germany			    &	Authors			& Cell tower		& 	15 min		    &	NUTS3 / weekly	 	        &	Yes			&	$\phi < 5$		&	Restricted		\\
Facebook~\cite{bonaccorsi2020economic,galeazzi2021human}				
& 	Social Media Access		&	Several countries	&	Company			& GPS			        &	?			    &	Administrative / 8 hours	&	Yes			&	Yes			&	Data4good	\\
Orange~\cite{pullano2020evaluating}	
&	Mobile Phone			&	France			    &	Authors			& Cell tower		& 	1 hour		    &	Municipality / weekly	 	&	Yes			&	?			&	Restricted		\\
Baidu~\cite{chinazzi2020effect,kraemer2020effect}	
&	Location Based Service	&	China               &	Company			& ?			        & 	?			    &	Province / daily	 		&	Yes			&	?			&	Restricted		\\
Baidu~\cite{rader2020crowding}	
&	Location Based Service	&	China               &	Company			& ?			        & 	?			    &	Prefecture / daily	 		&	Yes			&	?			&	Restricted		\\
Cuebiq HDR~\cite{pepe2020covid}	
&	Mobile Apps			    &	Italy				&	Authors			& GPS			    & 	5 min	        &	Province / daily	 		&	Yes			&	No	&	Public	\\
Tel\'efonica~\cite{Oidtman:2021bp} 
&	Mobile Phone			&	Colombia		    &	Authors			& Cell tower		& 	day		        &	Department / daily	 	    &	Yes			&	No		&	Restricted		\\
Google~\cite{rader2020crowding} 
& Location History          & Worldwide             &   Company           & 5km$^2$ area      & ?                 &   Municipality / weekly       &  (Yes)        &   Yes    & Restricted \\
Google~\cite{rader2020crowding} 
& Location History          & Italy                 &   Company           & 5km$^2$ area      & ?                 &   Province / weekly           &  (Yes)        &   Yes & Restricted \\ 
Google~\cite{Lemey:2021ch} 
& Location History          & Europe                &   Company           & (5km$^2$ area)    & ?                 &   Country / beweekly          &  (Yes)        &   Yes    & Restricted \\
Google~\cite{bassolas2019hierarchical,aguilar2020impact, Hazarie:2021jc} 
& Location History          & Intra-Urban           &   Company           & (5km$^2$ area)    & ?                 &   Weekly                      &  Yes          &   Yes    & Restricted \\
Safegraph~\cite{chang2021mobility}	
& Mobile Apps 			&  USA Metro areas  & Authors		& PoI 	& ?		&	Census Block/Weekly		& ?		& Yes 	& Data4Good \\

Safegraph~\cite{kang2020multiscale}	
& Mobile Apps 			&  USA  & Authors		& PoI 	& ?		&	Census Block/Daily		& Yes		& Yes 	& Public \\
\end{tabular}
}
\caption{{\bf Characteristics of the datasets used for modeling the epidemic spreding of COVID-19.} Note: the Google dataset is in principle also part of a Data4good initiative and accessible upon request, however we started the request procedure more than a year and a half ago and our demand is still ``under consideration''. Information between parenthesis has been found in a different paper on the same dataset.}
\end{table}
\end{center}

\end{document}